\RequirePackage{lineno} 
\documentclass[aps,prd,showpacs,groupedaddress,superscriptaddress,floatfix,amsmath,twocolumn]{revtex4}

\usepackage{graphicx} 
\usepackage{dcolumn} 
\usepackage{bm} 
\usepackage{epstopdf}
\usepackage{slashed}

\newcommand{\met}{\ensuremath{{\slash\kern-.7emE}_{T}}}
\newcommand{\vmet}{\ensuremath{\vec{\slash\kern-.7emE}_{T}}}

\newcommand{\mt}{\ensuremath{m_T}}

\newcommand{\uT}{\ensuremath{\vec{u}_T}}
\newcommand{\upara}{\ensuremath{u_{\parallel}}}
\newcommand{\uperp}{\ensuremath{u_{\perp}}}

\newcommand{\wen}{\ensuremath{W \rightarrow e \nu}}

\newcommand{\zee}{\ensuremath{Z \rightarrow ee}}

\newcommand{\ppbar}{\ensuremath{p\overline{p}}}

\newcommand{\pte}{\ensuremath{p_T^e}}
\newcommand{\ptnu}{\ensuremath{p_T^\nu}}

\newcounter{appendix}
\def\theappendix{\Alph{appendix}}
\def\appendix#1{
  \clearpage
  \addtocounter{appendix}{1}
  \addcontentsline{toc}{section}{\numberline{\theappendix}{#1}}
  \noindent{\bf\large Appendix \theappendix: #1}}

\newcounter {subsubsubsection}[subsubsection]

\newcommand\T{\rule{0pt}{2.6ex}}       
\newcommand\B{\rule[-1.2ex]{0pt}{0pt}} 

\def\lsim{\mathrel{\rlap{\lower4pt\hbox{\hskip1pt$\sim$}}\raise1pt\hbox{$<$}}}
\newcommand{\metsub}{\ensuremath{{\slash\kern-.5emE}_{T}}}

\begin{document}

\title{Measurement of the $\boldsymbol{W}$ boson mass with the D0 detector}

\hspace{5.2in} \mbox{FERMILAB-PUB-13-489-E}

\affiliation{LAFEX, Centro Brasileiro de Pesquisas F\'{i}sicas, Rio de Janeiro, Brazil}
\affiliation{Universidade do Estado do Rio de Janeiro, Rio de Janeiro, Brazil}
\affiliation{Universidade Federal do ABC, Santo Andr\'e, Brazil}
\affiliation{University of Science and Technology of China, Hefei, People's Republic of China}
\affiliation{Universidad de los Andes, Bogot\'a, Colombia}
\affiliation{Charles University, Faculty of Mathematics and Physics, Center for Particle Physics, Prague, Czech Republic}
\affiliation{Czech Technical University in Prague, Prague, Czech Republic}
\affiliation{Institute of Physics, Academy of Sciences of the Czech Republic, Prague, Czech Republic}
\affiliation{Universidad San Francisco de Quito, Quito, Ecuador}
\affiliation{LPC, Universit\'e Blaise Pascal, CNRS/IN2P3, Clermont, France}
\affiliation{LPSC, Universit\'e Joseph Fourier Grenoble 1, CNRS/IN2P3, Institut National Polytechnique de Grenoble, Grenoble, France}
\affiliation{CPPM, Aix-Marseille Universit\'e, CNRS/IN2P3, Marseille, France}
\affiliation{LAL, Universit\'e Paris-Sud, CNRS/IN2P3, Orsay, France}
\affiliation{LPNHE, Universit\'es Paris VI and VII, CNRS/IN2P3, Paris, France}
\affiliation{CEA, Irfu, SPP, Saclay, France}
\affiliation{IPHC, Universit\'e de Strasbourg, CNRS/IN2P3, Strasbourg, France}
\affiliation{IPNL, Universit\'e Lyon 1, CNRS/IN2P3, Villeurbanne, France and Universit\'e de Lyon, Lyon, France}
\affiliation{III. Physikalisches Institut A, RWTH Aachen University, Aachen, Germany}
\affiliation{Physikalisches Institut, Universit\"at Freiburg, Freiburg, Germany}
\affiliation{II. Physikalisches Institut, Georg-August-Universit\"at G\"ottingen, G\"ottingen, Germany}
\affiliation{Institut f\"ur Physik, Universit\"at Mainz, Mainz, Germany}
\affiliation{Ludwig-Maximilians-Universit\"at M\"unchen, M\"unchen, Germany}
\affiliation{Panjab University, Chandigarh, India}
\affiliation{Delhi University, Delhi, India}
\affiliation{Tata Institute of Fundamental Research, Mumbai, India}
\affiliation{University College Dublin, Dublin, Ireland}
\affiliation{Korea Detector Laboratory, Korea University, Seoul, Korea}
\affiliation{CINVESTAV, Mexico City, Mexico}
\affiliation{Nikhef, Science Park, Amsterdam, the Netherlands}
\affiliation{Radboud University Nijmegen, Nijmegen, the Netherlands}
\affiliation{Joint Institute for Nuclear Research, Dubna, Russia}
\affiliation{Institute for Theoretical and Experimental Physics, Moscow, Russia}
\affiliation{Moscow State University, Moscow, Russia}
\affiliation{Institute for High Energy Physics, Protvino, Russia}
\affiliation{Petersburg Nuclear Physics Institute, St. Petersburg, Russia}
\affiliation{Instituci\'{o} Catalana de Recerca i Estudis Avan\c{c}ats (ICREA) and Institut de F\'{i}sica d'Altes Energies (IFAE), Barcelona, Spain}
\affiliation{Uppsala University, Uppsala, Sweden}
\affiliation{Lancaster University, Lancaster LA1 4YB, United Kingdom}
\affiliation{Imperial College London, London SW7 2AZ, United Kingdom}
\affiliation{The University of Manchester, Manchester M13 9PL, United Kingdom}
\affiliation{University of Arizona, Tucson, Arizona 85721, USA}
\affiliation{University of California Riverside, Riverside, California 92521, USA}
\affiliation{University of Maryland, College Park, Maryland 20742, USA}
\affiliation{Florida State University, Tallahassee, Florida 32306, USA}
\affiliation{Fermi National Accelerator Laboratory, Batavia, Illinois 60510, USA}
\affiliation{University of Illinois at Chicago, Chicago, Illinois 60607, USA}
\affiliation{Northern Illinois University, DeKalb, Illinois 60115, USA}
\affiliation{Northwestern University, Evanston, Illinois 60208, USA}
\affiliation{Indiana University, Bloomington, Indiana 47405, USA}
\affiliation{Purdue University Calumet, Hammond, Indiana 46323, USA}
\affiliation{University of Notre Dame, Notre Dame, Indiana 46556, USA}
\affiliation{Iowa State University, Ames, Iowa 50011, USA}
\affiliation{University of Kansas, Lawrence, Kansas 66045, USA}
\affiliation{Louisiana Tech University, Ruston, Louisiana 71272, USA}
\affiliation{Northeastern University, Boston, Massachusetts 02115, USA}
\affiliation{University of Michigan, Ann Arbor, Michigan 48109, USA}
\affiliation{Michigan State University, East Lansing, Michigan 48824, USA}
\affiliation{University of Mississippi, University, Mississippi 38677, USA}
\affiliation{University of Nebraska, Lincoln, Nebraska 68588, USA}
\affiliation{Rutgers University, Piscataway, New Jersey 08855, USA}
\affiliation{Princeton University, Princeton, New Jersey 08544, USA}
\affiliation{State University of New York, Buffalo, New York 14260, USA}
\affiliation{University of Rochester, Rochester, New York 14627, USA}
\affiliation{State University of New York, Stony Brook, New York 11794, USA}
\affiliation{Brookhaven National Laboratory, Upton, New York 11973, USA}
\affiliation{Langston University, Langston, Oklahoma 73050, USA}
\affiliation{University of Oklahoma, Norman, Oklahoma 73019, USA}
\affiliation{Oklahoma State University, Stillwater, Oklahoma 74078, USA}
\affiliation{Brown University, Providence, Rhode Island 02912, USA}
\affiliation{University of Texas, Arlington, Texas 76019, USA}
\affiliation{Southern Methodist University, Dallas, Texas 75275, USA}
\affiliation{Rice University, Houston, Texas 77005, USA}
\affiliation{University of Virginia, Charlottesville, Virginia 22904, USA}
\affiliation{University of Washington, Seattle, Washington 98195, USA}
\author{V.M.~Abazov} \affiliation{Joint Institute for Nuclear Research, Dubna, Russia}
\author{B.~Abbott} \affiliation{University of Oklahoma, Norman, Oklahoma 73019, USA}
\author{B.S.~Acharya} \affiliation{Tata Institute of Fundamental Research, Mumbai, India}
\author{M.~Adams} \affiliation{University of Illinois at Chicago, Chicago, Illinois 60607, USA}
\author{T.~Adams} \affiliation{Florida State University, Tallahassee, Florida 32306, USA}
\author{J.P.~Agnew} \affiliation{The University of Manchester, Manchester M13 9PL, United Kingdom}
\author{G.D.~Alexeev} \affiliation{Joint Institute for Nuclear Research, Dubna, Russia}
\author{G.~Alkhazov} \affiliation{Petersburg Nuclear Physics Institute, St. Petersburg, Russia}
\author{A.~Alton$^{a}$} \affiliation{University of Michigan, Ann Arbor, Michigan 48109, USA}
\author{T.~Andeen} \affiliation{Northwestern University, Evanston, Illinois 60208, USA}
% ...
%
%
\collaboration{The D0 Collaboration\footnote{with visitors from
$^{a}$Augustana College, Sioux Falls, SD, USA,
$^{b}$The University of Liverpool, Liverpool, UK,
$^{c}$DESY, Hamburg, Germany,
$^{d}$Universidad Michoacana de San Nicolas de Hidalgo, Morelia, Mexico
$^{e}$SLAC, Menlo Park, CA, USA,
$^{f}$University College London, London, UK,
$^{g}$Centro de Investigacion en Computacion - IPN, Mexico City, Mexico,
$^{h}$Universidade Estadual Paulista, S\~ao Paulo, Brazil,
$^{i}$Karlsruher Institut f\"ur Technologie (KIT) - Steinbuch Centre for Computing (SCC),
$^{j}$Office of Science, U.S. Department of Energy, Washington, D.C. 20585, USA,
and
$^{k}$Thomas Jefferson National Accelerator Facility, Newport News, VA 23606, USA.
}} \noaffiliation
\vskip 0.25cm

\date{October 31, 2013}

\begin{abstract}
We give a detailed description of the measurement of the $W$ boson mass, $M_W$, performed on an integrated luminosity of $4.3\,\text{fb}^{-1}$, which is based on similar techniques as used for our previous measurement done on an independent data set of $1\,\text{fb}^{-1}$ of data. The data were collected using the D0 detector at the Fermilab Tevatron Collider.  This data set yields $1.68\times10^6$ \wen\ candidate events.  We measure the mass using the transverse mass, electron transverse momentum, and missing transverse energy distributions.  The $M_W$ measurements using the transverse mass and the electron transverse momentum distributions are the most precise of these three and are combined to give $M_W = 80.367 \pm 0.013\thinspace \text{(stat)} \pm 0.022\thinspace \text{(syst)\ GeV} = 80.367 \pm 0.026\,\text{GeV}$.  When combined with our earlier measurement on $1\,\text{fb}^{-1}$ of data, we obtain  $M_W = 80.375 \pm 0.023\,\text{GeV}$.  
\end{abstract}
\pacs{12.15.-y, 13.38.Be, 14.70.Fm}

\maketitle 

\tableofcontents

\section{Introduction}
\label{sec:intro}

The 1983 observation of the $W$ and $Z$ vector bosons~\cite{CERN83_1, CERN83_2, CERN83_3, CERN83_4} provided important evidence for the electroweak (EW) sector of the standard model (SM) of particle physics~\cite{GSW_1, GSW_2, GSW_3}. Increasingly precise measurements of the vector boson masses, with a precision of $2.1\,\text{MeV}$, corresponding to 2 parts in $10^5$ for the $Z$ boson mass~\cite{LEPEWWG08}, and their properties compiled over the course of the following 30 years have verified the structure of the electroweak theory, which has been further confirmed by the recent discovery of the Higgs boson with mass $125.7\,\text{GeV}$.
.

In the electroweak theory, for a given renormalization scheme~\cite{RenScheme}, there is a well-defined relationship between the EW boson masses, coupling constants, and the other EW parameters arising from radiative corrections. Precise measurements of these observables provide tests of this relationship and constrain the size of additional corrections from unobserved fields. In the {\it on-shell} scheme~\cite{OnShell}, the SM relationship can be written as

\begin{equation}
\label{eq:SMWmassPred}
M_W^2 \left(1-\frac{M_W^2}{M_Z^2}\right) = \left(\frac{\pi\alpha}{\sqrt{2} G_F }\right)(1+\Delta r),
\end{equation}
where $M_W$ and $M_Z$ are the masses of the $W$ and $Z$ bosons, $G_F$ is the Fermi constant, and $\alpha$ is the fine structure constant at zero momentum.  The quantity $\Delta r$ contains all radiative corrections including the running of $\alpha$ and of the SM $\rho$ parameter~\cite{PDG2012}. The renormalization of the $\rho$ parameter includes a large contribution from the virtual top (with mass $m_t$) and $b$ quark loop, whose one-loop effect can be written~\cite{tbLoop_1, tbLoop_2} as

\begin{equation}
\Delta r_{tb} \approx \frac{-3G_F M_W^2 m_t^2}{8\sqrt{2}\pi^2(M_Z^2 - M_W^2)},
\end{equation}
after neglecting terms of order $m_b^2/m_t^2$, where $m_b$ is the $b$-quark mass. The value of $\Delta r$ also depends logarithmically on the Higgs boson mass.

Using Eq.~\ref{eq:SMWmassPred} as a prediction of the $W$ boson mass, the theoretical uncertainty on the $W$ mass arising from higher-order corrections to $\Delta r$ is estimated to be 4 {$\rm MeV$} using the complete two-loop SM prediction~\cite{SMWmass}. This can be compared with the world-average uncertainty on the measured value of the $W$ boson mass of 23 {$\rm MeV$} before the measurement reported here and the recent result from the CDF Collaboration~\cite{CDFnew}. This 23 {$\rm MeV$} uncertainty results from a compilation of measurements from the four LEP experiments (ALEPH~\cite{AlephW}, DELPHI~\cite{DelphiW}, L3~\cite{L3W}, and OPAL~\cite{OpalW}) and from the D0~\cite{D0W_1, D0W_2, D0W_3} and CDF~\cite{CDFW} Collaborations at the Tevatron, including the earlier CDF~\cite{CDFNewW_1, CDFNewW_2} and D0~\cite{OurPRL} Run~II results.


The $W$ boson mass, and to some extent the top quark mass, are the limiting factors in our ability to tighten the constraints on new physics that couples to the EW sector. Improving the measurement of $M_W$ is, therefore, an important contribution to our understanding of the electroweak interaction. 

This article presents the details of a previously published measurement of the $W$ boson mass~\cite{OurNewPRL} using data taken with the D0 detector during the 2006 -- 2009 Fermilab $p\bar{p}$ Tevatron run, \textit{ie.} during part of the Tevatron Run~IIb, with a total integrated luminosity of 4.3 fb$^{-1}$ and the combination of that measurement with our previous result~\cite{OurPRL} based on 1.0 fb$^{-1}$ of integrated luminosity collected in 2002 -- 2006 (Run~IIa). Both results were obtained using similar analysis techniques.


\section{Measurement Strategy}
\label{sec:strat}

\subsection{Conventions}
A momentum vector $\vec{p}$, in the D0 standard coordinate system, is represented using a right-handed Cartesian coordinate system, $p_x$, $p_y$, $p_z$  where ${\hat z}$ is the direction of the proton beam and ${\hat y}$ points upward.  It is convenient to use a cylindrical coordinate system in which the same vector is given by the magnitude of its components perpendicular to the beam (transverse momentum) $p_T$, its azimuthal angle $\phi$, and $p_z$.  In spherical coordinates, the polar angle $\theta$ is sometimes replaced by the pseudorapidity $\eta\,=\,-\ln\tan[\theta/2]$. In this paper, by {\it electron} we mean electron or positron unless specifically noted. When refering to instrumental effects, sometimes it is convenient to define $\eta_{\rm det}$ as the pseudorapidity of the particle determined as if it had been produced at the center of the calorimeter.

\subsection{{\boldmath $W$} and {\boldmath $Z$} Boson Production and Decay}

$W$ and $Z$ bosons are produced at the Tevatron predominantly through valence quark-antiquark annihilation with a smaller contribution involving the sea. Gluons may be radiated from quarks in the initial state. These gluons usually have lower transverse momentum than the boson (soft gluons) but could be energetic enough to give rise to hadron jets. Consequently, the transverse momentum of the boson is typically small compared to its mass, but has a long tail extending to large $p_T$ associated with events having jets. Spectator partons in the proton and antiproton, which remain after the hard annihilation, hadronize into low-$p_T$ hadrons. Since the transverse momentum vectors of the initial proton and antiproton are zero, the sum of the transverse momenta of the recoiling particles must balance the transverse momentum of the boson.

We measure the decays of the $W$ boson in the electron channel \wen\ and at the same time measure $Z\to ee$ decays that provide an important calibration sample. The size of the $\zee$ sample is limited by its relatively small branching fraction $\text{BR}(\zee)/\text{BR}(\wen) = 0.31$. The electrons typically have transverse momenta of about half the mass of the decaying boson and are well-isolated in the calorimeter. Isolated high-$p_T$ electrons are dominantly produced by $W$ and $Z$ decays and allow us to select a clean sample of $W$ and $Z$ boson events. The D0 calorimeter (Sec.~\ref{sec:det}) is well-suited for a precise measurement of electron energies, providing a single electron energy resolution of 4.5\%, using the angular and energy spectra of electrons from $W$ boson decay when averaged over the electron geometric acceptance and momentum distribution in this analysis. The small tracking volume of the D0 detector limits the momentum resolution of tracks, therefore we do not use $W\to \mu\nu$ decays in this measurement.

\subsection{Event Characteristics}
\label{sec:eventcharacteristics}
For the process $\ppbar \rightarrow W + X \rightarrow e + \nu + X$, we select the electron by requiring $|\eta_{\rm det}|<1.05$ and use all other particles detected up to $|\eta_{\rm det}| \lesssim 4.2$ for the hadronic recoil measurement. We cannot detect recoil particles with $|\eta_{\rm det}| \gtrsim 4.2$, but their transverse momenta are generally small and can be neglected in the recoil system transverse momentum, $\uT$.

A candidate $W$ boson event is characterized by a measurement of the electron momentum $\vec{p}^{\,e}$ and $\uT$. The neutrino escapes undetected but the magnitude and direction of its transverse momentum are inferred from the event's missing transverse energy, $\,\vmet \equiv -(\vec{p}^{\,e}_T + \uT)$. The signature of a $W \rightarrow e \nu$ decay is therefore an isolated high-$p_{T}$ electron and large $\met$.

The signature of $Z \rightarrow ee$ decays consists of two isolated high-$p_{T}$ electrons. In a manner similar to candidate $W$ boson events, a candidate $Z$ boson event is characterized by a measurement of the two electron momenta and $\uT$. 

\subsection{Mass Measurement Strategy}
\label{sec:measurementstrategy}

Since we cannot reconstruct the longitudinal component of the neutrino momentum, we must resort to variables different from the invariant mass. To measure the $W$ boson mass, we use the following three kinematic variables: the $W$ boson transverse mass, \mt, the electron transverse momentum, $p_T^{e}$, and the neutrino transverse momentum, $p_T^{\nu}$ ($\met$).  
The $W$ boson transverse mass is calculated with the formula
\begin{equation}
\mt = \sqrt{2{p_T^e}p_T^{\nu}(1-\cos(\phi^e-\phi^{\nu}))},
\end{equation} 
where $\phi^e$ and $\phi^{\nu}$ are the azimuthal angles of $\vec{p}^{\,e}_T$ and $\vmet$, respectively.  

\begin{figure}[hbpt]
  \includegraphics[width=\linewidth]{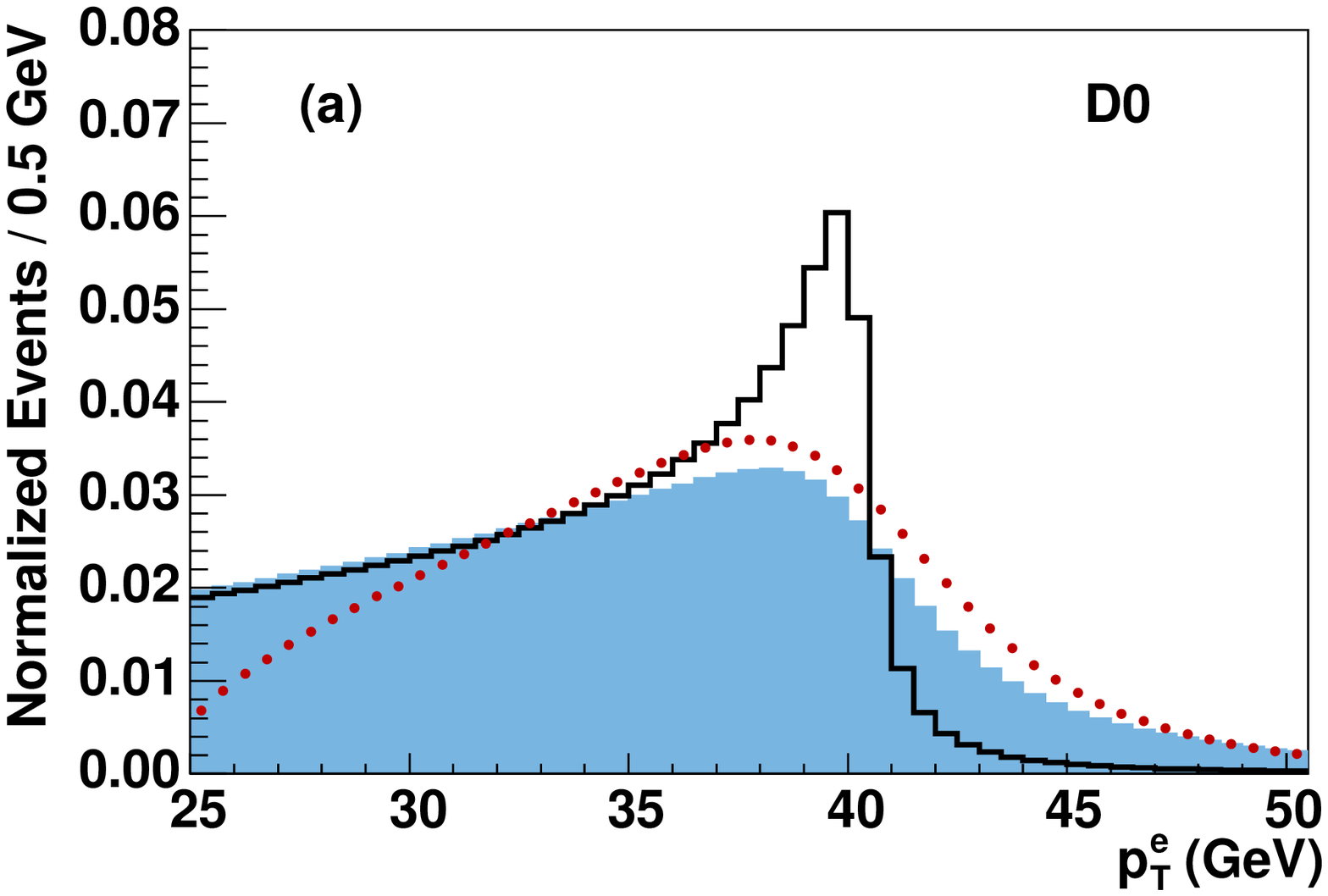}
  \includegraphics[width=\linewidth]{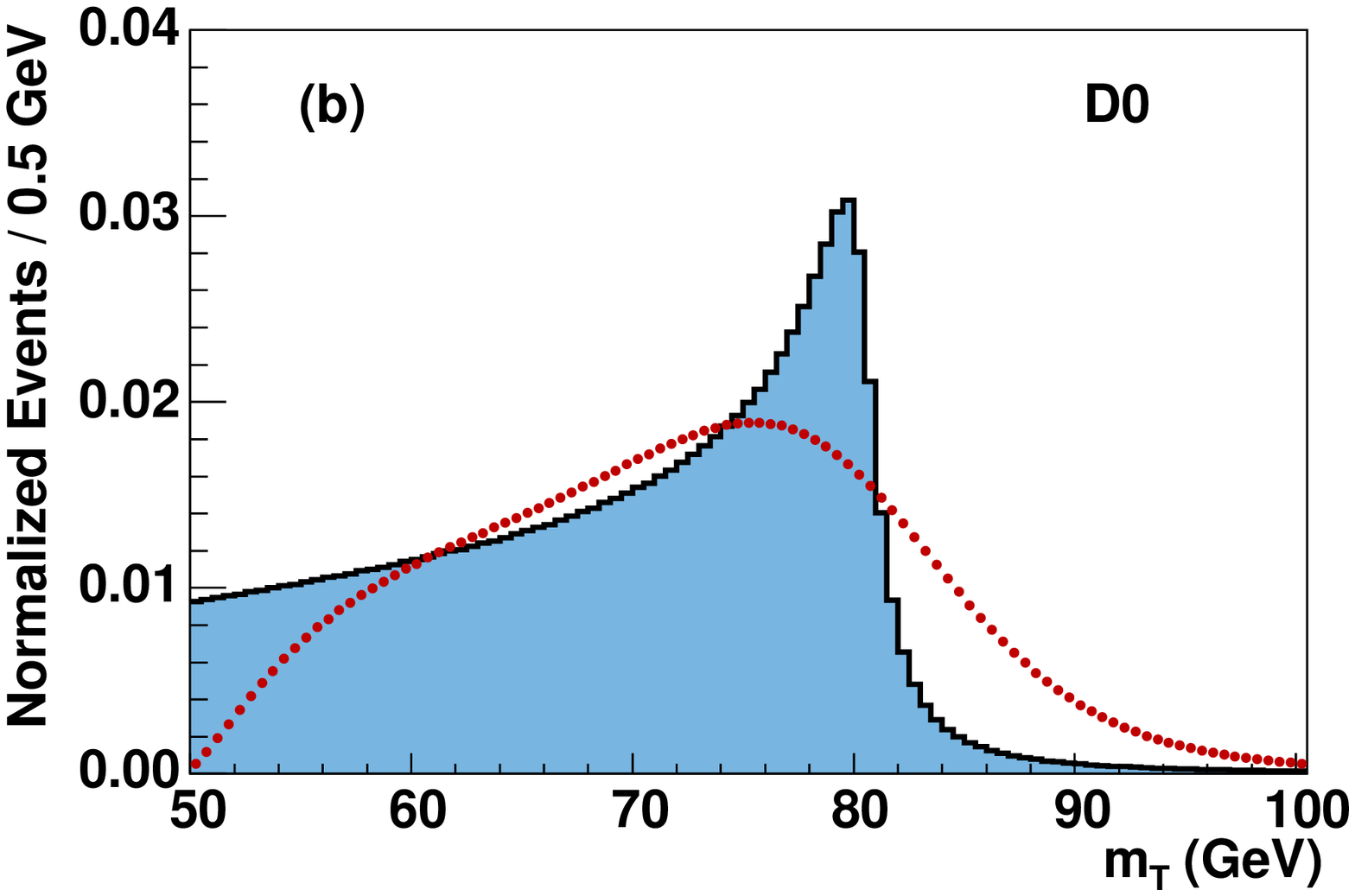}
  \caption{[color online] The (a) $p_{T}^{e}$ and (b) $\mt$ spectra for simulated $W$ bosons without detector resolution effects and $W$ boson transverse momentum $p_{T}^{W}=0$ (solid line), with the natural $p_{T}^{W}$ spectrum at the Tevatron (shaded area), and with the natural $p_{T}^{W}$ distribution and all detector resolution effects included (points).  All curves are normalized to unit area.\label{fig:shape}}
\end{figure}

The $\mt$ and $p_T^e$ measurements provide a powerful cross-check because of their different systematic uncertainties.  The shape of the $\mt$ distribution is dominated by the detector resolution (mainly the resolution due to the recoil system energy measurement), while the $\pte$ spectrum is affected by the transverse momentum of the $W$ boson, as well as by the recoil system and initial state radiation transverse momenta. This is illustrated in Fig.~\ref{fig:shape}. 

The $p_T^{\nu}$ measurement is sensitive to the same systematic uncertainties as both \mt~and $p_T^e$ and has poorer experimental resolution, but this measurement is still useful for a cross-check. These measurements are not fully correlated so we can combine them to improve precision.

The shapes of the distributions of these variables cannot be calculated analytically because of the various complex detector acceptance and resolution effects. The measurement of $M_W$ is obtained by a comparison of the spectra of the three different measurement variables with templates generated from a highly detailed Monte Carlo (MC) simulation with a series of $M_W$ hypotheses. This requires high statistics templates ($\approx\!10^9$ events) to characterize the different systematic uncertainties while ensuring that statistical fluctuations from the MC simulation are negligible. The detailed detector simulation (full MC) is too slow to generate many samples of this size, and it also does not reproduce the detector performance well enough to measure $M_W$ precisely.  To generate appropriate templates, a parametrized MC simulation (fast MC) has been developed to generate large samples on a reasonable time scale and to provide a detailed description of the detector performance. Here, $\zee$ events are used to determine the parameters, since both electrons from $Z$ boson decays are well-measured by the calorimeter and the $Z$ boson properties are well known. This allows a determination of the fast MC parameters, including details of the hadronic recoil system, from the data itself.  Since the $Z$ boson mass is known with high precision~\cite{LEPZ_1, LEPZ_2, LEPZ_3, LEPZ_4}, its value can be used to calibrate the energy scale of the electromagnetic (EM) part of the calorimeter.  Care must be taken to ensure that the calibrations using the $Z$ boson are valid at the lower average energy of the electrons from $W$ boson decay.  Once this has been established, the $M_W$ measurement is effectively a measurement of the ratio of $W$ boson and $Z$ boson masses.

A binned log-likelihood comparing collider data and simulated event distributions (a template) is computed for each of the $\mt$, $\pte$, and $\met$ observables. The log-likelihoods are calculated using the Poisson probability for bin $i$ with $m_i$ expected events from the template to have $n_i$ observed events from the data distribution. The total log-likelihood is formed from the sum over all bins:
\begin{equation}
-\ln{\mathcal{L}} =\sum_{i=1}^N \left[- n_i \ln{m_i} + m_i + \ln(n_i!) \right].
\label{eq:ll}
\end{equation}
Templates are generated for different hypothetical $M_W$ values in $10\,\text{MeV}$ intervals. This procedure gives a mass-dependent likelihood for each of the $m_T$, $p_T^e$, and $\met$ distributions.  We then measure $M_W$ using \textsc{Minuit}~\cite{minuit} by finding the $M_W$ value that maximizes the mass-dependent likelihood. The determination is performed separately for each of the three observables after the likelihood distribution is interpolated between the values at each discrete input mass.

Given the precision achievable in this analysis, we use a technique to avoid any bias which could arise from the knowledge of the current world average. To eliminate such bias, a blinded analysis procedure has been developed. The code that provides the template fits uses an unknown but recoverable offset in an interval of $[-2,\ 2]~{\rm GeV}$ around the $M_W$ value with which the templates have been generated. It therefore reports true differences between different mass fits, allowing systematic studies, while keeping the measured $M_W$ value unknown. The same offset is applied to the result of the fit to $\mt$, $\pte$, and $\met$ so that the relative agreement between the three observables is known before unblinding. Hence, the $M_W$ measurement reported in~\cite{OurNewPRL} and described here was reviewed and approved by the D0 Collaboration based on the studies performed before the resulting value for $M_W$ was known.   

\subsection{Systematic Uncertainties\label{sec:syst}}

The systematic uncertainties are determined using a large ensemble of pseudo-experiments simulated with the fast MC. Pseudo-experiments are generated in which a given parameter is varied independently in steps of multiples of $\pm 0.5 \sigma$ (where $\sigma$ is the one standard deviation uncertainty for the parameter under study) while holding all other parameters constant. For each variation, $M_W$ is determined using the standard fit comparing the distribution for each pseudo-experiment to that of the unmodified template(s). This yields a value $M_{W_i}$ for each variation $\delta_i$.  The set of $(\delta_i,\ M_{W_i})$ pairs is fitted to a straight line. For all systematic uncertainties in this measurement, we verify that the linear regime assumed by this procedure is a valid approximation. The slope of the line determines the systematic uncertainty in $M_W$ as in the usual error propagation:
\begin{equation}
\sigma_{M_W}^2(X) =  {\left(\frac{\partial M_W}{\partial X}\right)}^2 \sigma_X^2,
\label{eq:mwapprox}
\end{equation}
where $\sigma_X$ is the uncertainty in the determination of the parameter $X$ in the simulation.

This equation does not include correlations. In many cases we can safely assume that the parameters are uncorrelated.  When this is not true, the correlations are taken into account by diagonalizing the covariance matrix prior to the propagation of the uncertainty to the $M_W$. The diagonalization defines uncorrelated parameters and uncertainties. The above procedure is applied to the uncorrelated uncertainties to determine the uncertainties on the measured $W$ boson mass.  

\subsection{Additional Kinematic Variables}

\label{ssec:kinematics}
\begin{figure}[hbt]
\begin{center}
\includegraphics[width=0.3\textwidth]{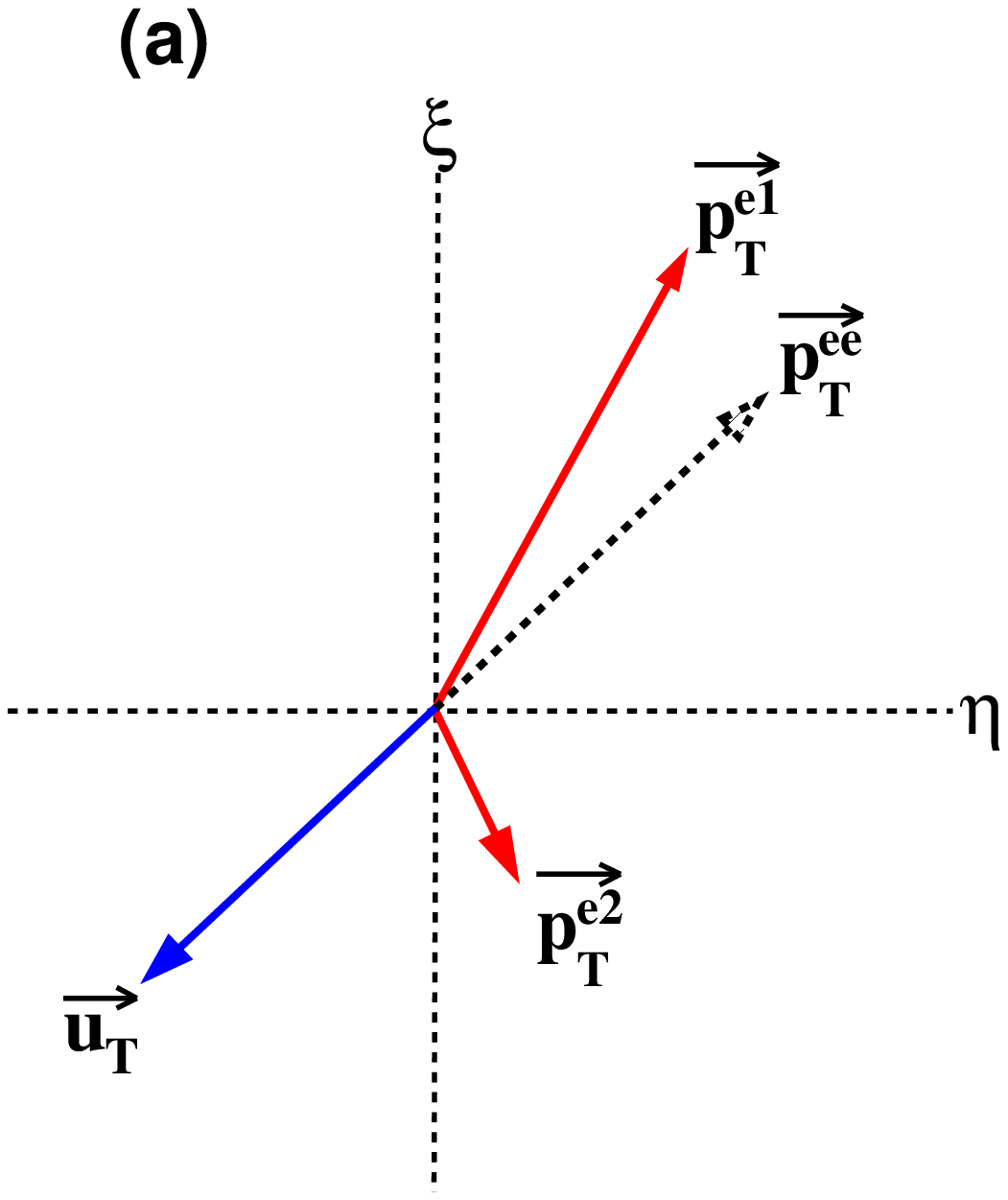}
\includegraphics[width=0.3\textwidth]{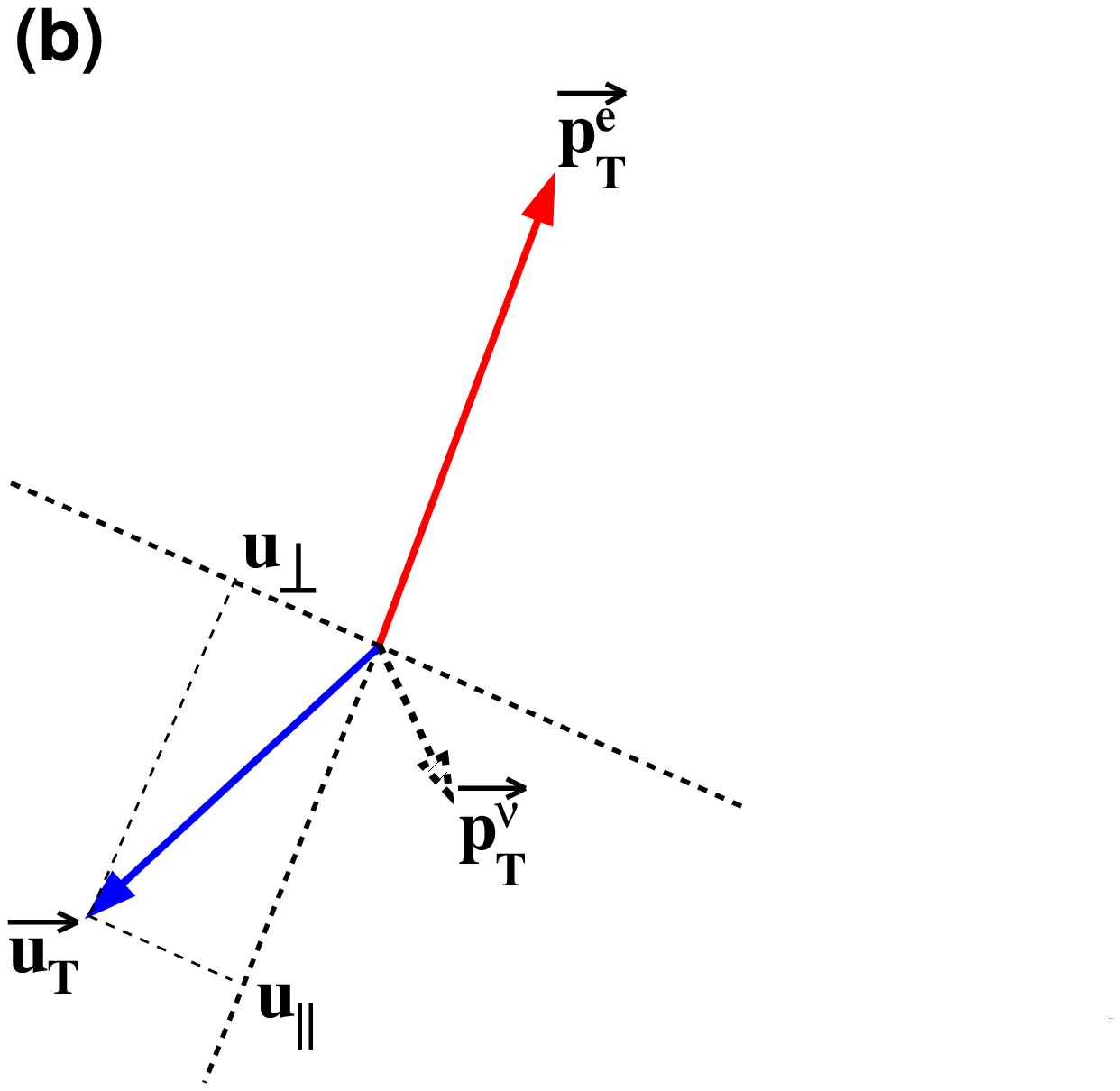}
\end{center}
\caption{[color online] (a)  Definition of $\eta$ and $\xi$ axes for $\zee$ events. (b)
Definition of $\upara$ and $\uperp$.  The variable $\upara$ is negative when opposite to the electron direction.}
\label{fig:eta_upara}
\end{figure}

In $\zee$ decays, the di-electron momentum is given by \mbox{$\vec{p}_{ee}=\vec{p}^{\,e_1}+\vec{p}^{\,e_2}$} and the di-electron invariant mass is \mbox{$m_{ee}=\sqrt{2E^{e_1}E^{e_2}(1-\cos\omega)}$}, where $\omega$ is the opening angle between the two electrons. When tuning the simulation and making comparisons with $\zee$ data, it is useful to define a coordinate system, first introduced by the UA2 experiment\cite{UA2}, in the plane transverse to the beams that depends only on the electron directions, not on electron energies.  We call the axis along the inner bisector of the two electrons the $\eta$ axis and the axis perpendicular to that, in the $(\vec{p}^{\,e_1}, \vec{p}^{\,e_2})$ plane, the $\xi$ axis. Figure~\ref{fig:eta_upara}(a) illustrates these definitions.

For $\wen$ decays, useful quantities are the projection of the recoil system transverse momentum onto the electron direction:
\begin{equation}
\upara = \vec{u}_T \cdot \hat{p}_T^e,
\end{equation}
and the projection on the direction perpendicular to the electron:
\begin{equation}
\uperp = \vec{u}_T \cdot (\hat{p}_T^e \times \hat{z}),
\end{equation}
where $\hat{z}$ is an unit vector in the $z$ direction. Figure~\ref{fig:eta_upara}(b) illustrates these definitions for $W$ boson events, but the definitions also apply for each electron from $Z \to ee$ events. 

The two variables $u_{\parallel}$ and $\uperp$ are useful to study the correlation between the recoil system and the electron direction.  Another variable, the scalar sum of all transverse energies (SET) measured by the calorimeter except those energies associated with electrons or with potential noise, reflects the total hadronic activity in the calorimeter.

\section{The D0 Detector}
\label{sec:det}

The D0 detector~\cite{RunIdetector} was built for the Run I of the Fermilab Tevatron Collider and upgraded~\cite{RunIIdetector} for the Run~II to the configuration relevant to the measurements described here.  It contains central tracking, calorimeter and muon subdetector systems.  The silicon microstrip tracker (SMT) detector located near the $p{\bar p}$ interaction point covers $|\eta_{\rm det}|<3$.  The central fiber tracker (CFT) surrounds the SMT and provides complete coverage out to $ |\eta_{\rm det}| \approx 1.7$.  A $1.9\, {\rm T}$ solenoid surrounds the central tracking system and gives a typical transverse momentum resolution of $10\%$ -– $16\%$ for tracks of $p_T = 40\,\text{GeV}$~\cite{muonid}.

\begin{figure}[ht]
\centering
\includegraphics [scale=0.65] {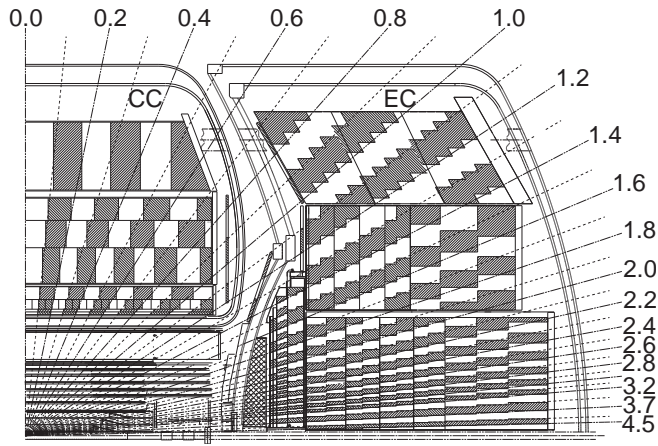}
\caption{Side view of one quadrant of the D0 detector, not showing the muon subdetector system. The calorimeter segmentation and tower definition are shown in both CC and EC. The lines extending from the center of the calorimeter denote the pseudorapidity ($\eta_{\rm det}$) coverage of cells and projected towers. The solenoid and tracking detectors are shown in the inner part of the detector.}
\label{fig:cal_quadratic_view}
\end{figure}

Three uranium liquid-argon calorimeters measure particle energies.  The central calorimeter (CC) covers $|\eta_{\rm det}|<1.1$ and two end calorimeters (EC) extend the coverage to $|\eta_{\rm det}| \approx 4.2$.  The CC is segmented in depth into eight layers.  The first four layers are used primarily to measure the energies of photons and electrons and are collectively called the electromagnetic (EM) calorimeter.  The remaining four layers (three fine hadronic (FH) layers and one coarse hadronic (CH) layer), along with the first four, are used to measure the energies of hadrons.  Most layers are segmented into $0.1 \times 0.1$ regions (cells) in $(\eta , \phi)$ space.  The third layer of the EM calorimeter is segmented into $0.05 \times 0.05$ regions. Between the central and end cryostats, the inter-cryostat detector (ICD) provides sampling of particles in the range $1.1 < | \eta_{\rm det} | < 1.4$ using scintillator pads. The calorimeter system is completed with central and forward preshower detectors located just before the central and forward cryostats up to $| \eta_{\rm det} | = 2.5$. Figure~\ref{fig:cal_quadratic_view} shows a cross sectional $r$-$z$ view of one quarter of the D0 detector, showing the calorimeter $\eta$ and depth segmentation, which indicates how the calorimeter system forms projective towers of size $0.1 \times 0.1$ in $(\eta , \phi)$ space.  

Muon trajectories are identified and measured outside the calorimeter system using a system  of proportional drift tube chambers, scintillation counters, and toroidal iron magnets.

The luminosity of $p {\bar p}$ collisions is monitored using two sets of 24 wedge-shaped scintillation counters, each placed on the face of one of the end calorimeters.  These counters are used to detect inelastic non-diffractive collisions~\cite{lumipaper}.

The D0 calorimeter is read out by a total of 47,032 electronic channels.  The electronic pedestal is measured frequently for each channel using special calorimeter pedestal runs during the quiet time  between stores when there is no beam in the Tevatron.  The energy measured for each channel in collider data is the energy recorded minus the pedestal.  The calorimeter readout uses zero suppression to avoid reading out noise.  If $\sigma_{\text{PED}}$ is defined as the root-mean-square variation of the pedestal of each channel about its mean, the criterion deciding whether or not to read out a channel is expressed in terms of its $\sigma_{\text{PED}}$. Normally, in zero-suppressed data, a cell is read out by the D0 electronic system only if its energy differs from the pedestal by more than 1.5$\sigma_{\text{PED}}$. The D0 electronic system also records data in which all the channels are read out with no zero suppression.  The D0 event reconstruction requires an energy deposit in a cell to exceed the pedestal by at least 4.0$\sigma_{\text{PED}}$ if it is to be considered the central cell of an energy cluster.   An adjacent cell with energy exceeding its pedestal by at least 2.5$\sigma_{\text{PED}}$ is considered to be part of this same cluster.   Cells with energy less than 2.5$\sigma_{\text{PED}}$ above  pedestal are not considered for reconstruction in normal (zero-suppressed) collider data. 

Events are selected for this analysis if they pass a single electron trigger requirement in the CC.  In this way, trigger and other efficiencies can be measured with $Z\to ee$ events using the tag and probe method: if the tag electron is required to satisfy the trigger, the probe electron is considered to be unbiased.  Each trigger is a combination of requirements at three trigger levels (L1, L2, L3).  At each succeeding level the trigger uses more detailed detector information and becomes more precise.

The trigger towers in the calorimeter are $0.2\times 0.2$ in $(\eta , \phi)$ space. The triggers used in this analysis require, at the L1 trigger level, at least one EM object, defined by two neighboring trigger towers~\cite{L1trigger}. The EM object must satisfy $E_T^{\text{L1}} > 19\,\text{GeV}$ and $|\eta^{\text{L1}}|<3.2$. Two different L2 trigger level requirements are used, depending on the period the data were taken. An early version of the trigger, {\em v15}, requires the EM object to be isolated if $19 < E_T^{\text{L1}} < 22\,\text{GeV}$, but makes no requirement above $22\,\text{GeV}$. A later version, {\em v16}, requires a more complex likelihood criterion based on the energy distribution in the L1-triggered EM trigger towers and in their neighboring towers if $19 < E_T^{\text{L1}} < 25\,\text{GeV}$, but no requirement above $25\,\text{GeV}$. At the L3 trigger level, the EM objects must satisfy $E^{\text{L3}}_T>25\,\text{GeV}$, $|\eta^{\text{L3}}|<3.6$, and a shower shape requirement. At higher instantaneous luminosities, the L3 threshold is increased to $E^{\text{L3}}_T>27\,\text{GeV}$ to cope with the higher trigger rate. For the trigger with the L3 threshold $E^{\text{L3}}_T>27\,\text{GeV}$, only the L2 likelihood criterion is used.

\section{Data Reconstruction}
\label{sec:dat}

\begin{figure}[hbpt]
\centering 
\includegraphics [width=\linewidth] {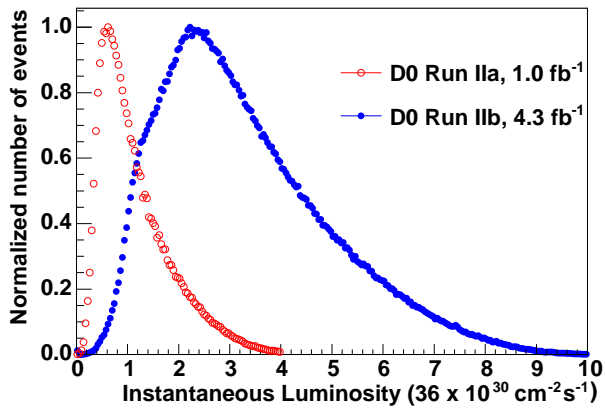} 
\caption{[color online] Instantaneous luminosity profiles for Run~IIa and Run~IIb. The instantaneous luminosity is given as a multiple of $36\times 10^{30}\,\text{cm}^{-2}s^{-1}$ since there were 36 $p\bar{p}$ bunch crossings per turn in the Tevatron Collider.}
\label{fig:lumiprofiles}
\end{figure}

The data sample for this Run~IIb measurement includes data with a total integrated luminosity of $4.3\, {\rm fb}^{-1}$ taken between June 2006 and June 2009. Figure~\ref{fig:lumiprofiles} compares the instantaneous luminosity ($L$) profile of this Run~IIb measurement with the profile of our previous measurement~\cite{OurPRL} using data recorded from 2002--2006 (Run~IIa). Here and in the rest of the paper, instantaneous luminosity is given as a multiple of $36\times 10^{30}\,\text{cm}^{-2}s^{-1}$ since there were 36 $p\bar{p}$ bunch crossings per turn in the Tevatron Collider. The Run~IIb instantaneous luminosity is significantly higher and much of the effort in preparing this measurement is dedicated to dealing with the multiple interactions (pileup) and calorimeter gain variation resulting from the high beam intensity in Run~IIb. For an instantaneous luminosity of $8\times 36\times 10^{30}\,\text{cm}^{-2}s^{-1}$, we expect an average of 10 simultaneous inelastic interactions per bunch crossing.

This high instantaneous luminosity results in extra $p{\overline p}$ interactions in the same beam crossing as the event of interest.  We measure the effect of this pileup by collecting $p{\overline p}$  interactions in random beam crossings which are labeled zero-bias (ZB) events.  There are also extra interactions not due to the hard parton-parton scattering of interest coming from spectator partons in the same $p{\overline p}$ collision as the hard scattering.  These extra interactions are studied using minimum-bias (MB) events, selected by requiring a coincidence between luminosity-monitor scintillation counters.  The selection requires zero or one reconstructed primary (hard collision) vertex. The number of multiple interactions accompanying an event of interest scales with instantaneous luminosity, while the contribution of spectator partons is independent of it.

\subsection{Electron Reconstruction}
\label{sec:elreconstruct} 

\begin{figure}[htbp]
\includegraphics[width=0.8\linewidth]{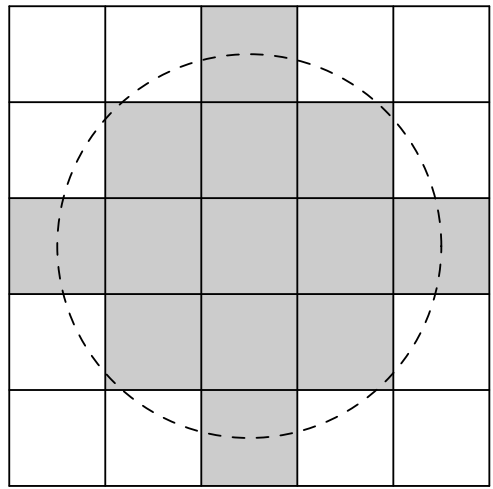}
\caption{The 13 calorimeter towers defined as the electron reconstruction cone. The cone is centered on the tower with the highest transverse energy. A circle of radius $\Delta R=0.2$ is shown for comparison.}
\label{fig:ewindow}
\end{figure}

The measured EM energy associated with an electron ($E^{\rm unc}_{\rm EM}$) in the central calorimeter is the sum of the energies in all EM cells whose centers lie in a cone of radius \mbox{$\Delta R = \sqrt{(\Delta\eta)^2 + (\Delta\phi)^2}=0.2$} centered on the tower with the highest transverse energy.  The definition of the electron energy reconstruction cone (13 towers) is shown in Fig.~\ref{fig:ewindow}. The total uncorrected energy $E^{\rm unc}_{\rm tot}(\Delta R)$ is the sum of the energies in all cells within a given cone of size $\Delta R$ centered on the central tower, over all layers of the calorimeter, including the hadronic calorimeter layers.

The identification of this cluster of EM energy as a candidate true electron is based on the following four parameters:

\begin{itemize}
\item{\bf EM fraction:} A true electron will deposit nearly all of its energy in the EM layers of the calorimeter.  Therefore the EM fraction 
\begin{equation}
f_{\text{EM}} \equiv {E^{\rm unc}_{\rm EM}(\Delta R<0.2) \over E^{\rm unc}_{\rm tot}(\Delta R<0.2)} 
\end{equation}
is expected to be close to 1.

\item{\bf Isolation:}  In an electron shower most of the energy is deposited in a narrow cone with little energy around it.  Therefore
\begin{equation}
f_{\text{iso}} \equiv { {E^{\rm unc}_{\rm tot}(\Delta R<0.4) - E^{\rm unc}_{\rm EM}(\Delta R<0.2)} \over {E^{\rm unc}_{\rm EM}(\Delta R<0.2)}}
\end{equation}
is expected to be close to zero. Isolation provides discrimination against hadronic showers, which tend to be wider.

\item{\bf HMatrix:}  The transverse and longitudinal shapes of an electron shower are well-modeled by MC simulations. Therefore, it is possible to determine a multivariate likelihood based on a set of variables whose correlations and variances allow the discrimination of electron showers. The variables used are:

\begin{itemize}
    {\bf HMatrix7} (used in the CC) is built from the following variables: EM fractions in layers 1, 2, 3, 4, shower transverse width in the $\phi$ direction, $\log(E^{\rm unc}_{\rm tot})$, and  $z_V$ (the production vertex $z$ coordinate).

    {\bf HMatrix8} (used in the EC) is built from the same variables as HMatrix7 plus the shower width in the direction perpendicular to the beam in the plane of third layer of the calorimeter (EM3).
\end{itemize} 

The inverse of the likelihood covariance matrix is used to determine a  $\chi_{\text{HM}}^2$ value for an EM cluster, which should be small if the cluster results from an electron shower~\cite{HMatrix_1, HMatrix_2}.

\item{\bf Track match:}  A track is reconstructed from SMT and CFT hits and is required to have $p_T>10$ GeV.  It is considered to be matched with an EM cluster if it is within $0.05$ in $\Delta \eta$ and within $0.05$ in $\Delta \phi$. Here, $\Delta\eta$ and $\Delta\phi$ are the distances between the cluster centroid, as determined by its cells in EM3 with weights proportional to the logarithm of the cell energy, and the extrapolation of the track to this layer of the calorimeter.  The quality of the match is determined by
\begin{equation}
\chi_{\rm TM}^2 \equiv {\Bigl({{\Delta \phi} \over {\sigma_\phi}}\Bigr)^2
     + \Bigl({{\Delta \eta} \over {\sigma_\eta}}\Bigr)^2},
\end{equation}
where $\sigma_\phi$ and $\sigma_\eta$ are the measured resolutions of $\Delta\phi$ and $\Delta\eta$. 
\end{itemize}
 
In the initial reconstruction, electromagnetic clusters are required to  have transverse energy $E^{\rm unc}_T>1.5$ GeV and EM fraction $f_{\rm EM}>0.9$.  If the cluster has a track matched to it, it is considered a candidate electron.

The energy of an electron $E^{e,{\rm unc}}$ is defined as the sum of the energies in all four electromagnetic calorimeter (EM1 to EM4) and first fine hadronic layer (FH1) cells in the 13 towers of the electron cone (Fig.~\ref{fig:ewindow}) centered on the tower with the highest transverse energy:

\begin{equation}
{E}^{e,{\rm unc}} = \sum_{i}E^{\rm unc}_{i}.
\end{equation}

\noindent
The FH1 layer is included to more fully contain the electromagnetic shower. The corrected electron energy $E^e$ is defined by applying the energy loss correction (Sec.~\ref{sec:deadmat}).

In this analysis, the direction of the electron is always taken to be the direction of the matched track:
\begin{displaymath}
\begin{array}{ll}
\theta^e = &  \theta_{\rm track}, \\
\phi^e =  & \phi_{\rm track}.
\end{array}
\end{displaymath} 
\noindent The track direction is determined with a resolution of $0.002\,\text{rad}$ in $\theta$ and $0.0004\,\text{rad}$ in $\phi$, which have a negligible impact on this measurement. The momentum of the electron, neglecting its mass, is given by
\begin{displaymath}
\vec{p}^{\,e} =  E^e \begin{pmatrix}
\sin \theta^e \cos \phi^e \\
\sin \theta^e \sin \phi^e \\
\cos \theta^e
\end{pmatrix},
\end{displaymath}
and the transverse energy of the electron is defined as $E^e_T = E^e \sin \theta^e$.  Corresponding to this definition, the uncorrected transverse energy of the electron is given by $E^{e,{\rm unc}}_T = E^{e,{\rm unc}}\sin \theta^e$. 
 
\subsection{Vertex Reconstruction}
\label{sec:vertex}

The coordinate of the $W$ boson production vertex along the beam line, $z_V$, is determined either using the standard D0 vertex algorithm (which uses a Kalman filter algorithm~\cite{ref:Ariel}), or is taken as the point of closest approach of the electron track to the beam line if this electron track vertex position differs by more than 2 cm from the point selected by the vertex algorithm.  For $Z$ boson events, $z_V$ is taken to be the average of the two points of closest approach of the electron tracks.  

\subsection{Uncorrected Missing {\boldmath $E_T$} and Recoil Reconstruction}

The uncorrected missing energy vector in the transverse plane is calculated by taking the vector sum
\begin{equation}
\vmet^{\,{\rm unc}} = -\sum_{i}E^{\rm unc}_{i} \mbox{sin}\theta_{i} \left( \begin{array}{c}
                                                  \mbox{cos}\phi_{i} \\
                                                  \mbox{sin}\phi_{i} \end{array} \right)
                 = - \sum_{i}\vec{E}^{\,i\ {\rm unc}}_{T},
\end{equation}

\noindent
where the sum runs over all calorimeter cells that were read out except cells in the coarse hadronic calorimeter and ICD. Here, the $E^{\rm unc}_{i}$ are cell energies, and $\phi_{i}$ and $\theta_{i}$ are the azimuthal and polar angle of the center of the cell $i$ with respect to the vertex.  

The recoil transverse momentum $\vec{u}_T$ for $W/Z$ boson events is calculated from the $\ \vmet$ and the electron transverse momentum:
\begin{equation}
 \uT^{\,{\rm unc}}\,=\,-\,\vmet^{\,{\rm unc}}\,-\,\sum_{e}{\vec{p}_{T}^{\,e\,{\rm unc}}}.
\end{equation}
The average energy deposition in the calorimeter cells away from the electron cluster is usually small. Thus, a hadronic energy scale would be dependent on specific details of the readout noise and suppression algorithms. Since these details are not correlated with the $W/Z$ event and vary with the run condition, we choose not to use a hadronic energy scale correction for the recoil $p_T$, and thus:
\begin{equation}
 \uT \equiv \uT^{\,{\rm unc}}.
\end{equation}

\subsection{SET Reconstruction}

The scalar sum of the transverse energies of all calorimeter cells is defined as:
\begin{equation}
 \mathrm{SET} = \sum_{i}E_{i}^{\rm unc} \sin\theta_{i},
\end{equation}
excluding cells inside the electron reconstruction cluster, from the coarse hadronic calorimeter and from the ICD detector.

\subsection{Corrected {\boldmath $\slashed{E}_T$} Reconstruction}

The corrected $\vmet$ is calculated from $\vec{u}_T$ and corrected $\vec{p}_T^{\,e}$. For $W\rightarrow e\nu$ events,
\begin{equation}
\vmet = -\vec{u}_T-\vec{p}_T^{\,e},
\end{equation}
and for $Z\rightarrow ee$ events,
\begin{equation}
\vmet = -\vec{u}_T-\vec{p}_T^{\,e_1}-\vec{p}_T^{\,e_2}.
\end{equation}

\subsection{Event Selection}
\label{sec:eventselection}

We select $\zee$ and $\wen$ candidate events using the decay electrons and the $\met$. The vertex is required to be within $|z_V| < 60\,\text{cm}$. The following electron requirements are applied to the reconstructed electron with the highest $p_T$ for $\wen$ candidate events and the two electrons with the highest $p_T$ for $\zee$ candidate events.

\begin{itemize}
\item  $f_{\text{EM}}>0.9$, $f_{\rm iso}<0.15$.
\item HMatrix7$\ <12$ in CC and HMatrix8$\ <20$ in EC (the EC electrons are used for tag and probe studies).
\item Regions near the edges of a calorimeter EM module in $\phi$ are excluded, see Sec.~\ref{sec:eff_phimod}.
\item $\pte>25$ GeV.
\item The associated track must have $p_{T}>10$ GeV, a track match with a probability of $P(\chi_{\rm TM}^2)>0.01$ (see Sec.~\ref{sec:elreconstruct}), and at least one SMT hit. No requirement is made on the number of CFT hits.
\end{itemize}

$Z \rightarrow ee$ candidate events are selected by requiring:
\begin{itemize}
\item At least one electron passes the trigger requirements of all three trigger levels.
\item Electron $|\eta_{\text{det}}|<1.05$,
except for studies of the electron efficiency, where one electron can be in the EC region $1.5<|\eta_{\text{det}}|<2.5$.
\item $u_{T}<15$ GeV.
\item $70<m_{ee}<110$ GeV.
\end{itemize}

$W \rightarrow e\nu$ candidate events are selected by requiring:
\begin{itemize}
\item The electron must pass the trigger requirements of all three trigger levels.
\item $\met > 25$ GeV.
\item Electron $|\eta_{\text{det}}|<1.05$.
\item $u_{T}<15$ GeV.
\item $50<m_{T}<200$ GeV.
\end{itemize}

After the selections, 54,512 candidate $\zee$ events remain with both electrons in the CC, which we use to determine the EM calibration, and 1,677,489 candidate $\wen$ events remain that are used to determine $M_W$.

\section{Uninstrumented Material Correction to the Electron Response}
\label{sec:deadmat}

\begin{figure}[hbpt]
  \centering 
  \includegraphics [width=\linewidth] {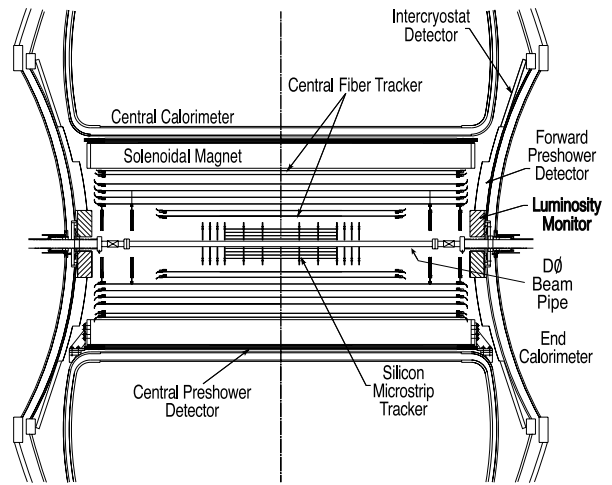} 
  \caption{Overview of the material in front of the CC. This drawing shows a cross sectional view of the central tracking system in the $x$~-~$z$~plane. Also shown are the locations of the solenoid, the preshower detectors, luminosity monitor, and the calorimeters.}
  \label{fig:Material}
\end{figure}

Figure~\ref{fig:Material} shows an overview of the material in front of the CC cryostat. An electron traveling from the interaction point to the CC at normal incidence encounters about 3.7 radiation lengths ($X_0$) of material before reaching the first active layer of liquid argon: $0.2\,X_0$ in the inner detector, $0.9\,X_0$ in the solenoid, $0.3\,X_0$ in the preshower detector plus $1.0\,X_0$ in the associated lead, and $1.3\,X_0$ in the cryostat walls plus related support structures. As a consequence of the uninstrumented material in front of the CC, the measured response to incident electron energy has a significant non-linear dependence on the true energy and the angle of impact. In this section we describe the derivation of the corrections to the electron response that are applied to data to account for the uninstrumented material. This correction is derived from a simulation of the detector response to electrons in which the shower description has been improved relative to the standard {\sc geant}3~\cite{GEANT3} description and the amount of uninstrumented material has been tuned. The uninstrumented material tuning  was derived with the $\zee$ data sample of the $1\, {\rm fb}^{-1}$ (Run~IIa) analysis~\cite{OurPRL} and re-validated for this analysis (Sec.~\ref{sec:EMfracValidation}). A comprehensive account of the calibration method can be found in~\cite{JanHabilitation, RclsaPhD}.

\subsection{Improvements in the Simulation of Electromagnetic Showers}
\label{sec:improve_shower}

Because of the large amount of material preceding the active layers of the calorimeter, a precise simulation of the electromagnetic shower is needed to ensure acceptable understanding of the electron energy reconstruction as a function of true energy and angle of incidence. Several improvements are needed to the standard {\sc geant}3 simulation to have a good description of the energy deposition and depth of the shower.

To improve the transport algorithm for low energy particles in the shower, we configure {\sc geant}3 to evaluate steps as small as $10^{-7}\, {\rm cm}$ in the tracking of particles~\cite{molierefootnote}.  We also force the maximum step length to be smaller than $10^{-1}\, {\rm cm}$. These modifications are chosen so that the Moli\`ere theory of multiple scattering is guaranteed to be valid in our simulation~\cite{moliere, molierebethe}.

The standard {\sc geant}3 parametrizations for bremsstrahlung and pair creation cross sections in matter are also insufficiently precise. We replace these with tables of cross sections from~\cite{Seltzer198595} and~\cite{hubbell:1023}, respectively.

Finally, the low energy cut-off for explicit simulation of $\delta$-rays was lowered from 1 MeV to 10 keV. This was necessary to obtain an adequate description for the local energy deposition of low energy electrons and photons, especially near the uranium--liquid argon boundary~\cite{slowgeantfootnote}.

\subsection{Observables Used for Tuning the Simulation}
\label{sec:observable_mat_tune}
To estimate the contribution of uninstrumented material, we exploit the segmentation of the calorimeter readout by studying the EM layer energy fractions, {\em i.e.}, the fraction of the measured electron energy deposited in each one of the layers EM1, EM2, EM3, EM4, and FH1. The depositions in EM4 and FH1 give contributions that are negligible in the tuning procedure.

\begin{figure}[hbpt]
\centering 
\includegraphics [width=\linewidth] {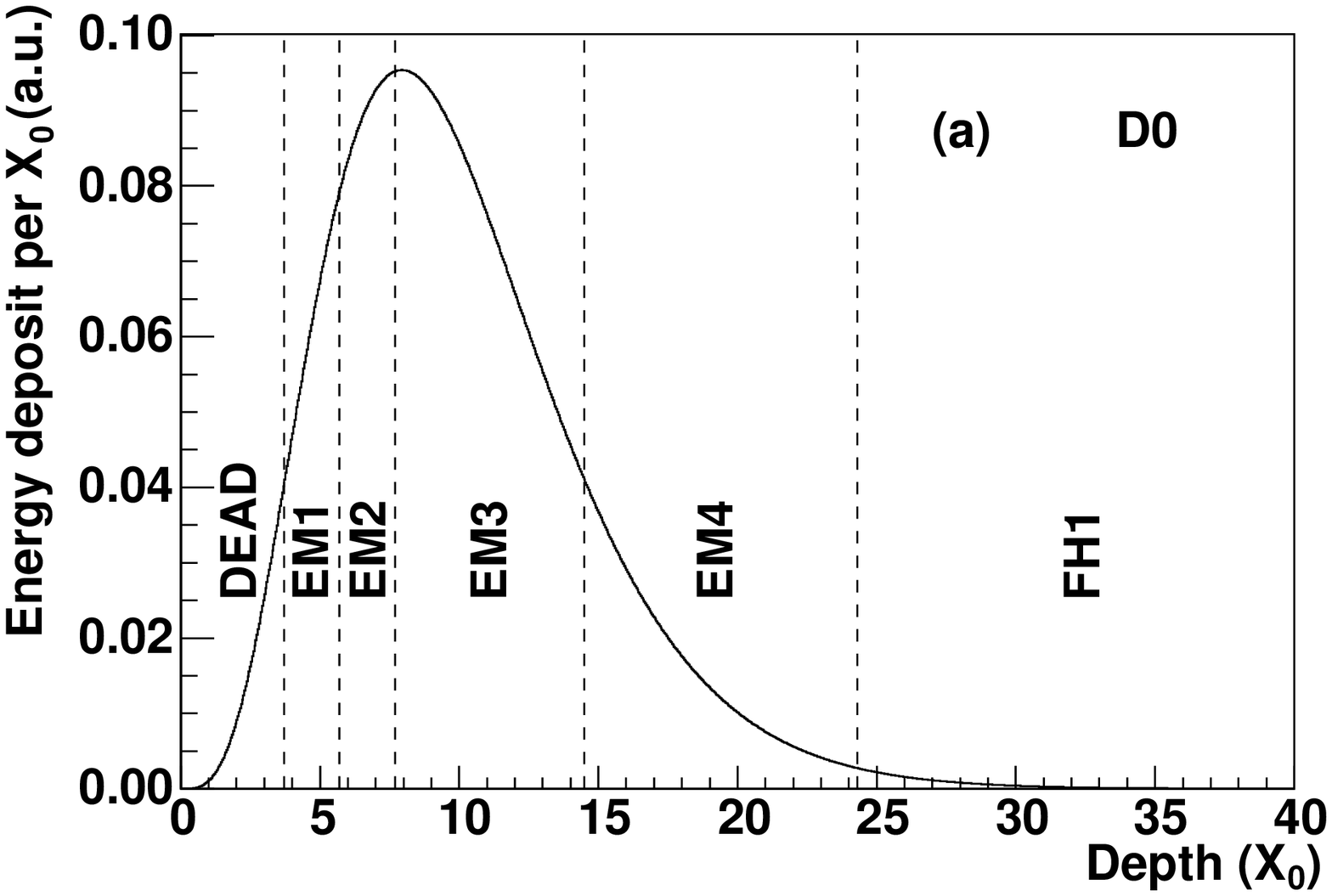} 

\includegraphics [width=\linewidth] {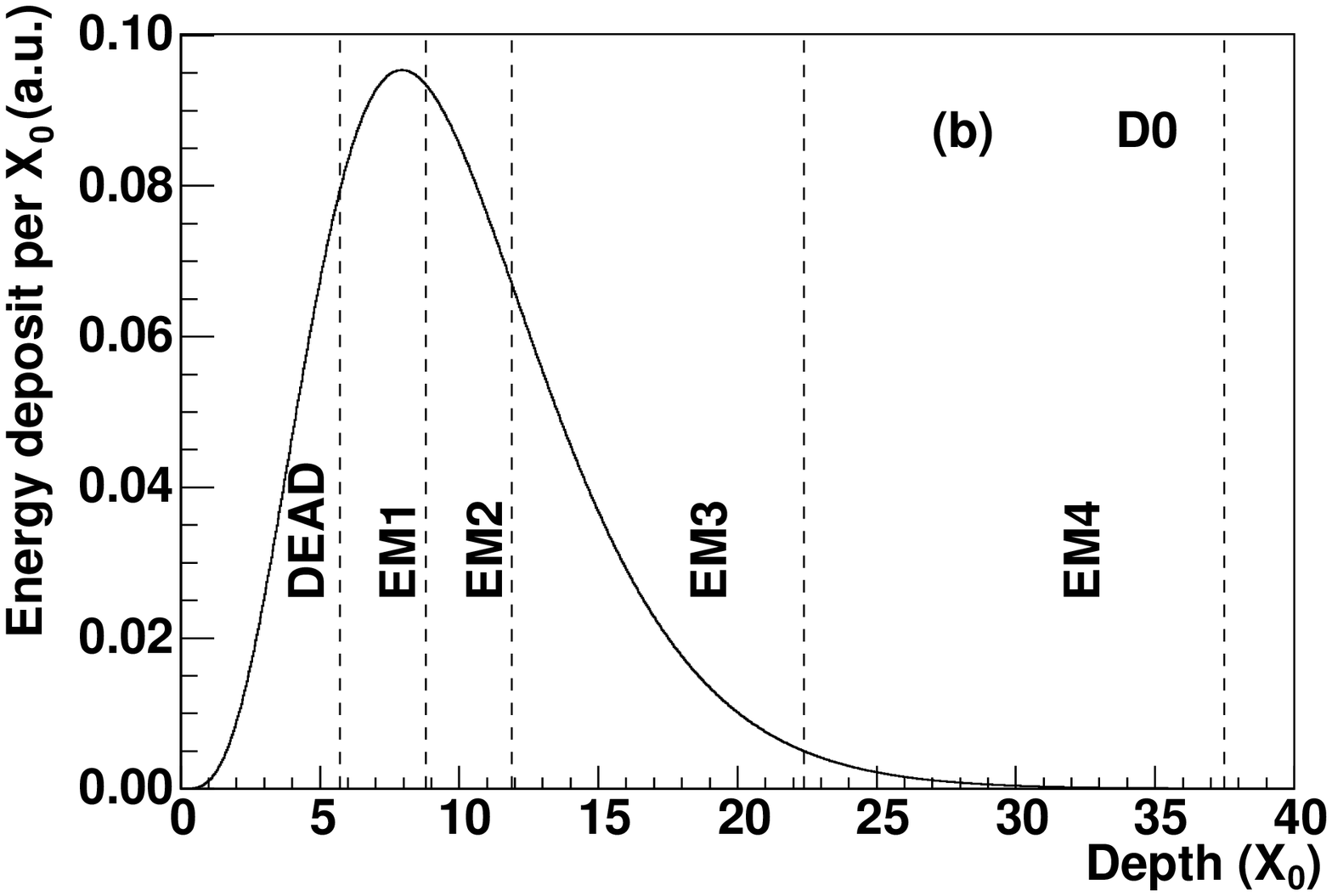} 

\caption{The average shower energy deposition profile (along the shower axis) of electrons with $E=45\,\text{GeV}$ simulated using the GFlash parametrization~\cite{GFlash}. The depth of each readout section of the central calorimeter is indicated for an electron with (a) normal incidence $\eta=0$ and with (b) non-normal incidence $\eta=1$.}
\label{fig:ShowerAverage}
\end{figure}

Electrons produced at different angles cross different amounts of uninstrumented material and the fraction of energy deposited in each layer will therefore be different, as shown in Fig.~\ref{fig:ShowerAverage}. We split the $\zee$ event sample into categories based on electron $\eta$. We define five bins of $|\eta|$ used as a measure of the angle of incidence on the uninstrumented material. The definition of the bins is given in Table \ref{table:StandardEtaBins}. We classify a $\zee$ event into one of 15~distinct categories shown in Table~\ref{table:StandardEtaCategories} according to the $|\eta|$~bins of the two electrons.  We do not distinguish between the leading and the subleading electron transverse momentum to avoid consideration of the calorimeter energy corrections that we are trying to determine.

\begin{table}
\begin{center}
\caption{Definition of bins in electron $|\eta|$ used for uninstrumented material studies.}
\label{table:StandardEtaBins}
\begin{tabular}{c|c}\hline\hline
Bin Number & $\eta$ range\\\hline
Bin 0 & $|\eta| < 0.2$ \\
Bin 1 & $0.2 \le |\eta| < 0.4$ \\
Bin 2 & $0.4 \le |\eta| < 0.6$ \\
Bin 3 & $0.6 \le |\eta| < 0.8$ \\
Bin 4 & $0.8 \le |\eta|$ \\\hline\hline
\end{tabular}
\end{center}
\end{table}

\begin{table}
\begin{center}
\caption{Definition of the di-electron $\eta$ categories for \zee\ events.}
\label{table:StandardEtaCategories}
\begin{tabular}{c|c}\hline\hline
Category & Combination of electron $\eta$ bins\\\hline
10       & 0, 0\\
11       & 0, 1\\
12       & 0, 2\\
13       & 0, 3\\
14       & 0, 4\\
15       & 1, 1\\
16       & 1, 2\\
17       & 1, 3\\
18       & 1, 4\\
19       & 2, 2\\
20       & 2, 3\\
21       & 2, 4\\
22       & 3, 3\\
23       & 3, 4\\
24       & 4, 4\\\hline\hline
\end{tabular}
\end{center}
\end{table}

\begin{figure}[hbpt]
\includegraphics [width=\linewidth] {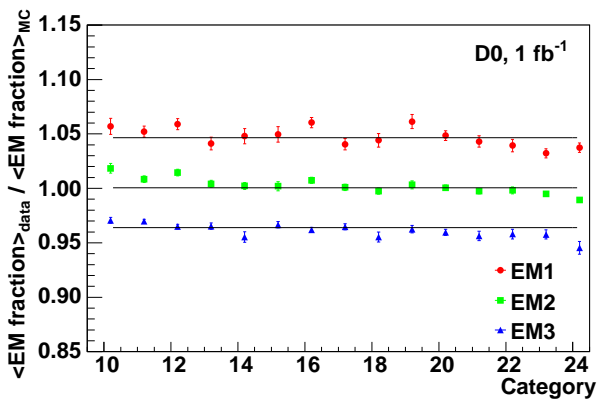}
\caption{[color online] The ratio of data to simulation for the means of the EM layer energy fraction distributions in \zee\ events for each of the first three EM layers and each of the 15~$\eta$~categories shown before the correction described in Sec.~\ref{sec:SystElecNonLin} is applied. Each of the three horizontal lines indicates the result of a fit of a common constant to the 15~data points from a given EM layer.}
\label{fig:EMFDMnoMat}
\end{figure}

We compare the mean of the EM layer energy fraction distribution for each layer in each category between $Z\rightarrow ee$ data and the full MC simulation using the improved shower simulation described in Sec.~\ref{sec:improve_shower}. As can be seen in Fig.~\ref{fig:EMFDMnoMat}, the agreement between the EM fraction from electrons produced at different angles is poor in each layer. The differences arise from inadequacies in the D0 material model included in the full MC.

\subsection{Improvements in the D0 Material Model}
\label{sec:SystElecNonLin}

\begin{figure}[hbpt]
\centering
\includegraphics [width=\linewidth] {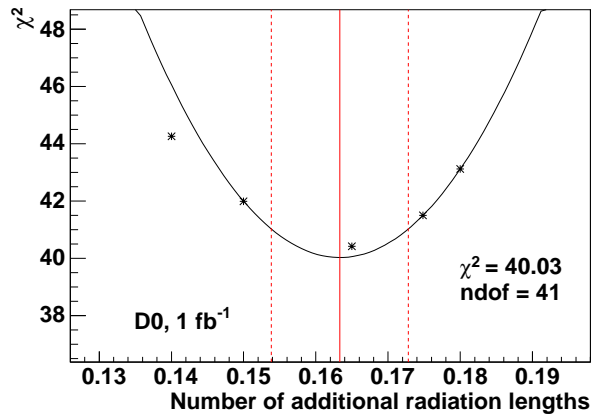}
\caption{[color online] Fit for $nX_0$, the amount of uninstrumented material (in radiation lengths) added to the nominal material in the improved simulation of the D0 detector.  The solid and dotted vertical lines show the best fit and one standard deviation uncertainties for $nX_0$.  This fit is performed with the $\zee$ data sample from our $1\, {\rm fb}^{-1}$ measurement~\cite{OurPRL}.}
\label{fig:FitnX0}
\end{figure}

As shown in Fig.~\ref{fig:EMFDMnoMat}, the data have a higher deposition in EM1 than the MC, so additional uninstrumented material must be added in front of the calorimeter to the detector model in the full MC.  We choose a relatively low atomic number material, copper, and add it to the simulation inside the solenoid.  The shape of the copper is a cylindrical shell with the same axis as the solenoid and uniform thickness. Along the $z$ direction, it extends over the length of the solenoid.  The shape of the missing material is driven by the observation that the materials in front of the central calorimeter have a geometry that is close to cylindrical. 

We use the improved {\sc geant}3 model described in Sec.~\ref{sec:improve_shower} to simulate the electrons from $Z\rightarrow ee$ events.  For these events, the thickness of the additional copper material is varied.  We then build a parametrized model of the mean EM layer energy fractions and the fluctuations around the average as a function of the copper thickness.  As shown in Fig.~\ref{fig:EMFDMnoMat}, we fit the ratio of the mean EM layer energy fraction in data to that in MC as a function of the $\zee$ event category to a constant for each of the first through third EM layers. We then form a total $\chi^2$ from the sum of the individual $\chi^2$ values from the three layer fits:
\begin{equation}
\chi^2 = \sum_{{\rm layer}(i)}\sum_{{\rm categ}(j)}\left[\frac{f^{\rm EML}_{ij} - \bar{f}^{\,\rm EML}_i}{\sigma_{ij}^{\rm EML}}\right]^2,
\end{equation}
where $f^{\rm EML}_{ij}$ (and $\sigma^{\rm EML}_{ij}$) are the data/full MC ratios of the mean EM layer energy fraction deposited by electrons with category $j$ at layer $i$ (and associated uncertainty), and $\bar{f}^{\,\rm EML}_i$ is the mean value of $f^{\rm EML}_{ij}$ for layer $i$. This is shown as a function of the thickness of the additional copper material in Fig.~\ref{fig:FitnX0}. The thickness of the cylinder is given as a multiple $nX_0$ of the thickness of one radiation length $X_0$ of copper.  This figure also shows the parabolic fit giving the minimum $\chi^2$ corresponding to the final thickness used in our tuned simulation, $nX_0 = 0.1633 \pm 0.0095$.  Because of the small energy deposit in EM4, we do not include it in our fits.

\begin{figure}[hbpt]
\centering
\includegraphics [width=\linewidth] {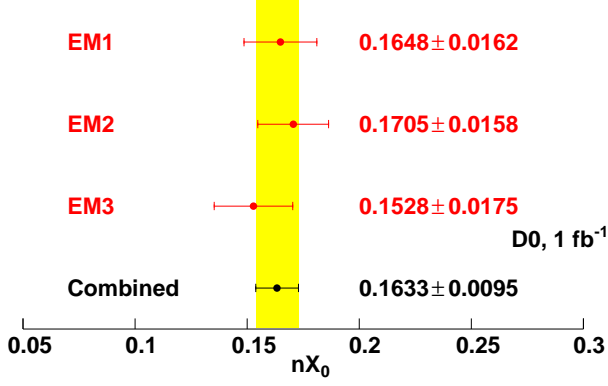}
\caption{[color online] Stability check: results of the fit for $nX_{0}$, performed separately for each of the three layers (EM1, EM2, and EM3). The result of the combined fit is also shown for comparison.}
\label{fig:PerLayerCheck}
\end{figure}

\begin{figure}[hbpt]
\includegraphics [width=\linewidth] {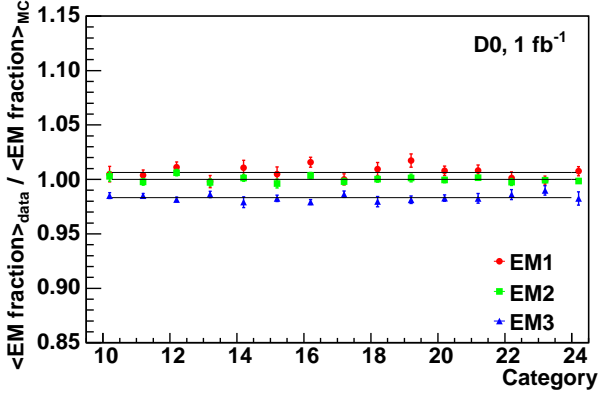}
\caption{[color online] The ratio of data to simulation for the means of the EM layer energy fraction distributions in \zee\ events for each of the first three EM layers and each of the 15~$\eta$~categories shown after the correction described in Sec.~\ref{sec:SystElecNonLin} is applied. Each of the three horizontal lines indicates the result of a fit of a common constant to the 15~data points from a given EM layer.}
\label{fig:EMFDMtuned}
\end{figure}

As a cross check, we repeat the fit for $nX_0$ separately for each of the three layers. The results are summarized in Fig.~\ref{fig:PerLayerCheck}. Good agreement is found between the overall fit and the results of the individual layers. The ratio of mean EM layer energy fraction in data to that in full MC after adding the missing material  is shown in Fig.~\ref{fig:EMFDMtuned}.  We interpret the deviations from unity as layer-intercalibration gain factors, which are applied during data reconstruction to have agreement with the detailed simulation.

\begin{figure}[hbpt]
  \includegraphics [width=\linewidth] {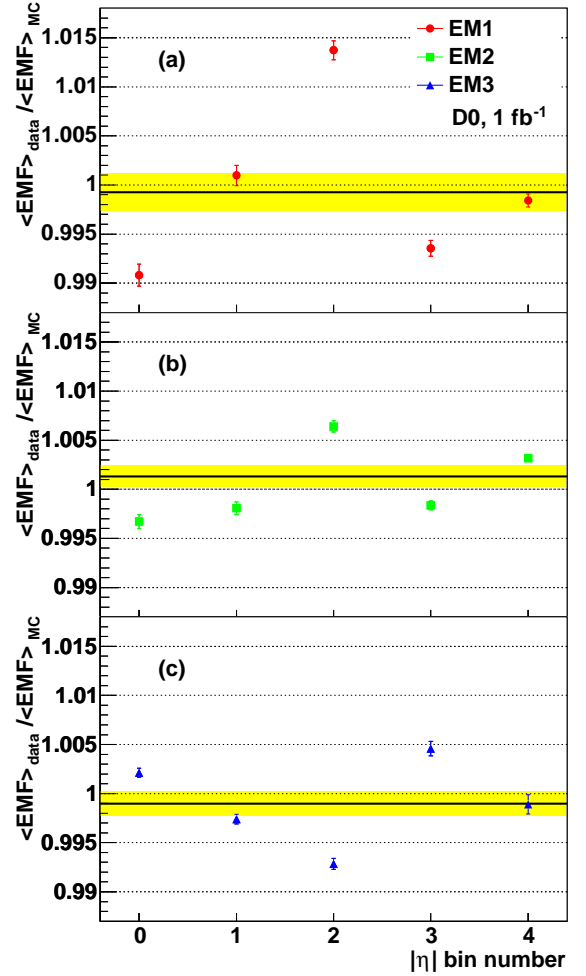}
  \caption{[color online] The data/full MC ratios for the means of the EM layer energy fraction distributions in $W\rightarrow e\nu$ events for the (a) EM1, (b) EM2, and (c) EM3 layers. The ratio is shown in five electron~$\eta$~bins. The thick horizontal lines indicate the average ratio across the central calorimeter and the yellow band represents the systematic and statistical uncertainty in the mean.}
  \label{fig:EMFDMW}
\end{figure}

Figure~\ref{fig:EMFDMW} shows the data/full MC ratio of the mean EM layer energy fraction for electrons from $W$ boson decays, using the same binning as in Table~\ref{table:StandardEtaBins}, after adding to the simulation the copper cylinder with thickness derived above and the layer intercalibration factors. Because of the larger number of $\wen$ events, it is possible to see non-statistical deviations from unity. These systematic deviations are an indication that the assumption of a cylindrical shape for the missing material is not perfect. Nevertheless, the mean values of the ratio across the central calorimeter are consistent with unity in EM1, EM2, and EM3.

\begin{figure}[hbpt]
  \includegraphics [width=\linewidth] {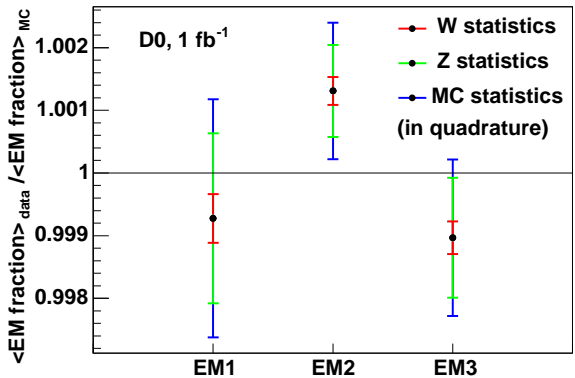}
  \caption{[color online] The mean data/full MC ratio for the means of the EM layer energy fractions, for electrons from $\wen$ decays, in each of the three first layers of the EM calorimeter. The innermost error bar (red) indicates the uncertainties from the $W$ boson sample size. The middle error bar (green) indicates the quadrature sum of the uncertainty from the $W$ boson sample size with the one from the $Z$ boson sample size, determined from the uncertainty in the thickness of the added material. Finally, the outermost error bar (blue) represents the quadrature sum of the two previous uncertainties with the one arising from the limited number of full MC events. In all three layers, the ratio is consistent with unity when all uncertainties are considered.} 
  \label{fig:EMFDMW_mean}
\end{figure}

Figure~\ref{fig:EMFDMW_mean} shows the mean values of the data/full MC ratio of the mean EM layer energy fraction for electrons from $W$ decays and the relative contributions for its uncertainty from the $W$ sample size, from the $Z$ sample size through the uncertainty in the thickness of the copper cylinder added to the simulation, and from the limited number of full MC events simulated with the improvements described in Sec.~\ref{sec:improve_shower}.

The precision of the measurement of the material in front of the calorimeter contributes directly to the energy measurement of the electron and therefore to the $W$ boson mass. Our measurement of $M_W$ depends critically on the assumption that the calibration made at the $Z$ boson mass is valid at the $W$ boson mass scale. A mismeasured material distribution would be the primary source of a non-linearity in this scaling. The uncertainty on the $W$ boson mass arising from the material tune is derived by varying the additional material by $\pm 1$ standard deviation (shown in Fig.~\ref{fig:FitnX0}) and recalibrating the EM calorimeter for each variation.  We build fast MC models of the response considering the combination of the material variation and impact of calibration procedure.  

The fast MC models resulting from $\pm 1$ standard variations in the additional material are used to generate $W$ boson events. The $m_T$, $\pte$, and $\met$ distributions from these events are fit to templates generated with the standard parameterization and the resulting $M_W$ is compared to the input mass. We find shifts of 4 MeV using the $m_T$ distribution for the fit, 6 MeV using the electron $\pte$ distribution, and 7 MeV for the $\met$ distribution.

\subsection{Energy Loss Corrections}
\label{sec:QElossCorr}
        
The average electron energy loss is recovered with correction functions determined using full MC samples of single-energy electrons with incident energies from 1 GeV to 135 GeV and applying the improvements described above. The precision of the corrections is therefore limited by the statistical precision of the full MC sample.  As will be discussed in Sec.~\ref{sec:elec_energy}, the final tuning of the electron energy response using $\zee$ events from the data fixes some imperfections in the energy-loss parametrization, for example, a global scale shift in the energy-loss function.

\begin{figure}[htbp]
  \includegraphics[width=\linewidth]{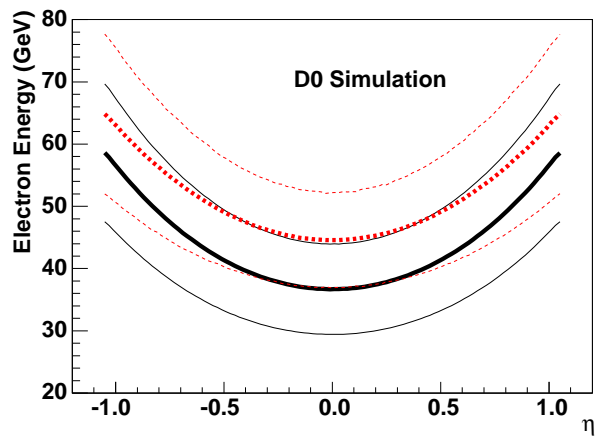}
  \caption{[color online] The mean electron energy versus $\eta$  for electrons from $W$ boson (black solid line) and $Z$ boson (red dashes) events. The thin lines indicate the one standard deviation bands of the energy distributions versus $\eta$. \label{fig:e_vs_eta_W_and_Z}}
\end{figure}

\begin{figure}[ht]
\centering
\includegraphics [width=\linewidth] {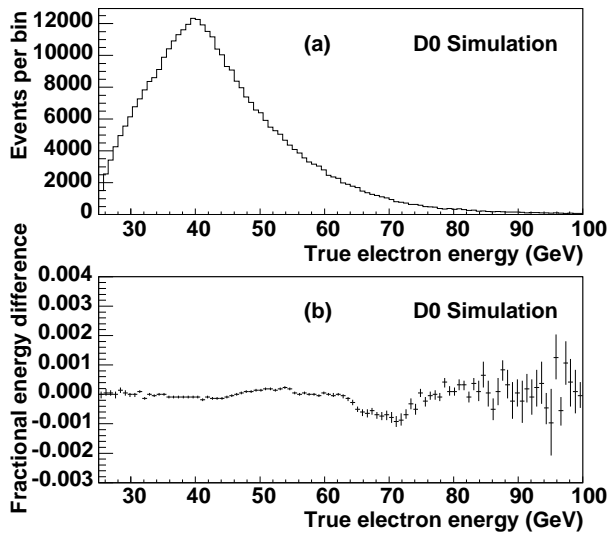}
\caption{(a) The true energy spectrum for electrons in simulated $W$ boson events that pass the full selection after reconstruction, and (b) the mean ratio of measured minus true energy to the true energy for electrons from $\zee$ events minus the same quantity for electrons from $\wen$ events as a function of true electron energy.}
\label{fig:EscaleMistakeBottomLine}
\end{figure}

Because of the difference between the $Z$ and $W$ boson masses, the electrons from $\zee$ decays populate one band in $E^e$ versus $\eta$ space and electrons from $W\to e\nu$ populate another band (see Fig.~\ref{fig:e_vs_eta_W_and_Z}). If the energy dependence of the energy loss correction is not correctly derived, the energy scale tuned on $\zee$ events will be slightly incorrect when applied to $\wen$ events. To estimate this effect, we calculate the mean difference between reconstructed and true electron energies, divide it by the true energy for electrons from $\wen$ events, and subtract the same quantity calculated using electrons from $\zee$ decays. The difference between the two averages reflects the imperfection of the energy loss corrections that cannot be corrected by the final tuning in the fast MC. The result is shown in Fig.~\ref{fig:EscaleMistakeBottomLine}. In order to estimate a systematic uncertainty for this imperfection in the energy loss corrections, we translate the difference between the corrections in $\wen$ and $\zee$ events as an electron energy shift in fast MC pseudo-experiments. After propagating the shift to the $W$ boson mass, we assign an uncertainty of 4 MeV for the fit with the $m_T$, $p_T^e$, and $\met$ observables.

\subsection{Validation of Analysis for {\boldmath $4.3 \text{fb}^{-1}$} Data Set}
\label{sec:EMfracValidation}

The uninstrumented material correction presented here is derived with the $\zee$ data sample of our $1\, {\rm fb}^{-1}$ analysis (Run~IIa). It is used again here, for the analysis of the Run~IIb data corresponding to $4.3\, {\rm fb}^{-1}$, because the distribution of EM layer energy fractions is essentially identical to the distribution of EM layer energy fractions in the Run~IIa measurement. There are two differences between the running conditions during Run~IIa and Run~IIb relevant to EM showers:

\begin{itemize}
\item Increased pileup in Run~IIb.
\item Insertion of an inner silicon tracking layer (L0) between Run~IIa and Run~IIb ($\approx$ 0.003 $X_0$).
\end{itemize}
The inclusion of L0 represents a small contribution to the total amount of uninstrumented material when compared to the CFT, solenoid, CPS, and cryostat, all of which remained unchanged throughout Run~II. 

\begin{figure}[hbpt]
  \centering
  \includegraphics[width=\linewidth]{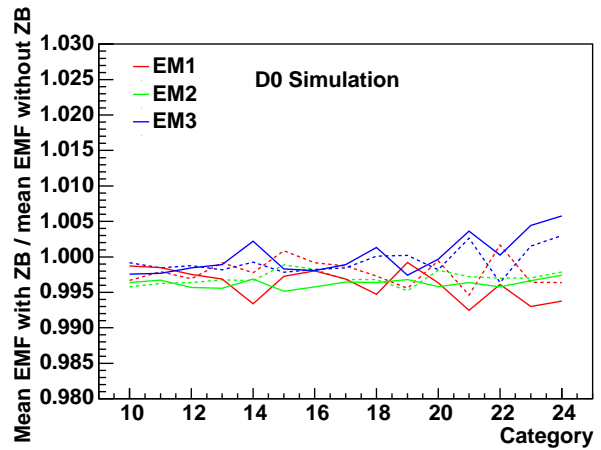}
  \caption{[color online] Each line represents the ratio of the mean EM layer energy fractions simulated with zero-bias overlay to the same sample simulated without overlay. It represents the contribution from extra $p\overline{p}$~interactions and noise to the mean EM layer energy fractions, which is determined separately for the Run~IIa sample (dotted lines) and the Run~IIb sample (continuous lines). The ratio between the continous to the dashed line is used as a correction factor to the EM layer energy fractions measured in Run~IIa when comparing them to the Run~IIb fractions (Fig.~\ref{fig:EMcmpab}).
    \label{fig:EMpileupab}}
\end{figure}

\begin{figure}[hbpt]
  \centering
  \includegraphics[width=\linewidth]{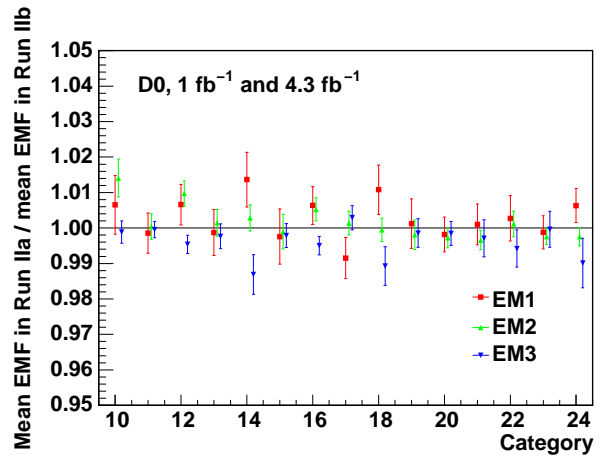}
  \caption{[color online] Ratio of the means of the EM layer energy fraction distributions in \zee\ events between the Run~IIa analysis and the present Run~IIb analysis, separately for each of the four EM layers and each of the 15 standard $\eta$~categories. 
    \label{fig:EMcmpab}}
\end{figure}

Figure~\ref{fig:EMpileupab} shows the contribution from extra $p\overline{p}$ interactions and noise to the mean EM layer energy fractions in $Z\rightarrow ee$ events, estimated separately for Run~IIa and Run~IIb.  Figure~\ref{fig:EMcmpab} shows the EM layer energy fraction distributions in $Z \rightarrow ee$ data for Run~IIa and Run~IIb, after correcting the Run~IIb data by the Run~IIa/Run~IIb ratio from Fig.~\ref{fig:EMpileupab}. The differences between Run~IIb and Run~IIa EM layer energy fractions are compatible with statistical fluctuations from the size of the $\zee$ data sample, with $\chi^2$ of 13.5, 23.0 and 22.0 for 15 degrees of freedom in EM1, EM2, and EM3, respectively.

\section{Generators for Full and Fast Simulation}
\label{sec:gener}

The initial step in constructing templates for extracting the $W$ boson mass is simulation of vector boson production and decay kinematics. The complete list of event generators used in this analysis is shown in Table~\ref{tab:generators}. We use the {\sc resbos}~\cite{resbos,resbos1,resbos2} program coupled with the {\sc CTEQ6.6} NLO parton distribution functions (PDFs)~\cite{cteq66} as the standard event generator. {\sc resbos} provides a good description of the dominant QCD effects, namely the emission of multiple gluons, that influence the shape of the boson~$p_T$ distribution at low boson~$p_T$. The $W$ boson $p_T$ spectrum has a significant impact on the generated $\pte$ and $\ptnu$ spectra. Its accurate description is an important ingredient of the $M_W$ measurement. 

The dominant effect from EW corrections on the $M_W$ measurement arises from radiation of a single photon from the final state charged lepton. This process is simulated by combining {\sc resbos} with {\sc photos}~\cite{photos}.  


\begin{table}
\begin{center}
\caption {{\label{tab:generators}} Event generators for $W$ boson and $Z$ boson
 processes used in this analysis. {\sc pythia} is
 used for the full MC closure test and for estimating PDF uncertainties.
 {\sc wgrad} and {\sc zgrad} are used only for estimation of QED theory
 uncertainty.}
\begin{tabular}{c | c c c}\hline\hline
{Tool} & {Process} & {QCD} & EW \\ \hline 
{{\sc resbos}} & $W$,$Z$ & NLO & - \\ 
{{\sc pythia}} & $W$,$Z$ & LO & QED FSR\\\hline
{{\sc wgrad}} & $W$ & LO & complete $\mathcal{O}(\alpha)$  \\
{{\sc zgrad}} & $Z$ & LO & complete $\mathcal{O}(\alpha)$ \\
{{\sc photos}}   &     &   & QED FSR \\\hline\hline
\end{tabular}
\end{center}
\end{table}

\subsection{QCD Corrections and Boson {\boldmath $p_T$}}
\label{sec:bosonpT}

{\sc resbos} uses the triple differential cross section $d^3\sigma/dp_T\,dy\,dM$ for $Z/\gamma^{*}$ and $W$ boson production, where $p_T$ is the boson transverse momentum, $y=\frac{1}{2}\ln[(E+p_z)/(E-p_z)]$ is the boson rapidity, and $M$ is the boson mass. The triple differential cross section is tabulated on a grid for discrete values of $p_T$, $y$, and $M$. They are calculated using the CSS $p_T$ resummation technique~\cite{b-css_1, b-css_2} for low boson $p_T$ matched to a fixed order calculation at high $p_T$. The resummation is performed in impact parameter space with Sudakov exponents calculated to NNLL precision and Wilson coefficients calculated to NLO precision. At large impact parameters, the perturbative calculation is modified by a phenomenological non-perturbative factor. In this measurement, we use the BNLY~\cite{g2} parametrization for the non-perturbative factor which is a function of three variables, $g_1$, $g_2$, and $g_3$.

The observed boson $p_T$ spectrum in this measurement is mostly sensitive to $g_2$ and has very limited sensitivity to the other non-perturbative parameters and scales in the cross section. Therefore, we take the uncertainty in $g_2$ as representative of the boson production model uncertainty. We use the world average for $g_2$~\cite{g2}, and the uncertainty is propagated using pseudo-experiments generated by varying $g_2$ within its quoted uncertainty. We find uncertainties of 2~MeV, 5~MeV, and 2~MeV for the $\mt$, $\pte$, and $\met$ fits, respectively.

\subsection{Electroweak Corrections}
\label{sec:EWcorrections}

In our fast MC, care is taken to model the EW corrections to $W$ boson production and decay as well as the detector response to the emitted photons. The most important correction is the real emission of final state photons, since it takes away some of the energy of the electron, and the invariant mass of the electron and neutrino will be smaller than the $W$ boson invariant mass, biasing the measurement.

As discussed above, we use {\sc photos} to simulate the leading effects of real photon emission. To estimate the uncertainties from this modeling, we explore the difference between the shower simulation done by {\sc photos} and the EW NLO calculation available in {\sc wgrad}~\cite{wgrad} and {\sc zgrad}~\cite{zgrad}. In the shower simulation done by {\sc PHOTOS}, a final state radiation (FSR) emission probability kernel is introduced that is accurate only in the collinear limit. In the NLO simulation done by {\sc wgrad} and {\sc zgrad}, all one-loop real and virtual contributions are considered, including interference terms, but a soft and a collinear cutoff are introduced to avoid infrared divergencies.  {\sc wgrad} and {\sc zgrad} cannot be used to measure $M_W$, since they do not include higher-order QCD corrections, but are adequate to estimate the purely EW uncertainties.

{\sc wgrad} allows both shower and EW NLO calculations. We generate pseudo-experiments using both options and fit them against templates prepared with {\sc photos}. The difference of the fitted $M_W$ is taken as a measure of the uncertainty and is found to be 5 MeV for the $m_T$, $p_T^{\ e}$ and $\met$ fits.

To estimate the uncertainty in the EW NLO calculation itself, we study the dependence of the measured $M_W$ on the soft and collinear cutoffs. No variation is observed by changing the collinear cutoff, but a non-negligible effect is seen when varying the soft cutoff. We consider the difference between the cutoff at $10$ MeV and at $800$ MeV as an estimation of the uncertainty due to higher-order corrections~\cite{wgradnote}. We find shifts of 2~MeV, 1~MeV, and 3~MeV for the $m_T$, $p_T^e$, and $\met$ fits, respectively.

Finally, an experimental scale is also present in the FSR simulation: the radius of the cone used as the boundary between photons whose energy is detected as part of the electron cluster or as part of the unclustered recoil. The simulation uses the value $\Delta R=0.3$ as standard, and we vary it by the size of a cell of the D0 calorimeter, between $\Delta R = 0.2$ and 0.4, to estimate the uncertainty coming from this experimentally introduced scale. We find uncertainties of 1~MeV, 1~MeV, 5~MeV for the $\mt$, $\pte$, and $\met$ fits, respectively.

\subsection{Parton Distribution Functions}
\label{sec:PDF}


The $M_W$ measurement is sensitive to PDF variations because of the limited detector acceptance and the thresholds on the selection of the decay products kinematics. In the ideal case of full pseudorapidity acceptance by the detector and no kinematic cuts, the lack of knowledge of the PDFs would introduce a negligible uncertainty on $M_W$. We determine the systematic uncertainty arising from the PDFs using {\sc pythia} and the CTEQ6.1 PDF set~\cite{cteq}, which is available at LO. We generate pseudo-experiments using the 40 members of the CTEQ6.1 error set, each of which corresponds to a one-sided uncorrelated variation of the effective $\chi^2_{\text{eff}}$ used for the PDF determination. The variation adopted in the CTEQ6.1 error set corresponds to $\Delta\chi^2_{\text{eff}} = 100$. Studies from the CTEQ collaboration show that a 90\% C.L. can be achieved with $\Delta\chi^2_{\text{eff}}$ between 100 and 225, depending on the specific experiment in the global analysis~\cite{cteq_1, cteq_2}.
 
The pseudo-experiments from each of the 40 members of the error set are compared to mass templates generated using the nominal set. Following the CTEQ prescription, we take the average of the two-sided variation $|M^{+}-M^{-}|/2$ as the estimate of the uncertainty for each uncorrelated combination of the PDF parameters. The total uncertainty is determined with the prescription: 
\begin{eqnarray}
\Delta M_W =\frac{1}{1.64}\sqrt{\sum_{i=1}^{20} \left( \frac{M^{+}_{i}-M^{-}_{i}}{2} \right)^{2}},
\end{eqnarray}
where the factor $1/1.64$ brings the coverage of the uncertainty to 68\% C.L.

The final PDF uncertainties are found to be 11~MeV, 11~MeV, and 14~MeV for the $\mt$, $\pte$, and $\met$ methods. These values are slightly larger than those in our Run IIa measurement~\cite{OurPRL}, which uses exactly the same prescription, due to the deterioration of our hadronic recoil resolution at higher luminosity.

\section{Determination of the Fast Simulation Parameters}
\label{sec:param}

As described in Sec.~\ref{sec:eventcharacteristics}, $W(Z)$ events are characterized by the measurements of the electron(s) and the hadronic recoil in the event.  Our fast simulation is designed to reproduce these measurements and their correlations starting from the four-vectors provided by an event generator (Sec.~\ref{sec:gener}).  The simulation consists of four parts: (1) simulation of the vertex $z$ coordinate, (2) simulation of the electron reconstruction and identification efficiency, (3) simulation of the electron energy measurement, and (4) simulation of the hadronic recoil energy measurement.  The vertex $z$ coordinate is needed to predict the detector regions with which the electrons interact when computing efficiencies and reconstructed energy. In our fast simulation, photons within the electron energy reconstruction cone (Fig.~\ref{fig:ewindow}) of a parent electron are merged back into the electron, treating the resulting electron plus photons system as the reconstructed electron. This procedure takes into account the reconstruction inefficiency induced by the photons as well as the probability of low energy photons to reach the calorimeter.  Photons far from electrons are reconstructed as part of the recoil system and are so described in our fast simulation.

Here, we describe the models used in the fast MC to simulate data and full MC.  Separate tunes are required for data and for full MC because our full MC does not describe our data with an accuracy sufficient to measure $M_W$. We perform the full measurement of $M_W$ twice: once using as input full MC and once with data. By treating the full MC events as data and using the same parametrized detector model, but with different parameters, we validate our experimental procedure. In our full MC measurement, we obtain a difference of our measured $Z$ mass from the input mass of $-3 \pm 4$ MeV and a difference of our measured $W$ mass from the input mass of $-2 \pm 5$ MeV from the fit to the $m_T$ distribution, $-2 \pm 5$ MeV from the fit to the $p^e_T$ distribution, and $+5 \pm 6$ MeV from the fit to the $\met$ distribution.  These uncertainties are statistical, reflecting the size of the full MC sample.

\subsection{Vertex Parametrization}

We select only events with vertex position $|z_V|<60$~cm and electrons with $|\eta_{\rm det}|<1.05$ for the final analysis. Since the electron $\eta_{\rm det}$ depends on the electron $\eta$ and the vertex position, we need a model that can be used in the fast MC to predict the vertex distribution. The beam shape is modeled as a product of a Gaussian bunch length with a Lorentzian shape set by the accelerator $\beta^*$ functions in both transverse directions. The parameters are determined from fits to the vertex distribution for randomly triggered beam crossings.

\ \

\subsection{Electron Efficiency Parametrization}
\label{sec:eff}

The $\mt$, $\pte$, and $\met$ distributions are modified by inefficiencies in electron identification that depend on the event kinematics and on the hadronic environment. These effects introduce biases in the measured $M_W$ which must be accounted for. We accomplish this by building an efficiency model in the fast MC that reproduces the inefficiencies effects.  In this section we discuss the components of the fast MC model used to predict the combined electron reconstruction and identification efficiencies.  We begin by giving an overview of the model, then discuss each of the components, and end with a discussion of the model validation.

The efficiency model begins by describing the effect of FSR photons in the electron reconstruction and identification efficiency. Then, from $\zee$ data, we determine the effect of known sources of inefficiency, such as those arising from the trigger system or from the HMatrix, isolation, EM fraction and track matching requirements. The collective effect from other sources of inefficiency, such as pile-up, is modeled using full MC simulation. In the last step, data control samples are used again to provide final corrections to the full MC model.  The final corrections are small because the full MC used as reference has collider data zero-bias events added to the simulated hard scatter. These zero-bias events are added to the low-level channel information without zero-suppression, allowing for the modeling of the impact of hadronic energy in the reconstruction and identification of electrons, which is the leading source of inefficiency in a high instantaneous luminosity environment.

The electron identification efficiency model must be multi-dimensional and must depend on all quantities that introduce biases in the reconstructed $M_W$.  In the ideal case, a single multi-dimensional efficiency would depend on all necessary variables and automatically include all correlations. However, the \zee\ control data sample is not large enough to establish a model by binning the efficiency in all relevant variables to derive a single function. Many of the dependencies are, however, largely uncorrelated with each other, and our full MC program can describe parts of the efficiency reasonably well.

The overall efficiency $\epsilon$ can be written as a product of several terms:
\begin{widetext}
\begin{eqnarray}
 \epsilon  &=&  \epsilon_{\rm trig}(p_T^e) 
  \, \times \, \epsilon_{\rm FSR}(X,\Delta R,\eta,E^e)
  \, \times \, \epsilon_{\rm trk}(z_V,\eta,p_T^e) 
  \, \times \, \epsilon_{\rm EMID}(\eta_{\rm det},p_T^e)
  \, \times \, \epsilon_{\phi_{\mathrm{mod}}}(\phi_{\mathrm{mod}}) \nonumber \\
  & &  \, \times \, \epsilon_{\phi}(\phi^e)
    \, \times \, \epsilon_{\mathrm{had}}(\text{SET},p_T^e,\eta_{\rm det},L,\upara)
    \, \times \, R_1(\text{SET},L)
    \, \times \, R_2(\upara),
\label{e-effi}
\end{eqnarray}
\end{widetext}
in which $\epsilon_{\rm trig}$ measures the trigger efficiency for recording events in the sample, $\epsilon_{\rm FSR}$ the efficiency arising from radiated photons, $\epsilon_{\rm trk}$ the efficiency of the track selection requirement, $\epsilon_{\rm EMID}$ the efficiency of the calorimetric requirements used in the electron selection, and $\epsilon_{\phi_{\mathrm{mod}}}$ the efficiency loss caused by the calorimeter module boundaries. The efficiency $\epsilon_\phi$ models the electron $\phi$ dependent efficiency and $\epsilon_{\mathrm{had}}$ the effect on electron finding arising from hadronic activity in the event.  The term $\epsilon_{\mathrm{had}}$ also describes the effect of multiple $\ppbar$ interactions on the electron identification.

Finally, $R_1(\text{SET}, L)$ is introduced to account for imperfections in the efficiency description in full MC at high instantaneous luminosity, especially the one related to track matching, while $R_2(u_{\parallel})$ is introduced to describe the fine details of the hard recoil (see Sec.~\ref{smearing_model}) in the electron identification and reconstruction efficiency that were not fully described by the hadronic energy dependent efficiency. The correction $R_1$ is derived from a comparison of the efficiency in data and full MC, while $R_2$ is derived from a comparison of the efficiency between data and fast MC in which all previously determined efficiencies are applied to the fast MC. We describe each of these efficiencies in the following sections. The overall normalization of the total efficiency does not enter this analysis because the fast MC yields are always normalized to the data or full MC yield.

\subsubsection{Trigger Efficiency}
\label{sec:TriggerEff}

Events used in this analysis must satisfy one of the single-electron triggers described in Sec.~\ref{sec:det}. For this analysis, a one-to-one correspondence between a run period and a specific trigger is enforced. To achieve correspondence, we choose the lowest $p_T$ threshold unprescaled trigger available for the period. The efficiency for any of these three triggers will be less than unity near the threshold because the measured energy differs between the trigger system and the offline reconstruction program.  The efficiency modeling these effects, $\epsilon_{\rm trig}$, is thus a function of electron $p_T$.

A tag and probe method is used with data $\zee$ candidate events to measure the trigger efficiency as a function of $p^e_T$.  We require one electron (the tag) in a $\zee$ event to pass all selection requirements including the trigger.  The other electron (the probe) is initially required to pass the full selection except a requirement regarding the trigger.  The efficiency is then determined from the rate of electrons passing the trigger whose efficiency is being measured. For this efficiency determination, we allow the tag electron to be in the EC to gain statistics, but the probe electron must be in the CC.

\begin{figure}[htbp]
   \includegraphics[width=\linewidth]{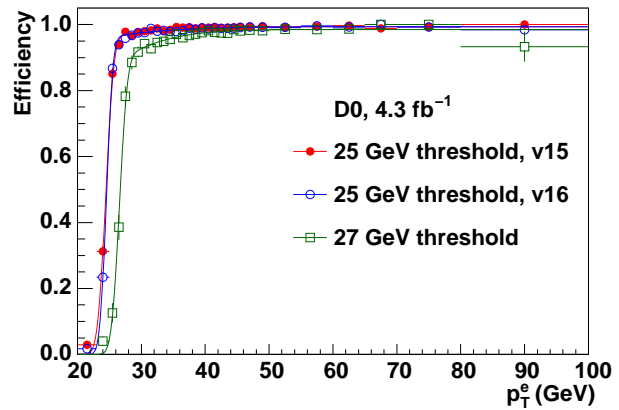}
   \caption{Trigger efficiency as a function of $p^e_T$ for the three triggers
   used.}  \label{fig:trigeff}
\end{figure}

The resulting measured efficiency is shown in Fig.~\ref{fig:trigeff} for each of the three triggers.  When simulating the trigger in the fast MC, a mix of the three efficiencies is used such that each replicates the frequency in data as determined from the integrated luminosity exposure for each trigger.  This efficiency is only used when using the fast MC to simulate events for comparison to collider data.  It does not apply to the full MC analysis.

\subsubsection{FSR Efficiency}
\label{sec:elec_fsr}

Radiated photons (FSR) close to or inside the electron reconstruction cone will affect the electron identification efficiency because of isolation, shower shape and track matching requirements.  To account for these effects, we introduce an electron efficiency $\epsilon_{\rm FSR}(X,\Delta R, \eta, E^e)$.  Here, $X$ is the fraction of the electron energy carried by the photon and $ \Delta R = \sqrt{\left[\phi(e)\,-\,\phi(\gamma)\right]^2\,+\,\left[\eta(e)\,-\,\eta(\gamma)\right]^2} $ measures the separation between the electron and photon.

\begin{figure*}[t]
\includegraphics [width=0.75\textwidth] {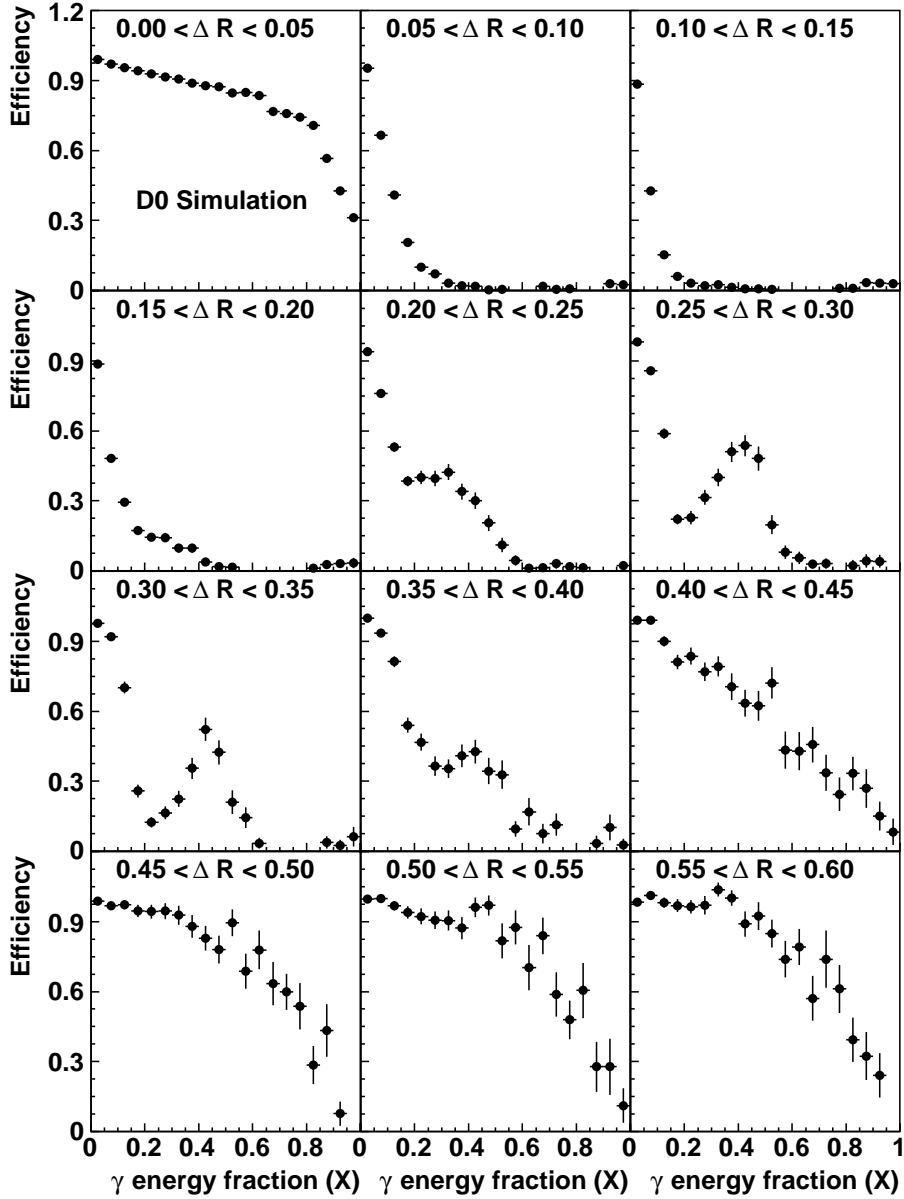}
\caption{ Electron identification efficiency for electrons accompained by FSR determined from full MC as a function of the fraction of the energy carried by the photon. Each pane corresponds to a different $\Delta R$ region, and the distributions are integrated over $\eta$ and $E^e$.}
 \label{fig:eideff_gammafrac}
\end{figure*}

The parametrization is derived by studying the electron reconstruction efficiency using two full MC samples: one with single electrons having the kinematics of those from $W\to e\nu$ decay that are accompanied by FSR photons, and a second sample that includes exactly the same events as the first one, except that the FSR photon energy has been added to the energy of the electron, and the photons themselves are removed.  Both samples have zero-bias event overlay, and the same zero-bias event is overlaid on a given $W$ boson event in each of the two samples.  The ratio of the electron yields in the first sample to that in the second sample defines this efficiency.  The efficiency is determined in bins of the four variables, $X$, $\Delta R$, $\eta$, and $E^e$. Figure~\ref{fig:eideff_gammafrac} shows examples of the electron reconstruction efficiency versus $X$ in twelve $\Delta R$ bins. The shapes of these efficiencies as functions of $X$ and $\Delta R$ are primarily a combination of effects of the photon distorting the cluster shower shape and cluster centroid position, causing the EMID or track match requirements to fail, and of the photon carrying sufficient energy that the electron fails either the track or cluster $p_T$ requirement.

The efficiency in the first three $\Delta R$ bins is mainly driven by the track matching requirement and, to a lesser extent, by the shower shape requirement. While the photon is still close enough to the electron for most of its energy to be deposited in the same reconstruction cone, the shower shape at large values of $X$ becomes too different from that expected for a single electron, and, more importantly, the calorimeter-based estimate of the cluster position deviates significantly from the track-based expectation. In intermediate $\Delta R$ bins, the photon is in the region that interferes with the cluster isolation requirement. The peak at intermediate values of $X$ separates the regimes in which the cluster is reconstructed around the electron or around the photon. In the last three $\Delta R$ bins, the photon is far away from the electron cone and does not directly interfere with the electron reconstruction.

\subsubsection{Track-Matching Efficiency}
\label{sec:eff_track}

\begin{figure}[htp]
\includegraphics [width=\linewidth] {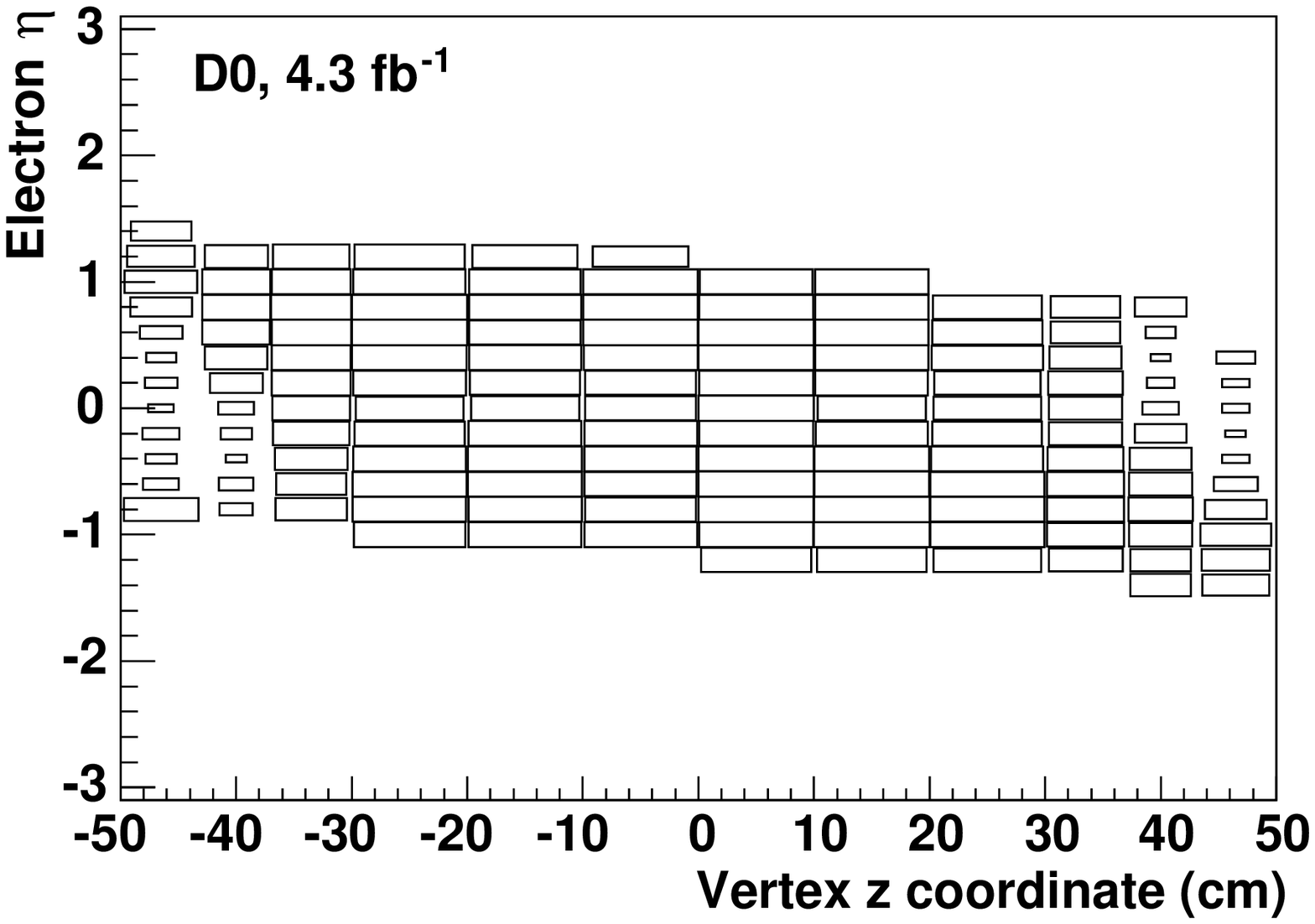}
\caption{Track-matching efficiency as a function of $z_{V}$ and $\eta$ in data. The efficiency is proportional to the area of the boxes.}
\label{fig:tight_trk_eff}
\end{figure}

\begin{figure}[htp]
\includegraphics[width=\linewidth]{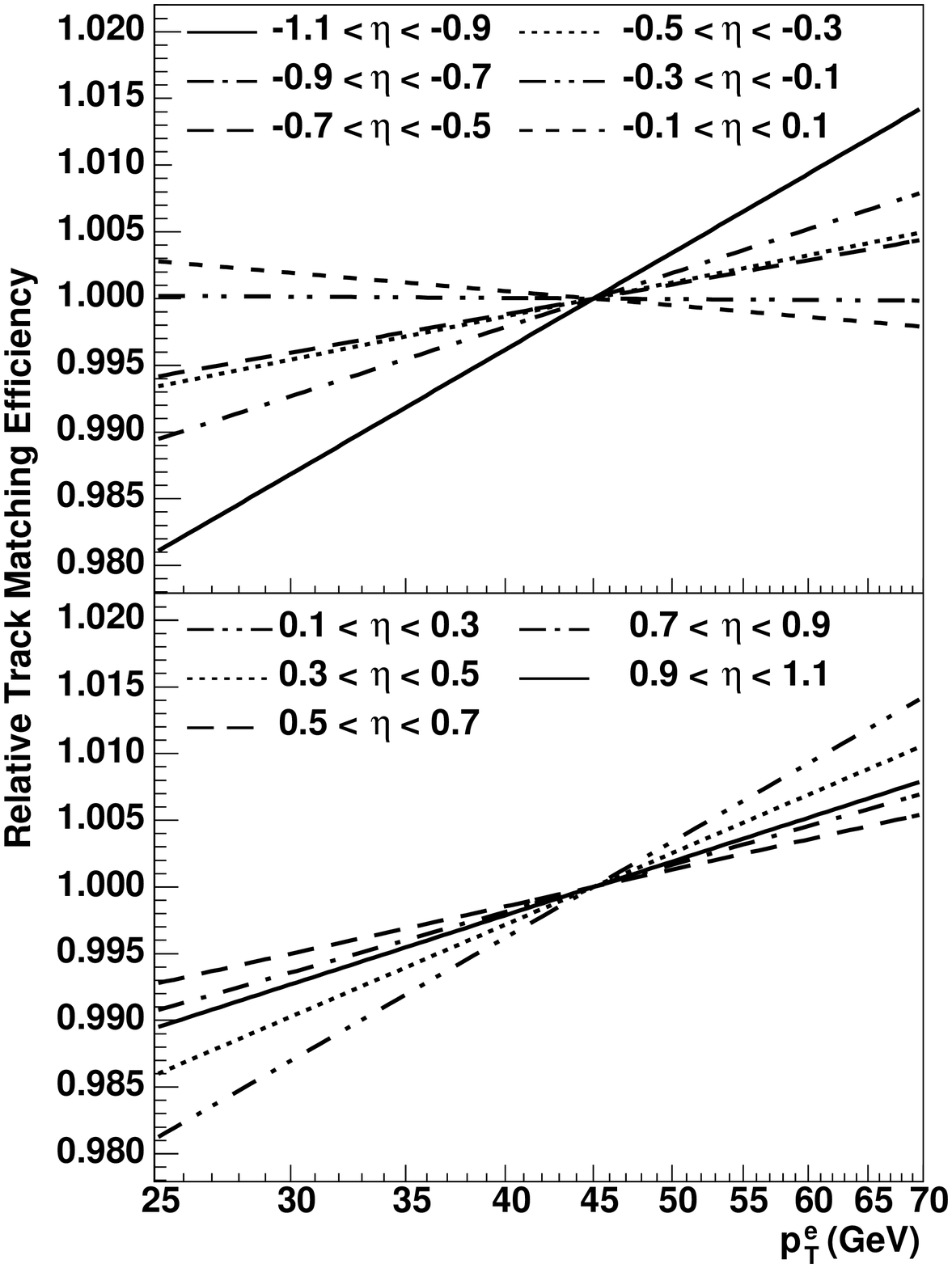}
\caption{The $\pte$-dependent perturbation over the track-matching efficiency in 11 bins of $\eta$ in full MC. The slopes change because electrons with higher energy are more easily matched to the calorimeter cluster. This efficiency perturbation is normalized at $p^e_T=45$ GeV.  The total track-matching efficiency is the product of the efficiencies shown in Fig.~\ref{fig:tight_trk_eff} and~Fig.~\ref{fig:tight_trk_eff_ptdep}.}
\label{fig:tight_trk_eff_ptdep}
\end{figure}

The track-matching efficiency $\epsilon_{\rm trk}(z_V,\eta,p_T)$ is
described as a product of two efficiencies, one expressed as a
function of $z_V$ and $\eta$ and the second expressed as a function of
$p_T$ and $\eta$.  The first of these is derived using the
tag and probe method applied to $\zee$ candidate events.  The probe electron
is initially required to pass all selections except the tracking requirements.
The resulting efficiency is shown in Fig.~\ref{fig:tight_trk_eff}.  Because
this is derived for both variables simultaneously, the correlations are
automatically included.  The second function describes the $p_T$ dependence and
correlation with $\eta$ of the track requirements. Because of the limited size of the $Z$ boson sample, the dependence is derived from
full MC.  It is modeled with an $\eta$-dependent logarithmic function $\epsilon_{\rm trk}(z_V,\eta,p_T) = \epsilon_{\rm trk}(z_V,\eta)\times \left[p_0(\eta) + p_1(\eta)\log(p_T)\right]$, and shown in Fig.~\ref{fig:tight_trk_eff_ptdep} for different $\eta$ regions. It is interpreted as a perturbation over the efficiency $\epsilon_{\rm trk}(z_V,\eta)$, without changing the relative normalization in each $\eta$ region.

\subsubsection{EM Identification Efficiency}

The efficiency accounting for the EM cluster finding, HMatrix, isolation, and EM fraction requirements is derived from $\zee$ data, again using the tag and probe method.  For this determination, the probe object is a track that passes the tracking requirements, and the invariant mass of the track and tag electron is required to be consistent with that of a $Z$ boson.

\subsubsection{Electron $\phi_{\mathrm{mod}}$ Efficiency}
\label{sec:eff_phimod}

The D0 calorimeter has 32 EM modules in the CC region. Each module is two cells wide, hence has a width of $2\pi/32\approx 0.2$~radian in $\phi$. Between any two adjacent modules, there is an uninstrumented region (crack) of width of $\approx 0.02$~radian in $\phi$. An intra-module $\phi$ variable $\phi_{\mathrm{mod}}$ is defined as the fractional part of $32\phi/2\pi$. This variable measures the angular position within any module as a fraction of module width (with $0 \leq \phi_{\mathrm{mod}} \leq 1$). Each of the EM1, EM2, and EM4 layers in an EM module consists of two readout cells. The central value $\phi_{\mathrm{mod}}=0.5$ corresponds to the inter-cell boundary in $\phi$ and values close to 0 and 1 are the module edges. The EM3 layer is segmented twice as finely in both $\eta$ and $\phi$ (0.05 radian wide). The $\phi_{\mathrm{mod}}$ values at 0.25, 0.5, and 0.75 correspond to the inter-cell boundaries of EM3.

\begin{figure}[htbp]
\centering
\includegraphics[width=\linewidth]{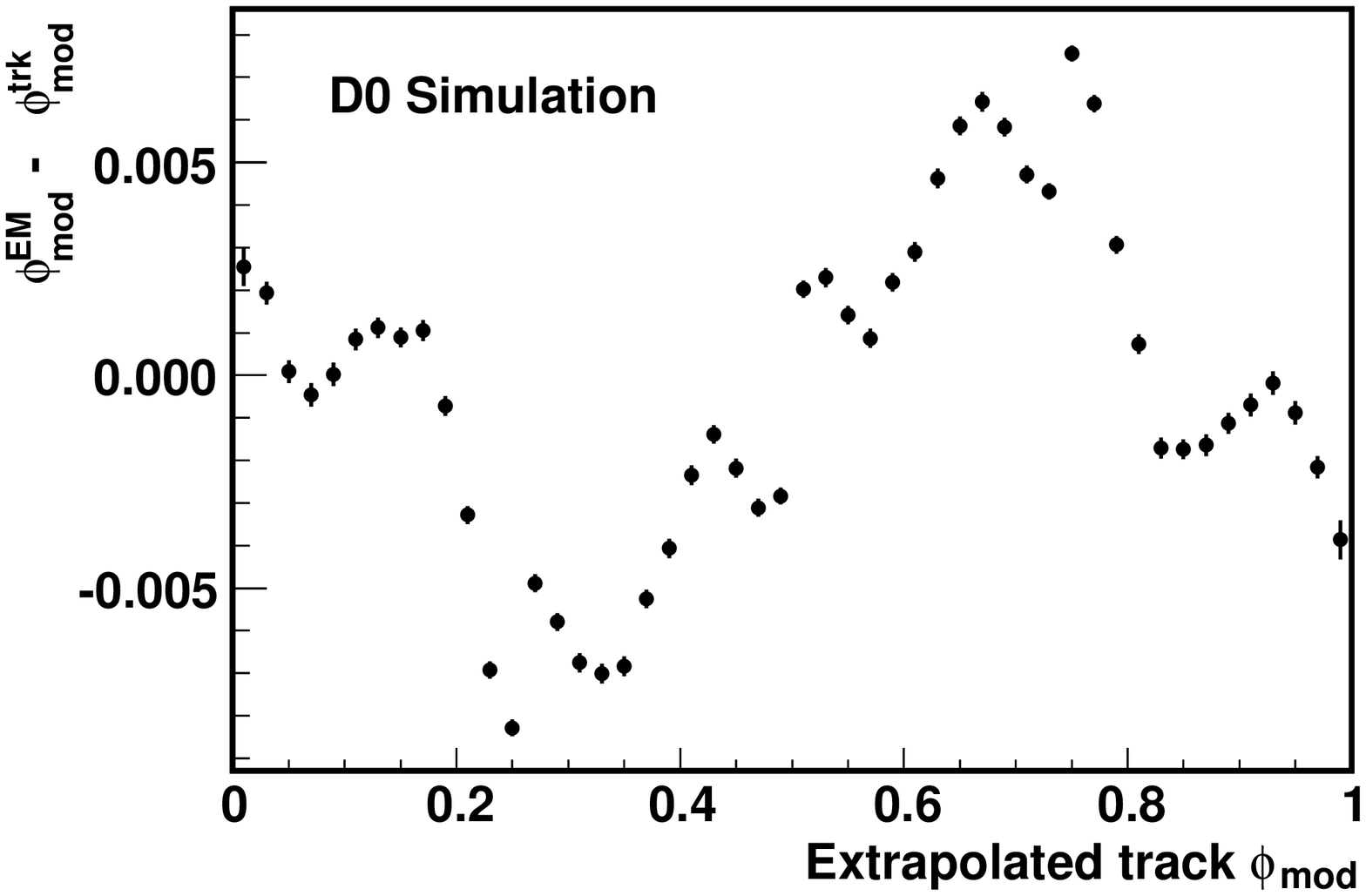}
\caption{Average difference between $\phi^{\rm EM}$ and $\phi^{trk}$ 
 in module units as a function of $\phi^{\rm trk}_{\mathrm{mod}}$ extrapolated to EM3.}
\label{fig:PhiModBias}
\end{figure}

Because of cell boundaries inside and uninstrumented regions outside an EM module, the reconstructed electron cluster center $\phi^{\rm EM}$ is biased away from these regions. Figure~\ref{fig:PhiModBias} shows the $\phi^{\rm EM}_{\mathrm{mod}}$ shift, $\phi^{\rm EM}_{\mathrm{mod}} - \phi^{\rm trk}_{\mathrm{mod}}$ as a function of $\phi^{\rm trk}_{\mathrm{mod}}$, which is calculated from the track $\phi$ extrapolated to the depth of EM3. Since $\phi^{\text{trk}}_{\text{mod}}$ is unbiased by calorimeter uninstrumented regions, we see a strong tendency for the $\phi^{\rm EM}_{\text{mod}}$ to move away from these regions, resulting in a bias in the EM cluster center. We also observe direction biases near the inter-cell boundaries with a complicated structure that arises from the sharing of shower energy among the EM3 cells used to define the calorimeter-based $\phi$ measurement.

\begin{figure}[htbp]
\centering
\includegraphics[width=\linewidth]{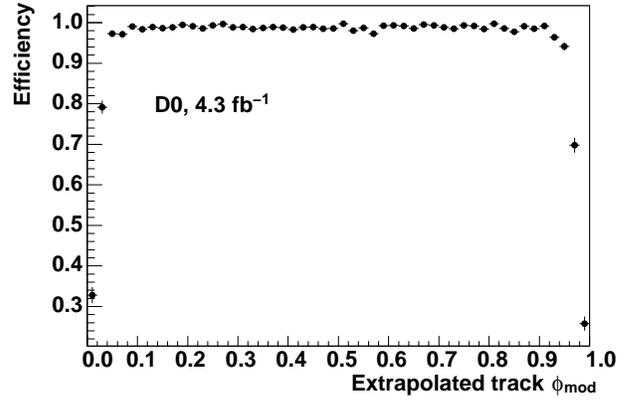}
\caption{Dependence of the electron reconstruction efficiency on the extrapolated 
  track $\phi_{\mathrm{mod}}$. }
\label{fig:PhiModEff}
\end{figure}

The $\phi_{\mathrm{mod}}$ efficiency is derived using the tag and probe method applied to $Z$ candidate events and is shown in Fig.~\ref{fig:PhiModEff}. The efficiency variation with $\phi^{\rm trk}_{\mathrm{mod}}$ is small except near the edges.  We therefore apply a fiducial requirement, $0.1\leq \phi^{\rm trk}_{\mathrm{mod}} \leq 0.9$, restricting the analysis to the region of stable $\phi_{\mathrm{mod}}$ efficiency.

\subsubsection{Hadronic Energy Dependent Electron Efficiency}
\label{sec:eff_set}

The efficiencies described thus far are directly related to kinematic properties of the electron and radiated photons. Indirect effects arising from the presence of hadrons in the same event have been accounted for through the presence of the recoil and additional $\ppbar$ interactions in events used to derive the efficiencies, but the independent effects of the hadronic energy are not specifically studied.  The hadronic energy dependent electron efficiency model accounts for the EM cluster reconstruction efficiency which is strongly affected by the presence of hadronic energy near the electron in the calorimeter cells. It also collectively describes any residual $\pte$ and SET dependency of the electron reconstruction and identification efficiency.

The hadronic efficiency $\epsilon_{\mathrm{had}}(\mathrm{SET},L,\upara,p_T^e,\eta_{\rm det})$ depends on five variables, the first three being direct measures of the hadronic energy. The instantaneous luminosity of the full MC event is taken from the zero-bias event overlaid on the hard scatter.  The use of SET  accounts for the impact of energy from additional interactions, and the use of $\upara$ accounts for the magnitude of the hard hadron recoil energy and its orientation with respect to the electron. The $p_T^e$ dependence arises because higher energy electrons are less affected by a fixed amount of nearby hadronic energy than lower energy electrons. Finally, the use of instantaneous luminosity $L$ accounts for the different behavior of the calorimeter read out at different instantaneous luminosity regimes.

The efficiency is derived in a multi-step process and uses the zero-bias-event SET and true $p_T$ of the electron in both full MC and fast MC as observables. These variables are chosen because they are not modified during the fast simulation and, thus, provide robust observables to describe the cluster reconstruction efficiency and its dependence on the hadronic energy, especially in high instantaneous luminosity environment. The first step is to create a version of the fast MC that has the zero-bias-event SET and electron true $p_T$ distribution reweighted to agree with the full MC distribution. This provides a high statistics target model for the fast MC.

In the next step, we compare the number of events in the original and reweighted fast MC in bins of $u_{\parallel}$, $\pte$, $\eta_{\text{det}}$, and $L$. Their ratio is taken as the initial estimate of the efficiency. In each bin, we compare the distribution of $\text{SET}/\pte$ between the original and the reweighted fast MC. The ratio is smoothed using a polynomial function and the average value is shifted to one, so that it can be interpreted as a perturbation over the initial estimate from the full MC. The hadronic efficiency is then the product of the initial estimate and the $\text{SET}/\pte$ perturbation in each bin.

\subsubsection{Electron $\phi$ Efficiency}
\label{sec:eff_phi}
\begin{figure}[htbp]
\centering
\includegraphics[width=\linewidth]{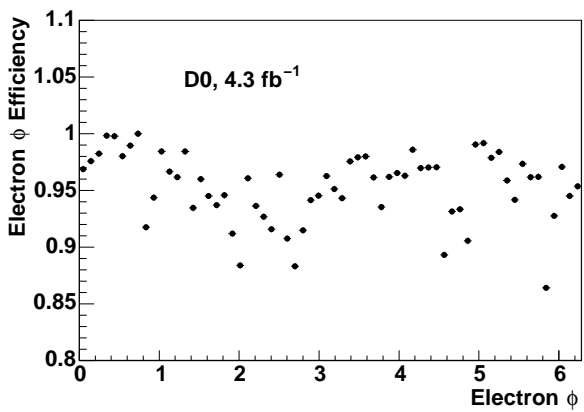}
\caption{The ratio data/fast MC of electron yield in $\wen$ events as a function of electron $\phi$ after all other effiencies have been applied to the fast MC. The ratio is used as a final efficiency correction. The maximum efficiency value is normalized to one.}
\label{fig:elec_phi_eff}
\end{figure}

The reconstructed electron $\phi$ distribution in $\wen$ events is not uniform. Once the $\phi_{\mathrm{mod}}$ induced effects are incorporated, we attribute the remaining overall $\phi$ dependence to small-scale imperfections in the detector, primarily inefficient tracker regions and calorimeter cells, which have no significant effect on the electron energy scale.  This efficiency is determined by dividing the $\phi$ distribution in data or full MC by that from the corresponding fast MC after including all other fast MC efficiencies. Figure~\ref{fig:elec_phi_eff} shows this efficiency for data $\wen$ events with the maximum efficiency value normalized to one.

\subsubsection{Monte Carlo Validation}
\label{sec:eff_valid}

\begin{figure}[htbp]
\includegraphics [width=\linewidth] {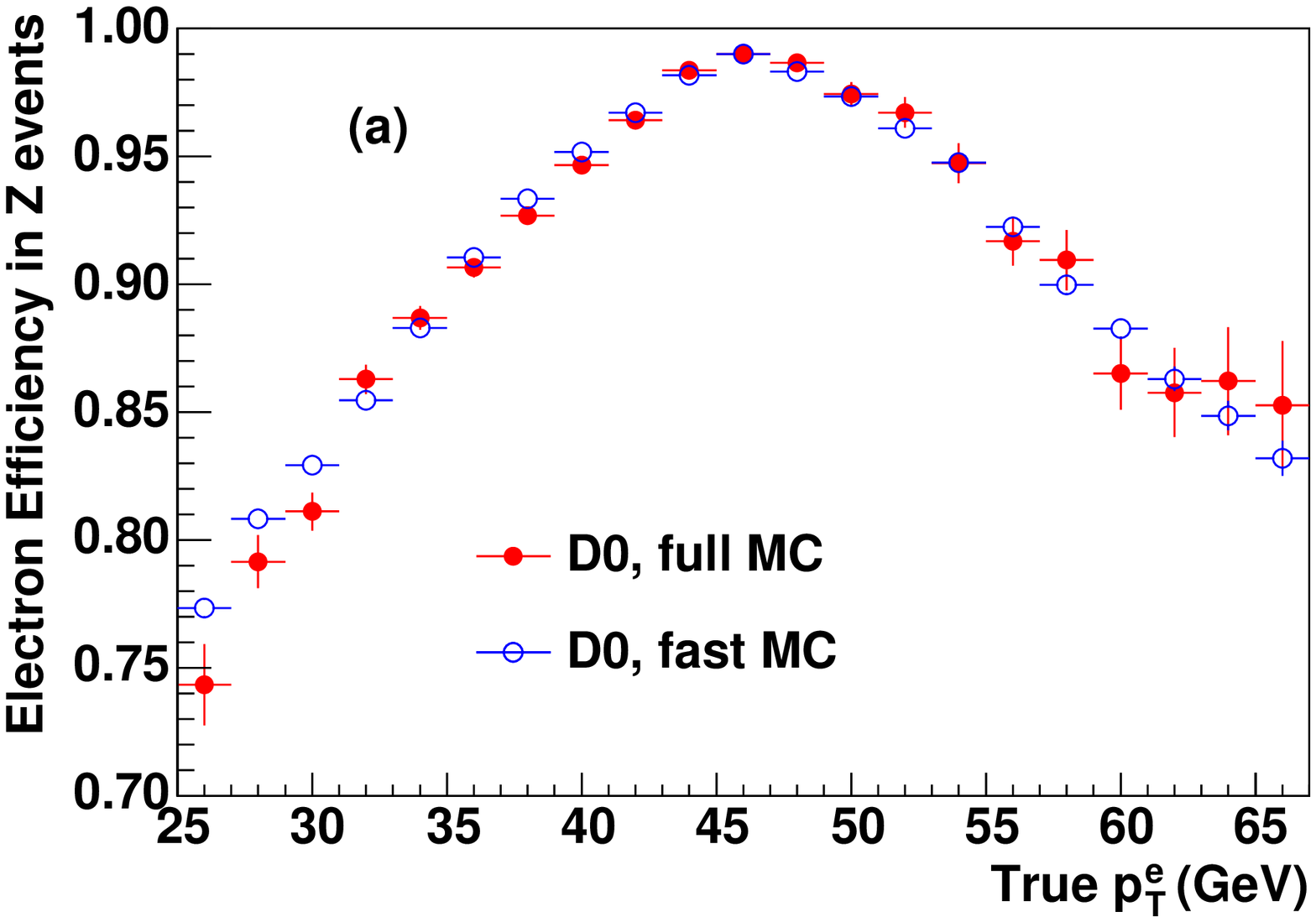}
\includegraphics [width=\linewidth] {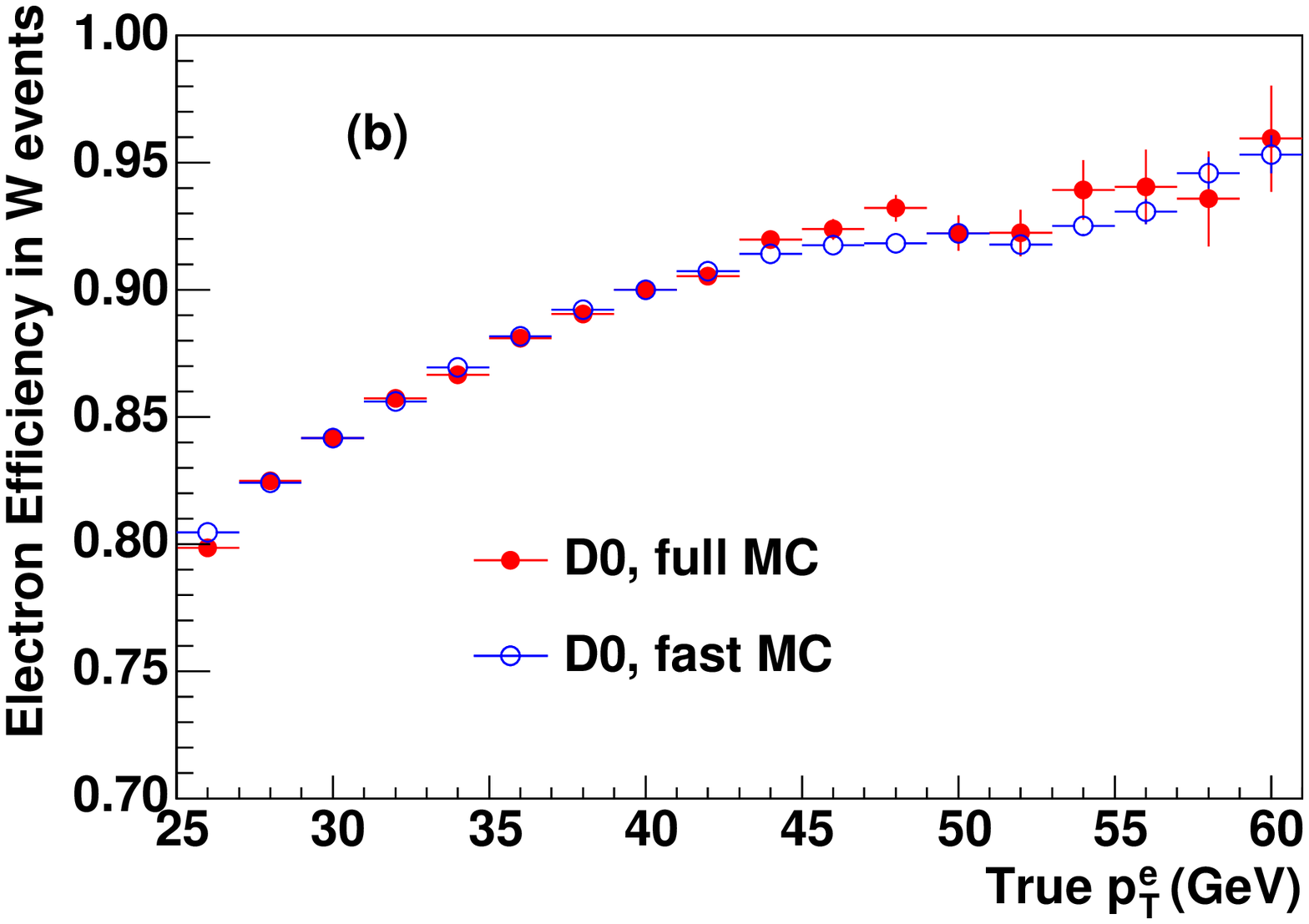}
\caption{[color online] The reconstruction and identification efficiency as a function of true $p_T^e$ in full MC and fast MC for electrons in (a) $Z\rightarrow ee$ and (b) $W\rightarrow e\nu$ events. In $\zee$ events, when the probed electron has high $\pte$, the other electron in the event is soft. When the soft electron is not properly identified, the $\zee$ event is not identified either. Thus, we observe a drop in the identification efficiency of $\zee$ events with high $\pte$ electrons, but not in $\wen$ events.}
\label{fig:TrueEptEff}
\end{figure}

\begin{figure}[htbp]
\includegraphics [width=\linewidth] {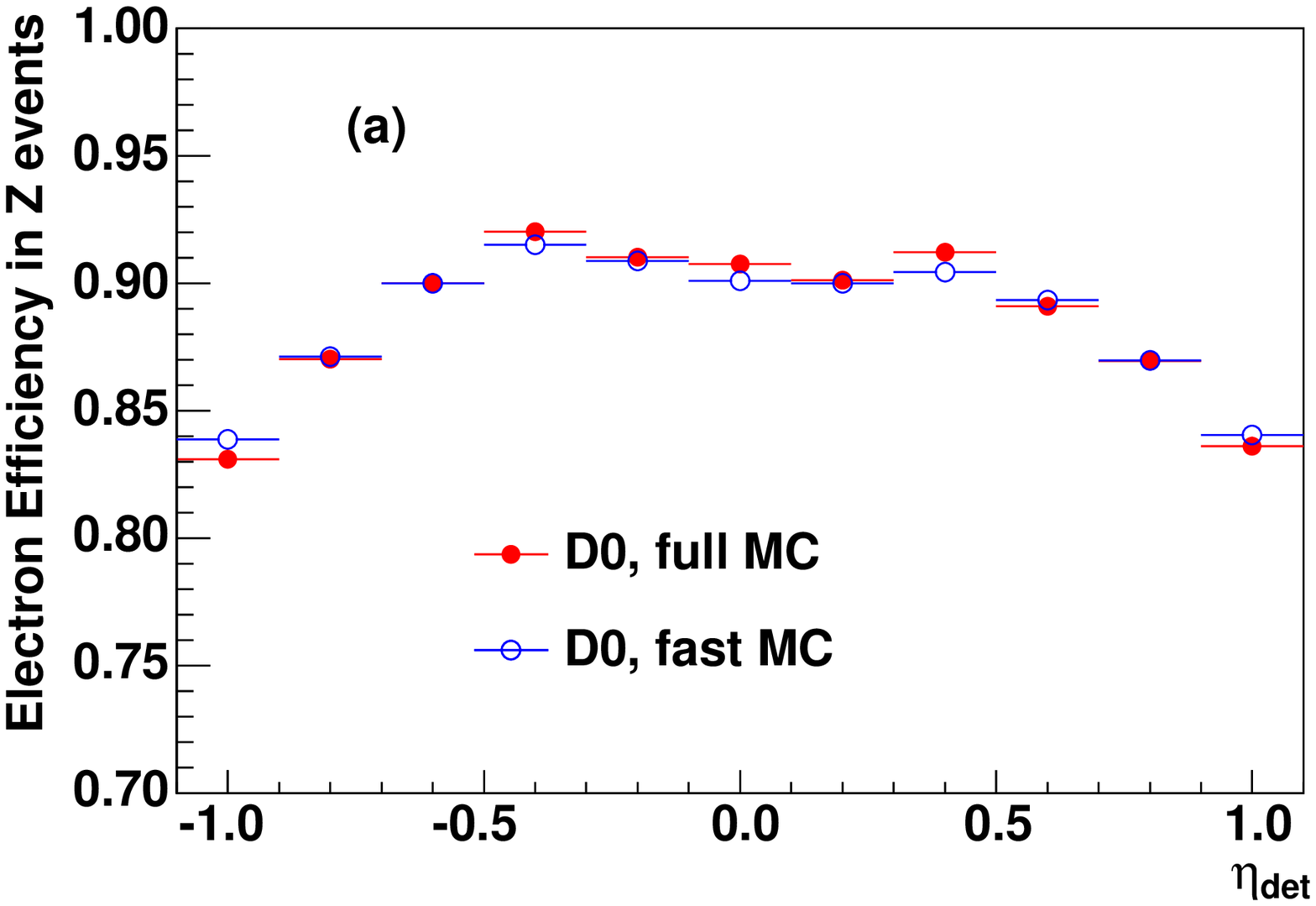}
\includegraphics [width=\linewidth] {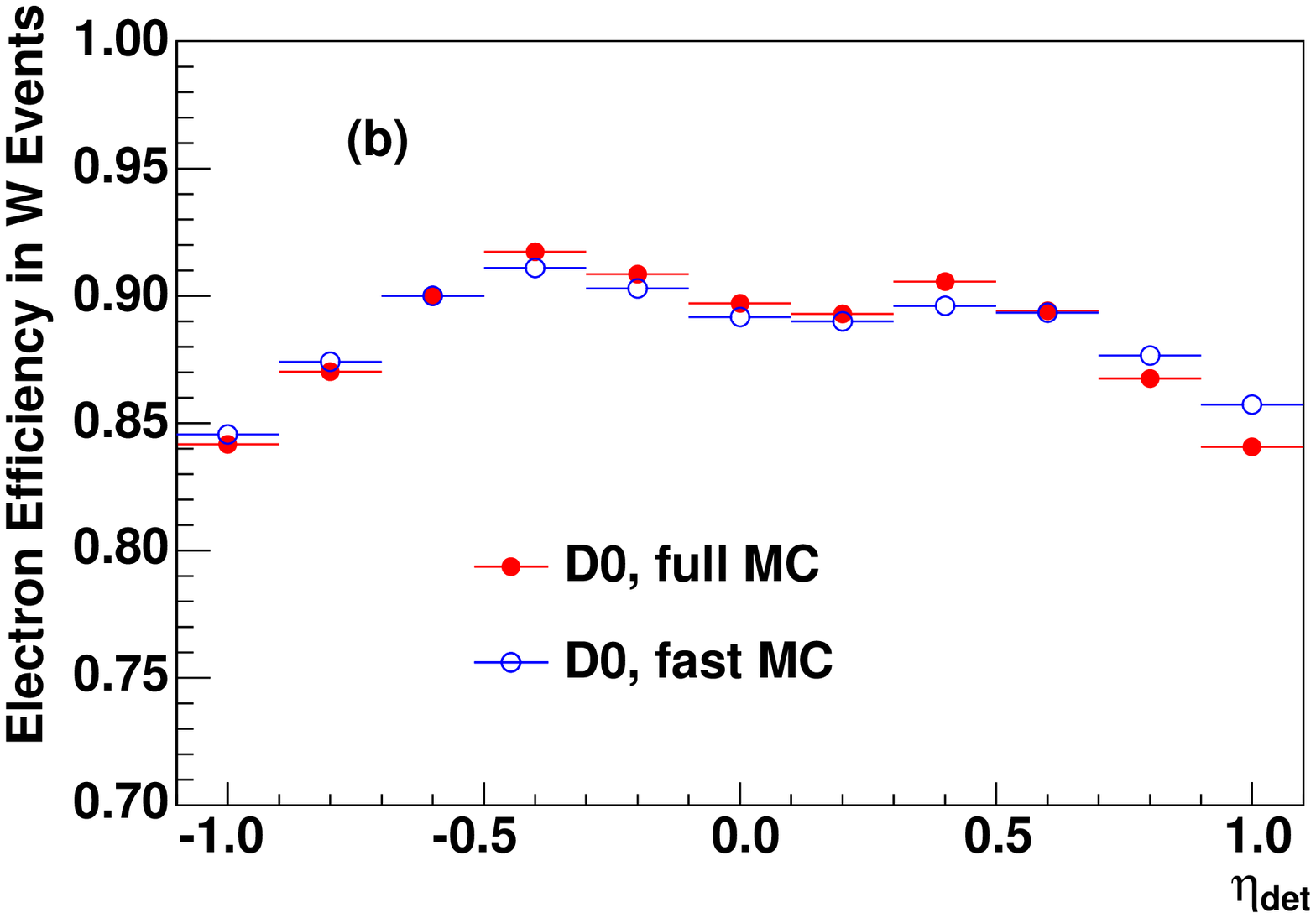}
\caption{[color online] The reconstruction and identification efficiency as a function of $\eta_{\rm det}$ in full MC and fast MC for electrons in (a) $\zee$ and (b) $\wen$ events.}
\label{fig:DetEtaEff}
\end{figure}

We validate our parametrized model derived from full MC using the generator level information by studying the efficiency as a function of the variables that are used to parametrize it. Figures~\ref{fig:TrueEptEff} and \ref{fig:DetEtaEff} show the comparison of the total electron efficiency, except for the trigger efficiency whose effect is not included in the full MC simulation, as a function of true $p_T^e$ and $\eta_{\rm det}$ in full MC and fast MC. For electrons in $Z\rightarrow ee$ events and in $W \rightarrow e\nu$ events, we observe that our efficiency model in fast MC accurately reproduces the efficiency in full MC.


We conclude that the full MC electron reconstruction and identification efficiency is well described by the parametrized model. This validates the strategy adopted for the derivation of the hadronic energy dependent efficiency (Sec.~\ref{sec:eff_set}). 

\subsubsection{Residual Efficiency Corrections\label{sec:DataHack}}

The efficiencies discussed thus far assume that the full MC can be used to accurately describe the efficiency dependencies. After applying the above efficiencies to the fast MC, we compare full MC or fast MC and data to derive two independent residual efficiency corrections $R_1(\mathrm{SET},L)$ and $R_2(\upara)$.

\begin{figure}[htbp]
\includegraphics[width=\linewidth]{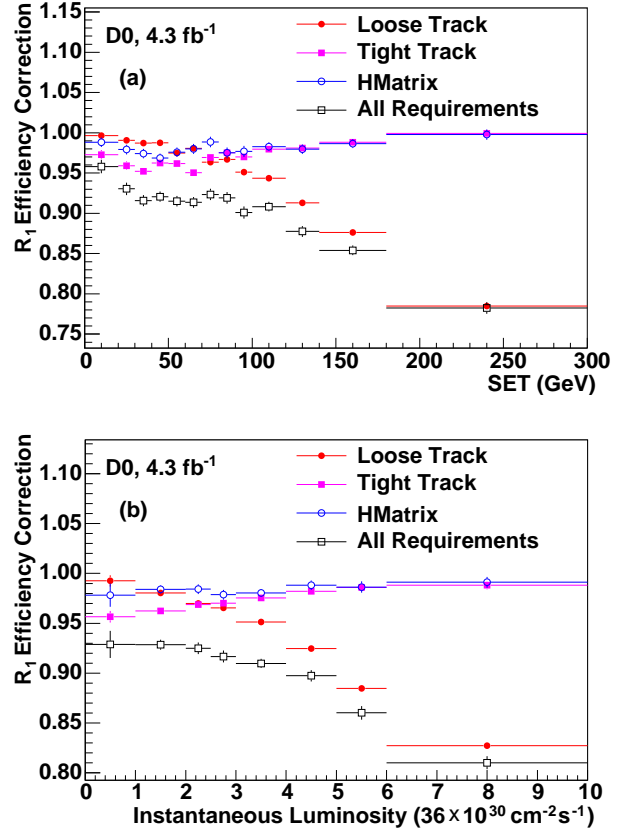}
\caption{[color online] Electron efficiency correction (data/full MC) as a function of (a) SET and (b) instantaneous luminosity. The contribution of each electron selection requirement to the efficiency correction is shown by the ratios derived from samples in which only the HMatrix (blue), loose track match (red), and tight track match (magenta) criterion is used (see Sec.~\ref{sec:eventselection} for the definition of each criterion).}
\label{fig:EffRatioSETLumi}
\end{figure}

The correction $R_1(\mathrm{SET},L)$ is derived by measuring the electron identification efficiency as a function of SET and $L$ in both $\zee$ data and full MC. The ratio of data and full MC efficiency defines this correction.  The ratios are shown in Fig.~\ref{fig:EffRatioSETLumi} for projections on the SET and $L$ axes. The correction $R_1(\mathrm{SET},L)$ is needed only for data analysis as a correction to the efficiency $\epsilon_{\mathrm{had}}$ derived previously by comparing full MC and fast MC.

To determine $R_1(\mathrm{SET},L)$, rather than directly counting the number of probe electrons, we study the $m_{ee}$ distribution in bins of the variables used to parametrize the correction.  Two $m_{ee}$ distributions are used, from a loose sample, when the probe electron is not required to satisfy the selection under study, and from a tight sample, when the probe electron is required to satisfy all the selection requirements.  The $\zee$ yield in each distribution is determined by fitting the distribution to $\zee$ signal and background components.

The second residual efficiency correction, $R_2(\upara)$, addresses imperfections in the $\upara$ dependency of the efficiency model. This is derived using the same technique of measuring the $m_{ee}$ distribution in bins of $\upara$, but taking the ratio of the efficiencies calculated in data to those derived in fast MC $\zee$ events.

\subsubsection[Systematic Uncertainty]{Systematic Uncertainty due to Efficiencies}
\label{sec:sys_eff}

The most significant efficiency-related uncertainty results from the adjustment of the final residual uncertainty correction $R_2$.  The resulting uncertainties on $M_W$ are 1~MeV, 2~MeV, and 5~MeV, respectively, for the $m_T$, $p^e_T$, and $\met$ methods.

\subsection{Electron Response Parameterization}
\label{sec:pmcs_elec}

The electron response model comprises three components: the response model, the resolution model and the underlying energy model.  A short introduction to the model is provided here, and detailed descriptions are then given in each of the following subsections.

The response model describes the average reconstructed electron energy for a
given electron true energy. We build a parametrized model for the contribution of radiated photons to the reconstructed electron cluster energy, since the energy from these photons is included in the electron energy after reconstruction.  We correct for residual luminosity and $\eta$ dependencies of the response that are not described by the data calibration. Finally, we use $\zee$ data and the measured value of the $Z$ boson mass~\cite{LEPZ_1, LEPZ_2, LEPZ_3, LEPZ_4} to calibrate the absolute energy scale.

The resolution model describes the fluctuations in the reconstructed electron energy. The EM calorimeter sampling resolution is modeled from a full MC sample that includes the improvements described in Sec.~\ref{sec:improve_shower}. This allows a detailed description of the dependence of the sampling term on the amount of uninstrumented material upstream of the calorimeter, as well as the energy and angular dependencies that it creates. We use $\zee$ data and the measured value of the $Z$ boson width~\cite{LEPZ_W_1, LEPZ_W_2, LEPZ_W_3, LEPZ_W_4} to calibrate the constant term of the resolution. Most of the noise fluctuations comes from fluctuations in the underlying hadronic energy inside the electron reconstruction cone. We therefore do not include an explicit noise term in the resolution model. 

The underlying energy model describes the average contribution of hadrons to the electron's reconstructed energy and its fluctuations.

\subsubsection{Photon Radiation Effects}
\label{sec:pmcs_elec_fsr}

Photon radiation from the $W$ boson decay electron can bias the $M_W$ measurement when the energy from radiated photons is not included in the reconstructed electron energy. This occurs if the radiated photon is separated from the electron and its energy is not counted in the electron energy, or if the photon shower is absorbed totally or partially by uninstrumented material in front of the calorimeter. 

The radiated photons arise either from FSR or during the interaction of the electron with material in front of the calorimeter (bremsstrahlung). The bremsstrahung energy loss is corrected in full MC by the electron energy loss correction (see Sec.~\ref{sec:deadmat}). The FSR energy loss is modeled in fast MC. 

\begin{figure*}[htp]
\includegraphics [scale=0.75] {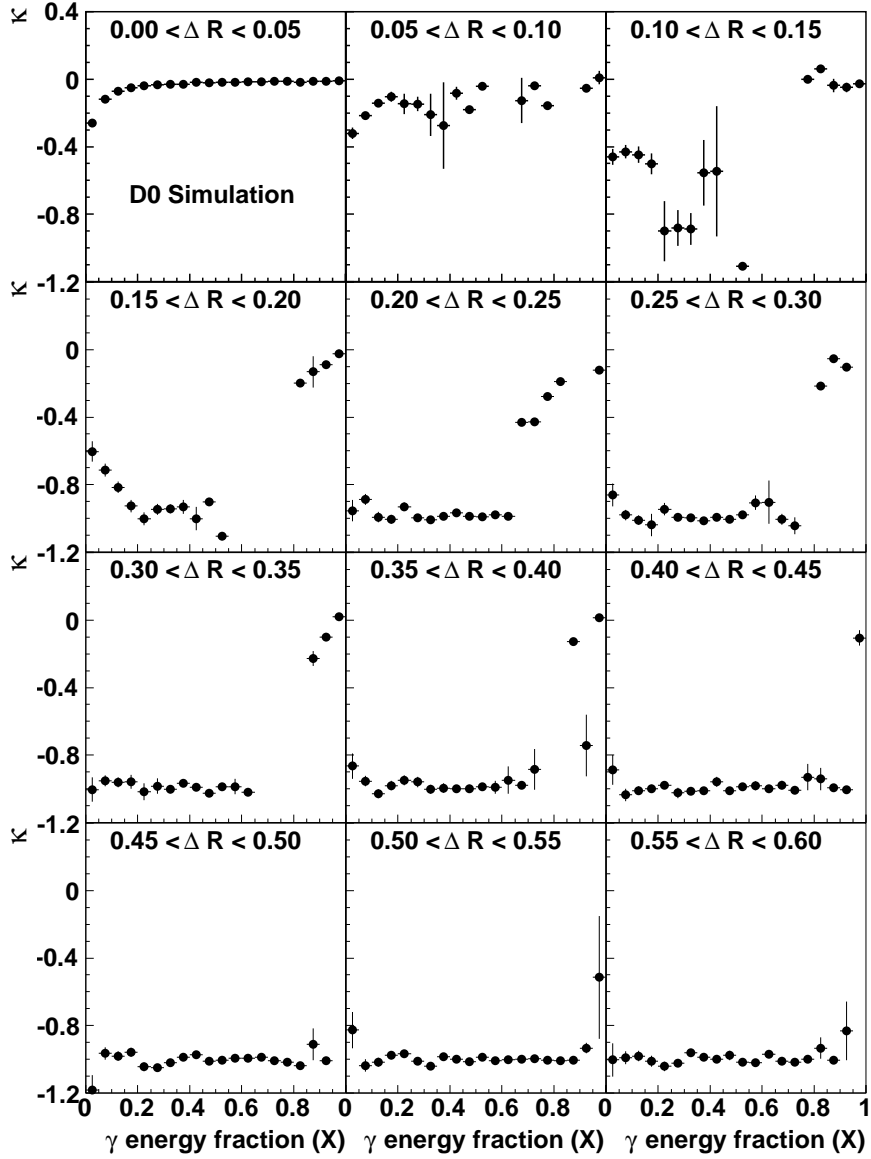}
\caption{ Electron energy correction determined from full MC as a function of fraction of the energy carried by the photon.}
\label{fig:eidloss_gammafrac}
\end{figure*}

The average contribution of FSR photons to the electron's reconstruction energy (for the same $\Delta R$ bins as Fig.~\ref{fig:eideff_gammafrac}) is given in Fig.~\ref{fig:eidloss_gammafrac}. The fraction of energy carried by the photon is denoted by $X$.  The vertical axis is the ratio $\kappa$, defined as the negative of the ratio of the difference of the reconstructed electron energy with and without FSR to the same difference for the true MC electron energy:

\begin{equation}
\kappa = - \frac{\displaystyle E^{e}_{\rm reco}\lbrack \mathrm{no}\,\mathrm{FSR} \rbrack -E^{e}_{\rm reco}\lbrack \mathrm{with}\,\mathrm{FSR} \rbrack}{\displaystyle E^{e}_{\rm true}\lbrack \mathrm{no}\, \mathrm{FSR} \rbrack -E^{e}_{\rm true}\lbrack \mathrm{with}\,\mathrm{FSR} \rbrack}
\end{equation}

At high $\Delta R$ we expect $\kappa = -1$ because the photon is well separated from the electron and does not contribute to the reconstructed electron energy. At low $\Delta R$, we expect negative values of $\kappa$ due to losses in the uninstrumented material, which decreases as $X$ increases. At intermediate $\Delta R$ and large values of $X$, $\kappa\approx 0$ since here the EM cluster is resconstructed around the photon. The final FSR energy loss parametrization is performed as a function of the same variables as the FSR efficiency: $\Delta R$, $X$, $\eta_{\rm det}$ and $E^{e}$.

\subsubsection{Dependence of the Calibration on the Instantaneous Luminosity}
\label{sec:gainlumi}

The $M_W$ measurement explores a much higher instantaneous luminosity regime than our previous measurement (Fig.~\ref{fig:lumiprofiles}), and we observe a significant dependence of the energy response on the instantaneous luminosity. Two opposite effects contribute to the change in energy response. The first is the extra energy in the calorimeter due to additional $p\bar{p}$ interactions, which causes an apparent increase in the response. The extra energy is correctly accounted for in the full simulation by overlaying data zero-bias events that have the same time and luminosity profile as the collider data and in the fast simulation by a parametrized model that will be described in Sec.~\ref{sec:ewindow}.

The second effect is due to a drop in the high voltage (HV) applied across the LAr gap that collects the ionization charge, causing an apparent reduction in the energy response if not corrected. The loss of HV occurs across the resistive coat on the signal boards~\cite{RunIdetector} that are used to deliver the HV to the LAr gaps. The resistitivity of this coat was measured {\em in situ}, at the temperature of liquid argon, to be of the order of 180 M$\Omega$ per square, with a large spread from one board to another. Whenever large currents flow through this coat, as is the case in high instantaneous luminosity operations, a sizable HV drop occurs and the ionization charge collected is reduced. In the CC, the detector modules extend across the full $\eta_{\rm det}$ range and the HV is delivered from the edges, at $\eta_{\rm det} = \pm 1.2$, making the drop most pronounced at the center ($\eta_{\rm det}=0$).

The average current from each calorimeter cell as a function of instantaneous luminosity is determined using the energy deposited in zero-bias events.  Using a simple resistive circuit model of the calorimeter HV distribution, the current is translated into an $\eta_{\rm det}$ and luminosity dependent model of the HV drop. We use measurements of the electron drift velocity as a function of the electric field~\cite{bib:Walkowiak} and the cell geometry to determine the fractional loss in response.  A final overall correction is derived from the instantaneous luminosity dependence of the $m_{ee}$ peak position measured in data.

We simulate single electrons at different energies, angles and luminosities, both with and without the tuned model of luminosity dependence, to parameterize the response change for electrons as a function of instantaneous luminosity and $\eta_{\rm det}$. For electrons at normal incidence, where the effect is maximal, the fractional change in response at an instantaneous luminosity of $L=120\times 10^{30}\, {\rm cm}^{-2}{\rm s}^{-1}$ is 0.42\%. A possible dependence on electron energy has been considered and found to be negligible.

\subsubsection{Dependence of the Calibration on electron {\boldmath $\eta$}}
\label{sec:EMscaleEtaAdj}

The procedure used to calibrate the EM calorimeter includes an equalization of the energy response of towers at different $\eta$ values. This procedure adjusts the gains until the position of the $Z$ boson mass peak in data is the same for any combination of $\eta$ values of the two electrons in a \zee\ event. This procedure does not account for the $\eta$ dependence of the underlying energy flow which implies that reconstructed $Z$ boson mass should have a small $\eta$ dependence. This is a small effect, but we take it into account in the measurement of $M_W$ by simulating this dependence in fast MC.

\begin{figure}[hbpt]
  \centering
  \includegraphics[width=\linewidth]{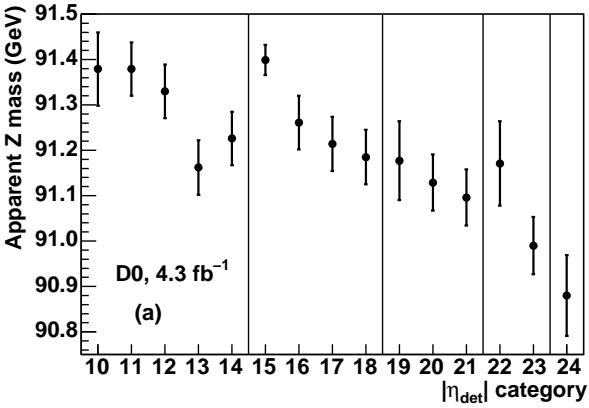}\\
  \includegraphics[width=\linewidth]{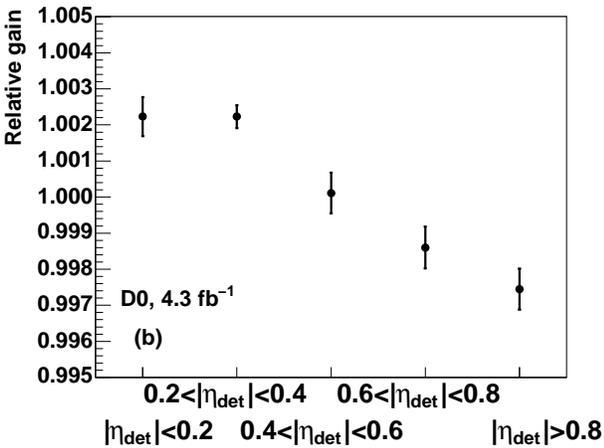}
  \caption{(a) Result of the $Z$ boson mass fit per $\eta_{\rm det}$ category prior to applying $\eta$-dependent corrections (Table~\ref{table:StandardEtaCategories}). (b) Result of the translation into one relative gain constant per $\eta_{\rm det}$ bin.
  \label{fig:EMscaleEtaAdj}}
\end{figure}

To derive an $\eta_{\rm det}$-dependent correction to the electron energy scale, we split our sample of CC-CC \zee\ events into 15~categories as defined in Table~\ref{table:StandardEtaCategories} (Sec.~\ref{sec:observable_mat_tune}). We use our standard procedures to fit for the $Z$~boson mass, separately for each category. These procedures use $m_{ee}$ templates produced using fast MC, in which the effect of the underlying energy is included. The results of these mass fits are summarized in Fig.~\ref{fig:EMscaleEtaAdj}.  We define one relative gain constant for each $|\eta_{\rm det}|$~bin (Table~\ref{table:StandardEtaBins}) and we translate the 15~mass values from Fig.~\ref{fig:EMscaleEtaAdj} into the values of the 5~relative gain constants. The world average value~\cite{LEPZ_1, LEPZ_2, LEPZ_3, LEPZ_4} of the $Z$~boson mass is used to translate energies into per-electron relative gains.

The results of the translation are shown in Fig.~\ref{fig:EMscaleEtaAdj}. They are used in fast MC for the simulation of the $\eta_{\rm det}$~nonuniformity in the calorimeter gains.

\subsubsection{Energy Response and Resolution}
\label{sec:elec_energy}

The reconstructed electron energy $E$ is simulated as:
\begin{equation}
\begin{split}
E=R_{\rm EM}(E_{0}, \eta_{\text{det}}, L)\,&\otimes\, \sigma_{\rm EM}(E_{0}, \eta) \\ &+ \Delta E(\text{SET}, L, \pte, \eta_{\text{det}}, u_{\parallel}),
\label{eqn:elec_energy1}
\end{split}
\end{equation}
where $E_0$ is the electron energy after the FSR simulation, $R_{\rm EM}\otimes \sigma_{\rm EM}$ is distributed as a gaussian with mean given by the energy response $R_{\rm EM}$, and width given by the energy resolution $\sigma_{\rm EM}$. The term $\Delta E$ describes the deposition of energy from hadronic showers inside the electron reconstruction cone. 

\begin{figure}[hbpt]
  \centering
  \includegraphics[width=\linewidth]{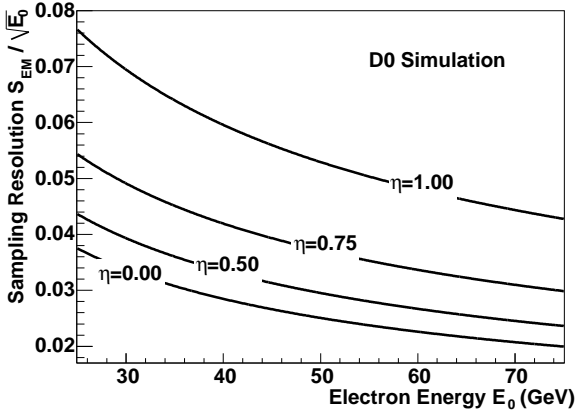}\\
  \caption{Sampling contribution $S_{\rm EM}/\sqrt{E_0}$ to the electron fractional energy resolution as a function of the electron energy $E_0$ for four different electron incident angles $\eta=0.0$, $0.5$, $0.75$, and $1.0$. The strong energy and angular dependencies are caused by the energy lost in the uninstrumented material before the calorimeter.
  \label{fig:samp_res}}
\end{figure}

The resolution of the EM calorimeter $\sigma_{\rm EM}$ is modeled as:
\begin{equation}
  \frac{\sigma_{\rm EM}} {E_0} = \sqrt{C^{2}_{\rm EM} + \frac{S^{2}_{\rm EM}}{E_0} + \frac{N^{2}_{\rm EM}}{E_0^{2}} }\ \ .
\end{equation}
in which $C_{\rm EM}$, $S_{\rm EM}$ and $N_{\rm EM}$ correspond to the constant, sampling and noise terms, respectively. Owing to the uninstrumented material in front of the calorimeter, the sampling term parameter $S_{\rm EM}$ depends on electron energy and incident angle, and is parametrized as:
\begin{equation}
S_{\rm EM} = S_0\exp\left[S_1\left(\frac{1}{\sin\theta} - 1\right)\right] +\frac{(S_2\eta+S_3)}{\sqrt{E_0}},
\label{eq:resolution_model}
\end{equation}
where,
\begin{equation}\nonumber
\begin{split}
 S_0=&0.15294\pm 0.00005\,{\rm GeV}^{1/2}\\\nonumber
 S_1=&1.543\pm 0.007\\\nonumber
 S_2=&-0.025\pm 0.001\,{\rm GeV}\\\nonumber
 S_3=&0.172\pm 0.002\,{\rm GeV}.
\label{eq:resolution_parameters} 
\end{split}
\end{equation}
The values of the smearing parameters $S_{0}$ to $S_{3}$ are determined from the improved simulation of the D0 detector, as discussed in Sec.~\ref{sec:deadmat}. The uncertainties quoted in the parameters are determined by propagating the uncertainty in the thickness of the cylinder added to the full MC simulation, which comes from the limited size of the $\zee$ sample used in the tuning procedure. Figure~\ref{fig:samp_res} shows the electron energy sampling resolution $S_{\rm EM}/\sqrt{E_0}$ for four different values of electron $\eta$. The strong energy and angular dependencies in Eq.~\ref{eq:resolution_model} are caused by the energy lost in the uninstrumented material before the calorimeter.

The value of $N_{\rm EM}$ is set to zero since the contribution of the noise term is small at the energies of electrons from $W$~boson and $Z$~boson decay and since the most important source of noise is already discribed in the fluctuations of $\Delta E$. The extraction of $C_{\rm EM}$ from the width of the $Z$~boson mass peak is discussed in Sec.~\ref{sec:ConstantTerm}.

In the vicinity of the $\phi$-module boundaries of the central calorimeter, the modeling of the electron energy response and resolution in fast MC is modified compared to the description above. The Gaussian resolution model is modified to include a {\it lossy tail} given by a Crystal Ball function~\cite{CrystalBall} for $\phi_{\text{mod}} < 0.2$ and $\phi_{\text{mod}} > 0.8$. In the same range, the loss in average response is modeled by a simple linear function. The parameters for both response and resolution modifications near the module boundary are determined from template fits to $\zee$ data.

The energy response for electrons in Eq.~(\ref{eqn:elec_energy1}) is modeled as:
\begin{equation}
\begin{split}
R_{\rm EM} = F_{\rm \eta-eq}(\eta_{\det}) &\times F_{\rm HV-loss}(L, \eta_{\det})\\ & \times \left(\alpha ( E_{0} - \overline{E_0}) + \beta + \overline{E_0} \right)
\label{eq:response}
\end{split}
\end{equation}
where $F_{\rm HV-loss}( L, \eta_{\det})$ implements the model of the luminosity dependence of the calorimeter gains due to the HV~loss that is discussed in Sec.~\ref{sec:gainlumi}, and $F_{\rm \eta-eq}(\eta_{\det})$ describes the $\eta$ nonuniformity discussed in Sec.~\ref{sec:EMscaleEtaAdj}. The parameters $\alpha$ and $\beta$ are referred to as scale and offset, and $\overline{E_0}=43$~GeV is a reference value for the energy of electrons in $\zee$ events. The values of $\alpha$ and $\beta$ are determined from $\zee$ events in collider data. The constant $\overline{E_0}$ is introduced to reduce the correlation between the
parameters $\alpha$ and $\beta$ to improve the stability of the numerical evaluation of the covariance matrix of the simultaneous fit for $\alpha$ and $\beta$.

The determination of the parameters of the energy response of the calorimeter to electrons is one of the most important steps in the measurement of $M_W$.  The scale and offset cannot be distinguished from one another to the precision required using only the $Z$~boson mass distribution. However, the different electron energies from $Z$~boson decays can be exploited to constrain the energy dependence of the energy response. The measured $m_{ee}$ is calculated from:
\begin{equation}
m_{ee}=\sqrt{2 E^{e_1} E^{e_2}(1-\cos\omega)},
\end{equation}
where $\omega$ is the opening angle between the two electrons.

Substituting and expanding in a Taylor series with $\beta \ll  E^{e_1} + E^{e_2}$ gives (ignoring $\overline{E_0}$) 
\begin{equation}
m_{ee} = \alpha m^0_{ee} + \beta f_Z^0 + \mathcal{O}(\beta^2),
\label{eq:fder}
\end{equation}
where $f_Z$ is a kinematic variable defined as:
\begin{equation}
\label{eq:fzdef}
f_Z^0 = \frac{(E_0^{e_1}+E_0^{e_2})(1-\cos\omega)}{m^0_{ee}}
\end{equation}
in which quantities with a zero subscript or superscript are calculated with all corrections except the $\alpha$ and $\beta$ correction.

Equation~\ref{eq:response} relates the observed $m_{ee}$ to the scale, offset and the true energies of the electrons. By varying both the scale and offset, templates of the two dimensional distribution of $m_{ee}$ versus $f_Z$ in the fast MC are compared to the equivalent distribution in data.  The final values of $\alpha$ and $\beta$ used are found by maximizing the likelihood formed during the comparison.

\begin{figure}[ht]
 \centering
 \includegraphics[width=\linewidth]{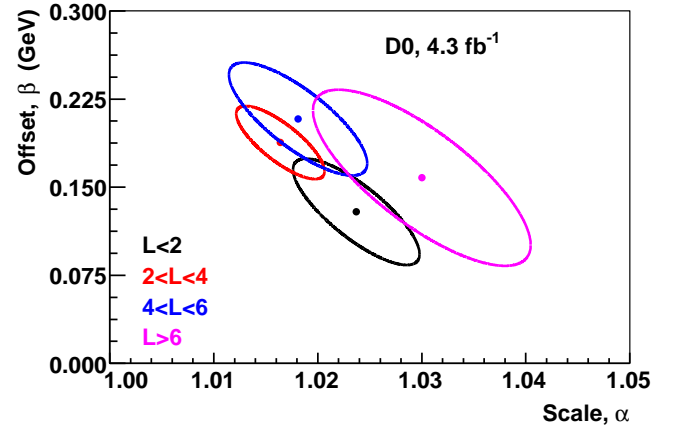}
 \caption{[color online]  Central values and one standard deviation contours of the fits for electron energy scale and offset to the collider data. Instantaneous luminosity is given in units of $36\times10^{30}\, {\rm cm}^{-2}{\rm s}^{-1}$.}
 \label{fig:EMscalePlots}
\end{figure}

The scale and offset are determined separately in different bins of instantaneous luminosity, which is expressed in units of $36\times10^{30}\, {\rm cm}^{-2}{\rm s}^{-1}$. Table~\ref{tab:EMscaleFits} summarizes the scale and offset parameters, along with the correlation coefficients from the fits. The fit results are shown in Fig.~\ref{fig:EMscalePlots}. The results from the fits for each of the bins in instantaneous luminosity agree well with each other.  This shows that our model of the underlying energy flow into the electron cone (Sec.~\ref{sec:ewindow}) and the model of the luminosity-dependence of the calorimeter gains (Sec.~\ref{sec:gainlumi}) are correctly accounting for the luminosity dependence of the detector response to electrons. Rather than defining one luminosity-averaged set of parameters for the scale and offset, we use the different values per bin in luminosity, because there is no loss in statistical power, {\em i.e.}, the systematic uncertainty on $M_W$  due to the electron energy scale is not increased by splitting into luminosity bins.

\begin{table*}[htp]
\begin{center}
\caption{Results of the fits for electron energy scale and offset to the collider data.
\label{tab:EMscaleFits}}
\begin{tabular}{c|c|c|c|c}\hline\hline
              & $0<L<2$               & $2<L<4$                 & $4<L<6$              & $L>6$               \\ \hline
$\alpha$      & $1.0237 \pm 0.0043$   & $1.0164 \pm 0.0030$     & $1.0181 \pm 0.0047$  & $1.0300 \pm 0.0074$ \\
$\beta$ (GeV) & $0.129 \pm 0.032$     & $0.188 \pm 0.022$       & $0.208 \pm 0.034$    & $0.158 \pm 0.053$   \\
Correlation   & $-0.796$              & $-0.786$                & $-0.783$             & $-0.764$            \\\hline\hline
\end{tabular}
\end{center}
\end{table*}

The dominant systematic uncertainty on our measurement of $M_W$ is the precision with which we measure the mean electron energy response.  The uncertainties in the energy scale and offset are individually large but they are highly correlated.  We propagate the correlated uncertainties in the scale and offset parameters to our measurement of $M_W$ and obtain uncertainties of 16 MeV on $M_W$ using the $m_T$ and  $\met$ distributions and 17 MeV for the $p^e_T$ distribution.

\subsubsection{Determination of the Constant Term}
\label{sec:ConstantTerm}

For data, unlike the full MC, the constant term $C_{\rm EM}$ is also important.  It arises primarily from residual channel-to-channel calibration differences and describes an energy-independent contribution to the fractional energy resolution.  Thus, its main impact is felt at high electron energies where the sampling term is suppressed by its approximate $1/\sqrt{E}$ behavior.  The value of $C_{\rm EM}$ is extracted from the width of the $Z$~boson mass peak with the sampling term modeled as described above. The value of $C_{\rm EM}$ is determined using template fitting to the $m_{ee}$ distribution.  The best fit value for $C_{\rm EM}$ is
\begin{eqnarray*}
  C_{\rm EM} & = & (1.997 \pm 0.073)\% ,
\end{eqnarray*}
which is in good agreement with our determination in Run~IIa and with the Run~II design goal of 2\%.

In order to propagate the uncertainty from the electron energy resolution model to $M_W$, we use fast MC pseudo-experiments in which we vary the sampling resolution function parameters by their uncertainty (Eq.~\ref{eq:resolution_parameters}). For each of these fast MC pseudo-experiments, we fit the constant term to account for the correlation between the two components of the resolution model. Using the procedure described in Sec.~\ref{sec:syst}, we estimate the uncertainty to be 2~MeV for the $M_W$ measurement using $m_T$ and $p^e_T$ and 3~MeV for $\met$.

\subsubsection{Electron Cone Effects}
\label{sec:ewindow}

To reconstruct an electron, we must define an electron reconstruction cone (Fig.~\ref{fig:ewindow}). The energy in this cone arises not only from the electron, but also from hadronic recoil, spectator parton interactions, and additional $p\bar{p}$ collisions.  There are also effects from the suppression of electronic noise. These bias both the reconstructed electron energy and the reconstructed recoil energy.  Extra energy is given to the electron from the recoil and it is excluded from the reconstruction of $u_T$.  The additional energy added to the electron cone is denoted by $\Delta E$ (Eq.~\ref{eqn:elec_energy1}, Sec.~\ref{sec:elec_energy}), while the additional transverse energy subtracted from the recoil in the electron cone is denoted by $\Delta u_{\parallel }$ (Eq.~\ref{eqn:duparadef}) in Sec.~\ref{sec:recoil}.

The value of $\Delta u_{\parallel }$ is not equal to $\Delta E \sin\theta^e$ for two reasons:
\begin{itemize}
\item The energy loss due to uninstrumented material in front of the calorimeter is corrected for the electron, but not for the recoil.
\item Zero suppression has different effects near a large concentrated energy ($\Delta E$) compared to a small diffuse background energy ($\Delta u_{\parallel }$).
\end{itemize}

To study electron cone effects, we construct a $\Delta u_{\parallel }$ library by recording the energy deposition in random cones from $W\rightarrow e\nu$ events in collider data and in full MC. These random electron reconstruction cones are selected in such way to avoid any electron energy contribution.  Events in this library sample the same luminosity profile as the data used to measure $M_W$. For each electron in the fast MC simulation, we simulate its $\Delta u_{\parallel}$ by selecting a random cone from the library based on the electron's $\eta$, $\eta_{\text{det}}$, and $\upara$, as well as on the event's SET and luminosity.

To model the change in the electron energy $\Delta E$ associated to a given $\Delta u_{\parallel}$, we perform a dedicated full MC simulation in which we extract the electron and FSR photon energies separately from the hadronic recoil particles energies in each cell, and generate three $W \rightarrow e\nu$ full MC samples based on the same full detector simulation of each $W\rightarrow e\nu$ event:

\begin{itemize}
\item {\it Electron only}: contains only the electron and FSR photons.
\item {\it No electron}:  contains everything except the electron and FSR photons, {\it i.e.}, the hard recoil, spectator parton interactions, and additional $p\bar{p}$ interactions. 
\item {\it Full sample}: contains the complete event.
\end{itemize}

\begin{figure*}[ht]
\includegraphics[width=\linewidth]{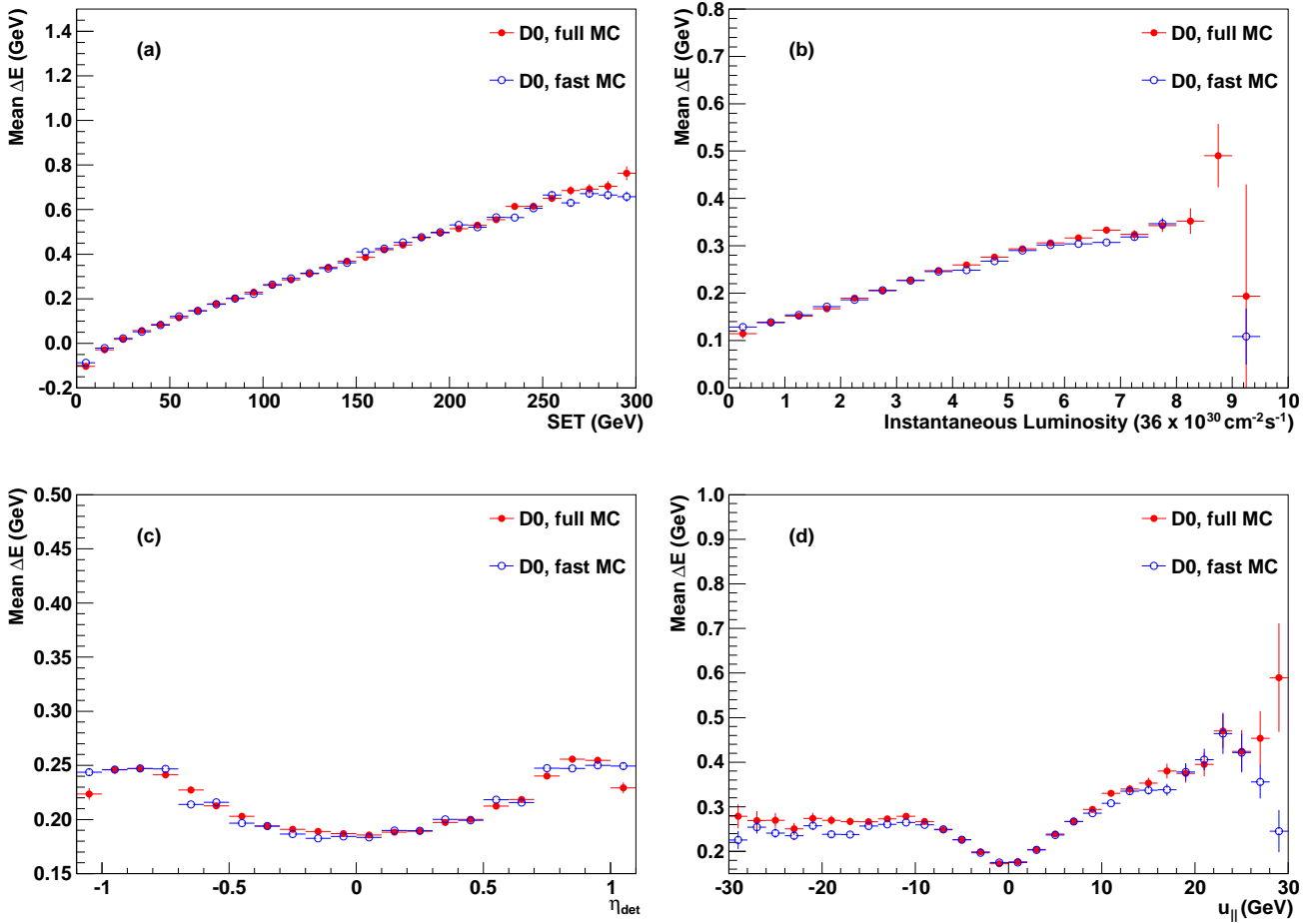}
\caption{[color online] Mean $\Delta E$ as a function of (a) SET, (b) instantaneous luminosity, (c) $\eta_{\rm det}$, and (d) $u_{\parallel }$ comparing full and fast MC. }
\label{fig:compare_mean_de_set_lumi_deteta_upara}
\end{figure*}

For a given reconstructed electron cluster in the {\it full sample}, the value of $\Delta u_{\parallel}$ can be determined by the sum, in the sample with {\it no electron}, of the energies over the calorimeter cells that compose the cluster. The value of $\Delta E$ corresponding to this $\Delta u_{\parallel }$ is determined from the difference of the reconstructed electron energy in the {\it full sample} to the one in the sample with {\it electron only}. The relationship between $\Delta u_{\parallel}$ and $\Delta E$ is strongly dependent on SET, $L$, $p^e_T$, $\eta_{\rm det}$, and $u_{\parallel }$ and those variables are used to parametrize the model in the fast MC. Figure~\ref{fig:compare_mean_de_set_lumi_deteta_upara} shows the comparison between the $\Delta E$ distribution in full MC and the one in fast MC, which uses the $\Delta u_{\parallel}$ library and the parametrized model for the relationship with $\Delta E$.

\begin{figure}[htbp]
 \includegraphics[width=1.1\linewidth]{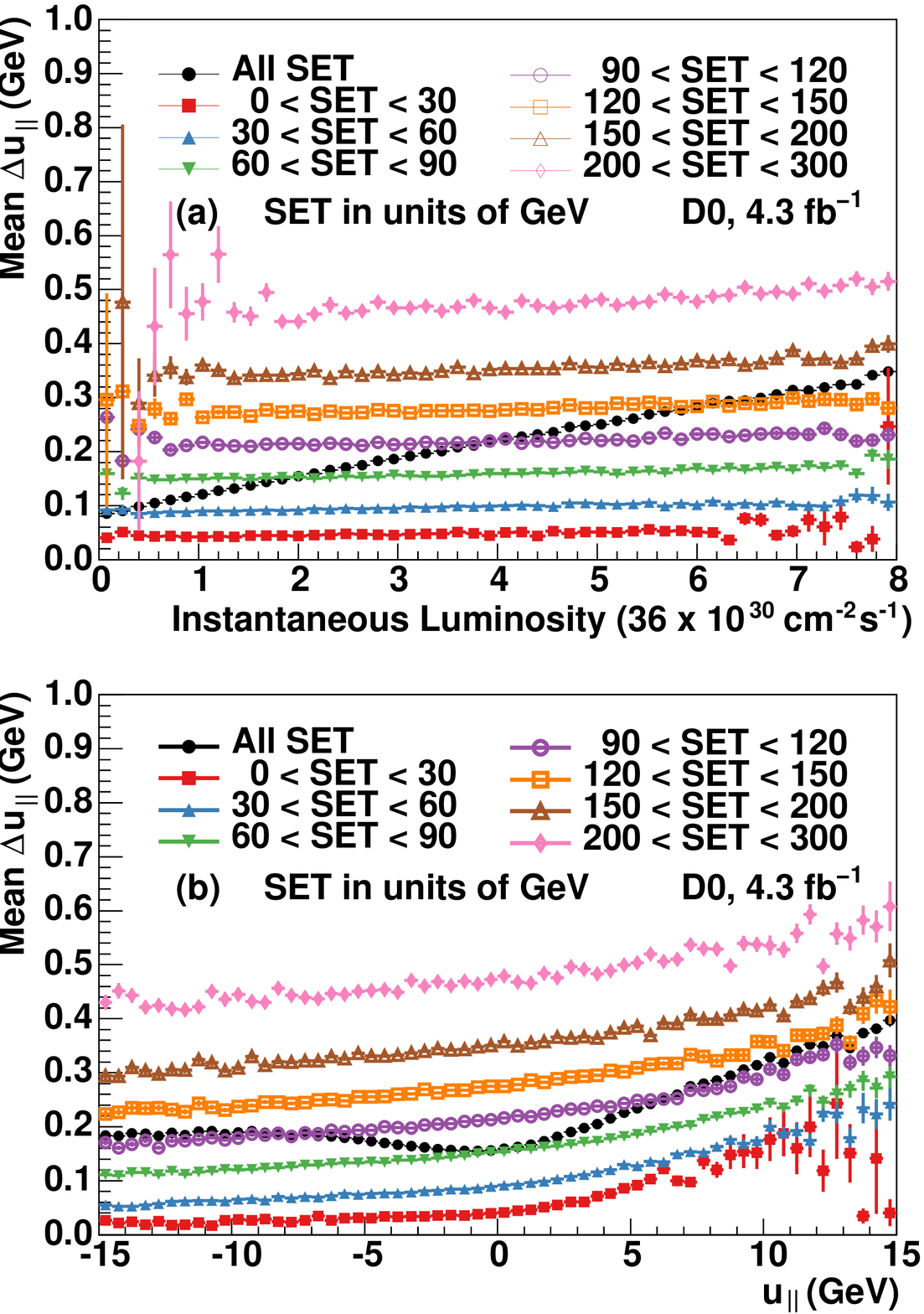}
 \caption{[color online] (a) Mean $\Delta u_{\parallel }$ as a function of $L$ separately for sub-samples with different SET.
Instantaneous luminosity $L$  is given in units of $36 \times 10^{30} cm^{-2} s^{-1}$. (b) Mean $\Delta u_{\parallel }$ as a function of  $u_{\parallel }$ separately for various bins of SET.}
\label{fig:dupara_combo}
\end{figure}

The $\Delta u_{\parallel}$ library determined in the studies of the electron cone effects also provide information for the recoil system, as discussed in Sec.~\ref{sec:recoil}. We show the dependence of the mean $\Delta u_{\parallel }$ ($\langle \Delta u_{\parallel } \rangle $) on $L$ in Fig.~\ref{fig:dupara_combo}(a) for various bins of SET. In a given bin of SET, there is almost no dependence of $\langle \Delta u_{\parallel } \rangle$ on $L$, while, for the full SET range, the strong dependence on $L$ comes only from the correlation between SET and $L$. We show the dependence of mean $\langle \Delta u_{\parallel } \rangle$ on $u_{\parallel }$ in Fig.~\ref{fig:dupara_combo}(b) for various bins of SET. In a given bin of SET, the $\langle \Delta u_{\parallel } \rangle$ always increases with increasing $u_{\parallel }$ as the recoil gets closer to the electron cone. Our interpretation is that, at a fixed SET, the soft recoil component is fixed and we can study the hard recoil contribution which is controlled by $u_{\parallel }$. For the full SET range, the dip around $u_{\parallel} \approx 0$ happens because, in a high pileup environment, a small $u_{\parallel}$ almost always implies small SET.

\subsection{Hadronic Recoil Parameterization}
\label{sec:recoil}

\begin{figure}[hbtp]
\centering
\includegraphics[scale=0.5]{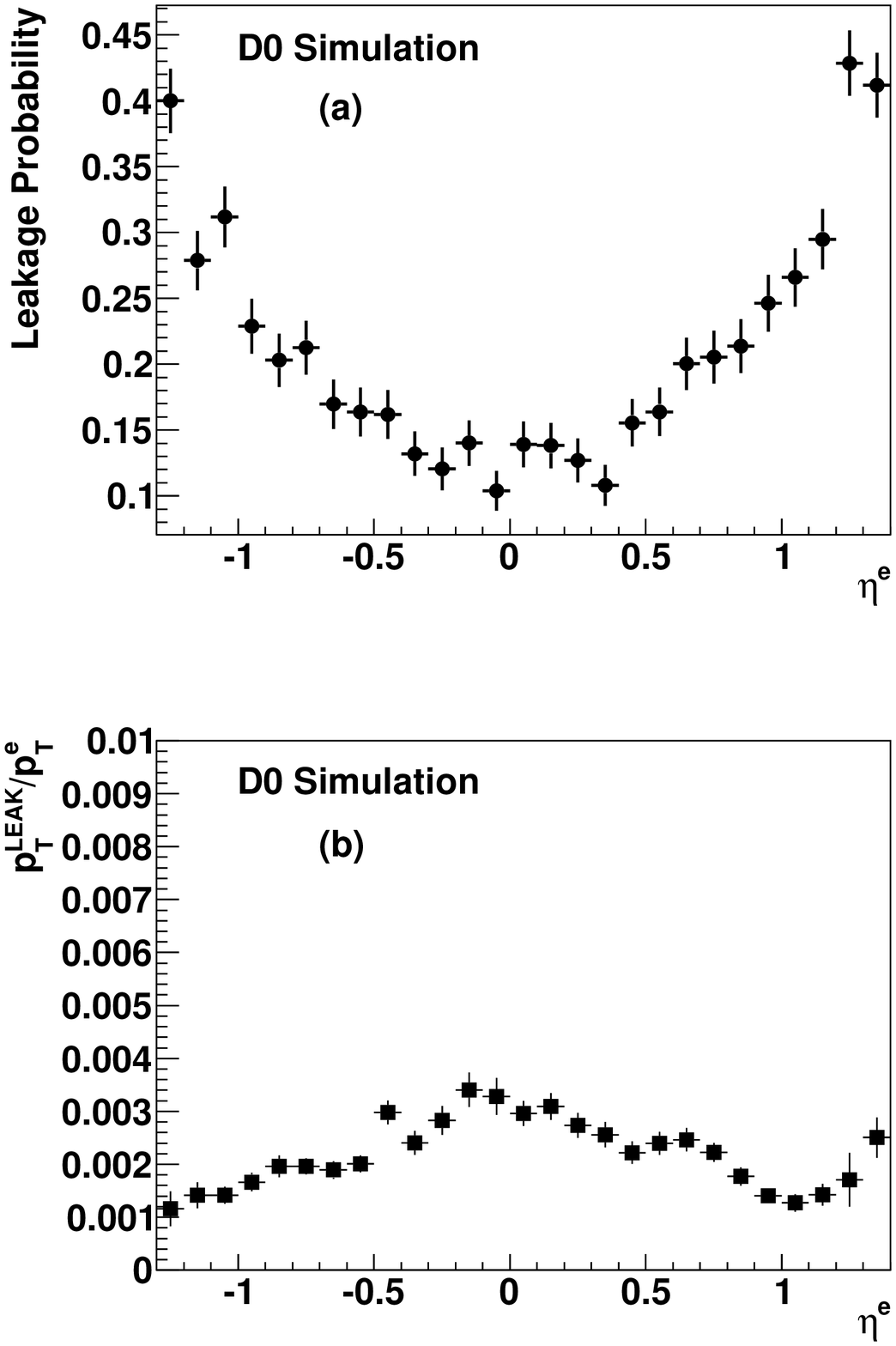}
\caption{(a) The fraction of electron showers that leak outside the reconstruction cone in the CC, and (b) the fraction of the electron transverse momentum that is added to the recoil system for clusters without in-cone FSR photons, both as a function of electron $\eta$. }
\label{fig:leakage}
\end{figure}

The hadronic recoil simulation in the fast MC uses a multi-component model that can be decomposed into:
\begin{equation}
 \vec{u}_T = \vec{u}^{\rm ~HARD}_T + \vec{u}^{\rm ~SOFT}_T + \vec{u}^{\rm ~ELEC}_T + \vec{u}^{\rm ~FSR}_T,
\label{eqn:utdef}
\end{equation}

\noindent 
where $\vec{u}_T^{\rm ~HARD}$ is the dominant part of the recoil balancing the vector boson, $\vec{u}_T^{\rm ~SOFT}$ describes the zero-bias and minimum-bias contribution, $\vec{u}^{\rm ~ELEC}_T$ models the hadronic energy in the electron cone and electron energy leakage out of the cone, and $\vec{u}^{\rm ~FSR}_T$ is the out-of-cone electron FSR contribution. The contribution of out-of-cone photons to the recoil transverse momentum, $\vec{u}^{\rm ~FSR}_T$, is parametrized as a function of the photon pseudorapidity and energy, derived from a dedicated full MC simulation. The third component, $\vec{u}^{\rm ~ELEC}_T$, is defined as:
\begin{equation}
\vec{u}_T^{\rm ~ELEC}= -\Delta u_{\parallel} \widehat{p_T^e} + \vec{p}_T^{\rm~LEAK},
\label{eqn:duparadef}
\end{equation}
where $\widehat{p_T^e}$ is an unit vector in the direction of $\vec{p}_T^{\,e}$ and $\Delta u_{\parallel}$ is discussed in Sec.~\ref{sec:ewindow}. The value of $\vec{p}_T^{\rm~LEAK}$, which describes the energy leakage from the electron reconstruction cone due to calorimeter shower development, is determined using single electron full MC as a $\eta^e$-dependent fraction of $\pte$. Figure~\ref{fig:leakage} shows the fraction of electron showers that leak outside the reconstruction cone and the fraction of their energy that is added to the recoil system. The electron shower leakage is parametrized independently for electrons with and without in-cone FSR, since the photon shower contributes to the total energy leaked.

\subsubsection{Hard and Soft Recoil Models}
\label{smearing_model}

The hard recoil model is derived from a special sample of $Z\rightarrow \nu\nu$ full MC events generated with {\sc pythia} without simulation of multiple parton interactions and without overlay of zero-bias events. The generated events are processed through the full chain of the detector simulation and reconstruction. Since the neutrinos escape undetected, all the energy measured in the detector can be attributed to the recoil alone. To obtain kinematics similar to $Z\rightarrow ee$ events, both neutrinos from a $Z$ boson decay are required to have $|\eta|<1.3$.

The model simulates the magnitude (${u}_T^{\nu\nu}$) and direction ($\phi$) of the reconstructed hard recoil as a function of the negative of the generator-level transverse momentum of the vector boson, $\vec{p}_T^{\,V}$.  The model is parametrized using two variables, the relative transverse momentum
\begin{equation}
R = \frac{u^{\nu\nu}_T-p_T^{V}}{p_T^{V}},
\end{equation}
and the angular resolution
\begin{equation}
\displaystyle \Delta\phi = \phi(\vec{u}^{\,\nu\nu}_T)-\phi(\vec{p}_T^{\,V}), \mbox{with}\ \
(|\Delta \phi|<\pi).
\end{equation}

\begin{figure}[hbtp]
\centering
\includegraphics [scale=0.4, angle=270] {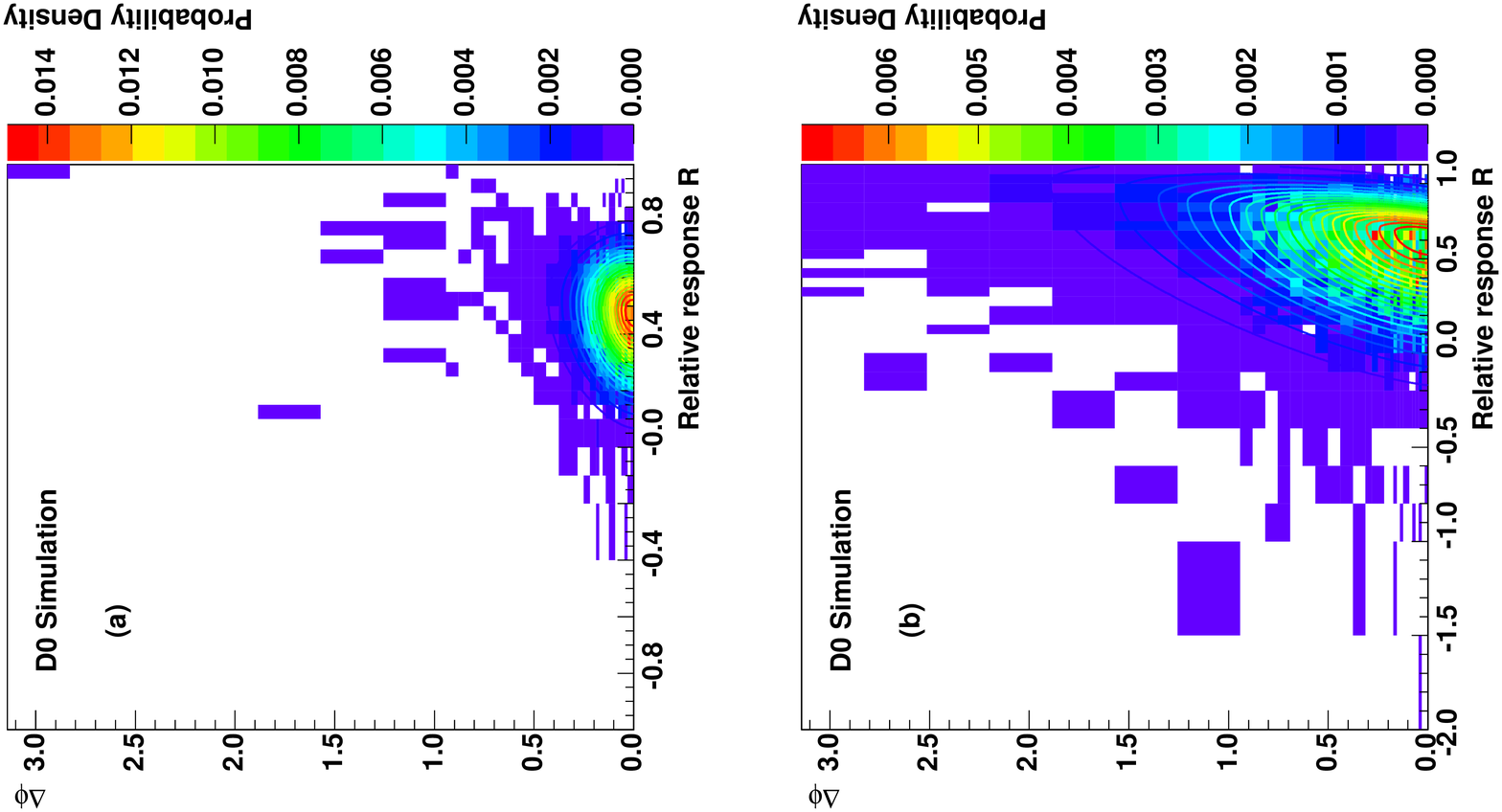}
\caption{[color online] The distribution of the recoil relative transverse momentum and $\Delta\phi$ resolutions for 
full MC (boxes) and fit (contours) for (a) $4.5 < p_T^V < 5.0\,\text{GeV}$ and for (b) $18 < p_T^V < 20\,\text{GeV}$.}
\label{fig:fit:combo}
\end{figure}

\noindent
The $Z\rightarrow \nu\nu$ sample is divided into 32 bins of $p_{T}^V$. For each
bin the distribution of $R$ versus $\Delta\phi$ is smoothed to obtain a continuous probability
density $P(R,\Delta\phi)$.  The smoothing function is a product of a
log-normal distribution in $R$ with a normal distribution in $\Delta\phi$. Two
examples of such probability density functions are shown in
Fig.~\ref{fig:fit:combo} for $4.5 < p_T^V < 5.0\,\text{GeV}$ and for $18 < p_T^V < 20\,\text{GeV}$.
The correlation between $R$ and $\Delta\phi$ is described by assuming that the mean of the log-normal distribution
has a linear dependence on $\Delta\phi$. The smoothing fits are shown in
Fig.~\ref{fig:fit:combo} as colored contours. From
these, the simulated $R$ and $\Delta\phi$ values for a fast MC event are chosen by randomly sampling the probability density corresponding to the
boson $p_T$.

The hard recoil model described thus far applies to full MC $Z\to\nu\nu$ events.  To correct for imperfections in the simulation, additional smearing parameters are introduced and applied to the component $u^{\nu\nu}_\parallel = u^{\nu\nu}_T\cos(\Delta\phi)$ in the direction of $\vec{p}_T^{\,V}$ to give the corrected recoil denoted by $u_{\parallel}^{\rm HARD}$:
\begin{eqnarray}
u_{\parallel}^{\rm HARD}/p_T^V = (r_0+r_1 e^{-p_T^V / \tau_{\rm HAD}}) (\overline{R}(p_T^V) + 1)\nonumber \\
     + \sigma_0 (u^{\nu\nu}_\parallel/p_T^V - \overline{R}(p_T^V) - 1).
\end{eqnarray}
The perpendicular component 
\begin{equation}
\nonumber u^{\rm HARD}_\perp = u^{\nu\nu}_T\sin(\Delta\phi)
\end{equation}
remains unmodified. The mean values $\overline{R}(p_T^V) = \langle{(u^{\nu\nu}_\parallel}-p_T^V)\big/{p_T^V}\rangle$ are determined from the smoothed distributions for $(R,\Delta\phi)$.  The smearing parameters $r_0$, $r_1$, $\tau_{\rm HAD}$, and $\sigma_0$ are determined as described below.

The soft recoil is modeled from the measured recoils in collider data minimum-bias and zero-bias events. In addition to being selected by the minimum-bias trigger, the minimum-bias events are required to have zero or one reconstructed primary vertex.  The zero-bias events are sampled to give the instantaneous luminosity distribution observed in the data. We create lists of the magnitude and direction of recoil in the minimum-bias and zero-bias events, and for a given fast MC event, the simulated soft recoil is created by taking one $\vec{\mathrm{u}}_T$ value from each of the minimum-bias and zero-bias lists and combining them to give the soft recoil
\begin{equation}
\vec{u}_{T}^{\rm SOFT} = \sqrt{\alpha_{\rm MB}}\  \vec{u}_T^{\rm MB} + \vec{u}_T^{\rm ZB},
\end{equation}
where $\alpha_{\rm MB}$ is a parameter that controls the soft recoil resolution.

We determine values for the five parameters $r_0$, $r_1$, $\tau_{\rm HAD}$, $\sigma_0$ and $\alpha_{\rm MB}$ by fits comparing data (or full MC) to the fast MC simulation using a method first used by the UA2 collaboration \cite{UA2}.  The momentum imbalance between the $p_T$ of the dielectron system and the recoil $u_T$ in $Z\to ee$ events is projected on the bisector $\hat\eta$ of the electron and positron directions
\begin{equation}
  \eta_{\text{imb}} \equiv (\vec{p}_T^{\ ee}+\vec{u}_T)\cdot\hat{\eta} 
\end{equation}
as shown in Fig.~\ref{fig:eta_upara}.  The bisector is chosen to reduce the dependence between the electron energy scale and the hadronic recoil, because the bisector is independent of fluctuations in the measured electron energies. The $\eta_{\text{imb}}$ distributions are made in bins of reconstructed $p_T^{ee}$ for both data (or full MC) and fast MC. The five parameters are determined by constructing separate fast MC samples with varying values of the parameters and finding the parameter values that minimize the $\chi^2$ difference between the mean (as functions of $r_0$, $r_1$ and $\tau_{\rm HAD}$) and RMS (as functions of $\sigma_0$ and $\alpha_{\rm MB}$) of $\eta_{\text{imb}}$ for data and fast MC distributions.  The fits using the mean and the RMS are performed independently.

\subsubsection{Fit Results}
\label{fitting_results}

The results from the minimization of the mean $\eta_{\text{imb}}$ as a function of $p_T^{ee}$ for collider data are
\begin{eqnarray}\nonumber
 r_0 &=&1.047\pm 0.008,\\\nonumber
 r_1 &=&2.07 \pm 0.39,\\\nonumber
\tau_{\rm HAD} &=&2.51\pm 0.32\ {\rm GeV},
\end{eqnarray}
and the results from the minimization of the RMS are
\begin{eqnarray}\nonumber
\sigma_0 &=&1.238 \pm 0.040,\\\nonumber
\alpha_{\rm MB} &=&0.633 \pm 0.064.
\end{eqnarray}
The corresponding two correlation matrices are:
\begin{equation}\nonumber
\bordermatrix{
           & r_0   & r_1   & \tau_{\rm HAD}\cr
 r_0       & 1   &  0.30 & -0.49 \cr
 r_1       & 0.30  &  1  & -0.90 \cr
\tau_{\rm HAD} & -0.49 & -0.90 &  1},
\end{equation}
and
\begin{equation}\nonumber
\bordermatrix{
             & \sigma_0 & \alpha_{\rm MB}\cr
 \sigma_0    & 1      & -0.68 \cr
 \alpha_{\rm MB} & -0.68   &  1 }.
\end{equation}

\begin{figure*}[htbp] 
\includegraphics [width=0.8\linewidth] {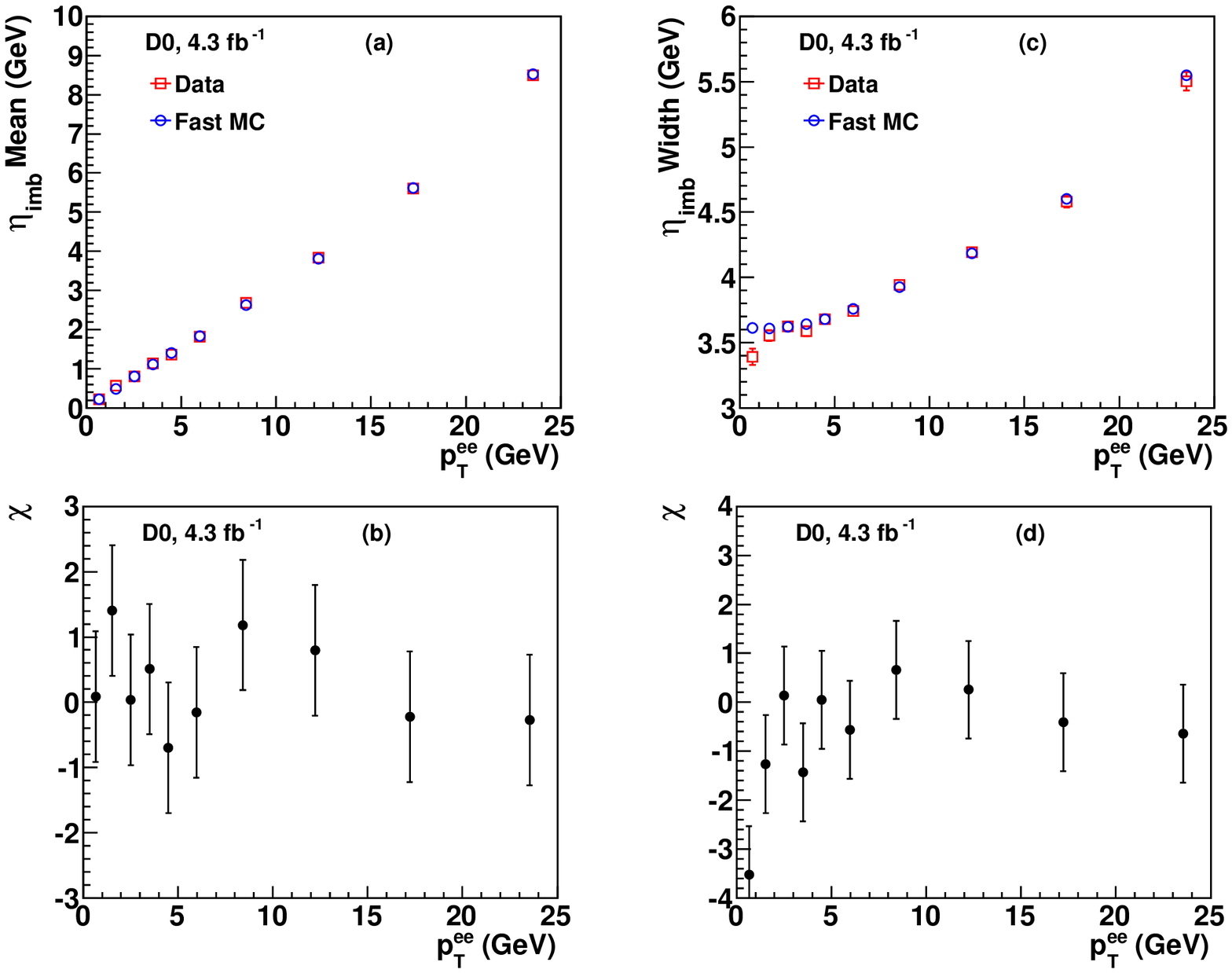}
\caption{[color online] Data and fast MC comparison of the (a) mean and (c) width of the $\eta_{\text{imb}}$ for the ten different bins in $p_{T}^{Z}$. The $\chi$ value per $p_{T}^{Z}$ bin for the (b) mean and (d) width of the $\eta_{\text{imb}}$.}
\label{fig:imbal.combo}
\end{figure*}

\noindent Figure~\ref{fig:imbal.combo} shows the comparison of the mean and the width of the $\eta_{\text{imb}}$ momentum imbalance distributions between data and fast MC for the ten different $p_{T}^{ee}$ bins. The quantity $\chi$ is defined as the ratio of the difference between data and fast MC divided by the uncertainty in the data for each bin.

\subsubsection{Recoil Modeling Systematic Uncertainties}
\label{sec:recoiluncertainty}

The size of the $\zee$ sample determines the statistical precision of the five smearing parameters. We use pseudo-experiments, as described in Sec.~\ref{sec:syst}, to propagate their uncertainties to the measured $M_W$ and determine the recoil modeling systematic uncertainty.  We find uncertainties of $5$~MeV, $6$~MeV and $14$~MeV for the $m_T$, $p_T^e$ and $\met$ results.

\section{Backgrounds}
\label{sec:background}
There are three significant backgrounds in the $\wen$ sample, whose shapes need to be added to the fast MC templates before comparing to the data distributions:

\begin{itemize}
\item $Z\to ee$ events in which one electron is not detected in a poorly instrumented region of the detector. 
\item Multijet events (MJ) in which a jet is misidentified as an electron and $\,\met$ arises from misreconstruction.
\item $W\to \tau\nu \to e\nu\nu\nu$ events.  
\end{itemize}

\noindent
The $\zee$ component is estimated directly from the $\wen$ data sample, the MJ component using a matrix method, and the $W\to\tau\nu$ from simulation. The subsections below provide detailed description of their determination.

\subsection{$\zee$ Background}
\label{sec:zee_bkgd}


$Z\rightarrow ee$ events are present in the $W\rightarrow e\nu$ sample when there is substantial $\met$ from mismeasurement of energy. We directly estimate the $\zee$ contamination from the $W\rightarrow e \nu$ sample, selecting events that pass the full $W$ sample selection, modified to include selection of an additional reconstructed cluster chosen to indicate that the selected event is likely a $Z$ boson decay.  Most often the second cluster is in the inter-cryostat region (ICR), which is outside the electron acceptance in this analysis and has poor sampling of the event energy flow since the ICD is not included in $\met$ reconstruction. The $\zee$ background from events where neither electron is in the ICR is negligible.

Since we cannot directly identify electrons in the ICR, we estimate the number of $\zee$ events using electrons reconstructed as jets in this region and electron tracks candidates. The jet is required to have a matched track such that the invariant mass of this track and the electron is consistent with the $Z$ boson mass. To estimate the absolute number of $\zee$ events in the $\wen$ sample, we count the number of candidates passing the $W$ plus the additional jet selection ($N(e,\text{jet})$) and use:


\begin{equation}
N({\zee} \hspace{1mm} \mbox{background})  = {N(e,\text{jet})\over  {\epsilon'_{\text{jet}} \times A(e,\text{trk}) }},
\label{eqn:zeebkgdformula1}
\end{equation}
where $\epsilon'_{\text{jet}} = \epsilon_{\text{jet}} \times A(e,\text{jet})/A(e,\text{trk})$ is the relative efficiency to find a jet given the presence of a matching track and $A(e, \text{trk})$ is the track acceptance in the invariant mass window $70 < m_{e,\text{trk}} < 110\,\text{GeV}$, both measured in data control samples. The fraction of $\zee$ background events in the $\wen$ candidate sample is found to be (1.08$\pm$0.02)\%. The uncertainty is dominated by the precision with which the efficiency $\epsilon'_{\text{jet}}$ is determined and by the limited number of jet objects reconstructed in the ICR consistent within the $\zee$ mass window.

\subsection{Multijet Background}
\label{sec:mj_bkgd}

\begin{figure}
  \centering 
  \includegraphics[width=\linewidth]{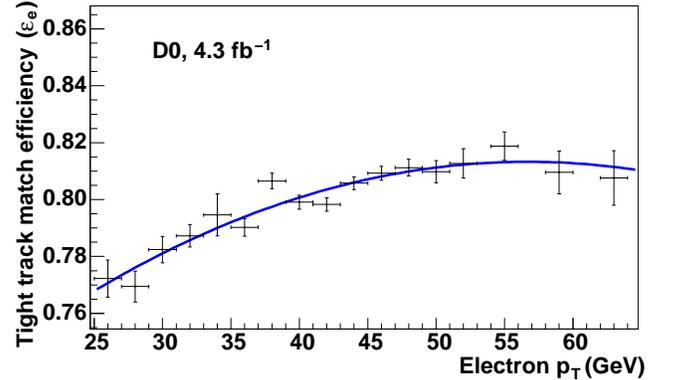}
  \caption{[color online] Tight track match efficiency as a function of the electron $\pte$ measured relative to the loose track match requirement.\label{bkgd:eff}}
\end{figure}

\begin{figure}
  \centering 
  \includegraphics[width=\linewidth]{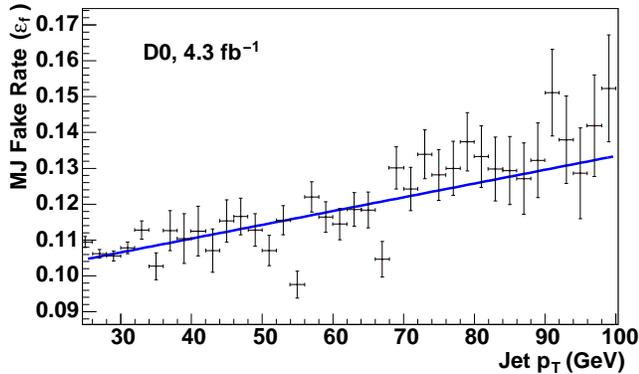}
  \caption{[color online] Probability of a jet object that passes the loose track match requirement to pass the tight track match requirement.\label{bkgd:fr}}
\end{figure}

The MJ background is determined using a loose sample obtained by only requiring that the matched track is within $0.05$ in $\Delta \eta$ and within $0.05$ in $\Delta \phi$ from the EM cluster (Sec.~\ref{sec:elreconstruct}), instead of using the standard track matching, which contains track quality requirements (Sec.~\ref{sec:eventselection}). This sample contains all events satisfying the standard selection requirements, but has a significantly higher contamination from MJ background than the standard sample. The probabilities for electron candidates in $\wen$ events ($\epsilon_e$) and in MJ events ($\epsilon_f$) to pass the complete matching requirements given that they already satisfy the loose match requirement are determined in control samples. The probability for real electrons is determined from $\zee$ data using tag and probe, and the probability for electron candidates in MJ events is determined from data dijet events. They are parametrized as a function of electron $p_T$ and can be seen in Figs.~\ref{bkgd:eff} and~\ref{bkgd:fr}. The loose sample event yield, $N_L$, the standard sample event yield, $N$, and the two probabilities are then used to determine the MJ background yield in each bin $i$ of a distribution by solving the system of equations
\begin{equation}
\begin{split}
  N_{L}^{(i)} & =  N_{W}^{(i)} + N_{\rm MJ}^{(i)}, \\
  N^{(i)} & =  \epsilon_e^{(i)} N_{W}^{(i)} + \epsilon_f^{(i)} N_{\rm MJ}^{(i)},
\end{split}
\end{equation}
for the MJ background, given by $\epsilon_f N_{\rm MJ}$. The contribution from MJ events is found to be $(1.02\pm0.06)$\% of the selected $W\to e\nu$ candidate sample. The uncertainty is dominated by the precision with which the tight track match efficiency is determined.

\subsection{$W\to\tau\nu$ Background}
\label{sec:tau_bkgd}

The $W\to\tau\nu\to e\nu\nu\nu$ contribution is determined from a simulation of the process using {\sc resbos} for event generation, {\sc tauola}~\cite{b-tauola_1, b-tauola_2, b-tauola_3, b-tauola_4} for $\tau$ lepton decay, and fast MC for detector simulation. Because the electrons arise from a secondary decay, their momenta are lower than for electrons from $\wen$ decays and their distribution is broader. The background contribution from $W\to\tau\nu$ decays is found to be $(1.668 \pm 0.004)$\%, with the uncertainty dominated by the uncertainty in the $\tau\rightarrow e\nu\nu$ branching ratio~\cite{PDG2012}. The uncertainty in the $M_W$ measurement arising from incorporating the $W\to\tau\nu\to e\nu\nu\nu$ events as background instead of a $M_W$ dependent signal is small.  

\begin{figure*}
  \centering 
  \includegraphics[keepaspectratio,width=\textwidth]{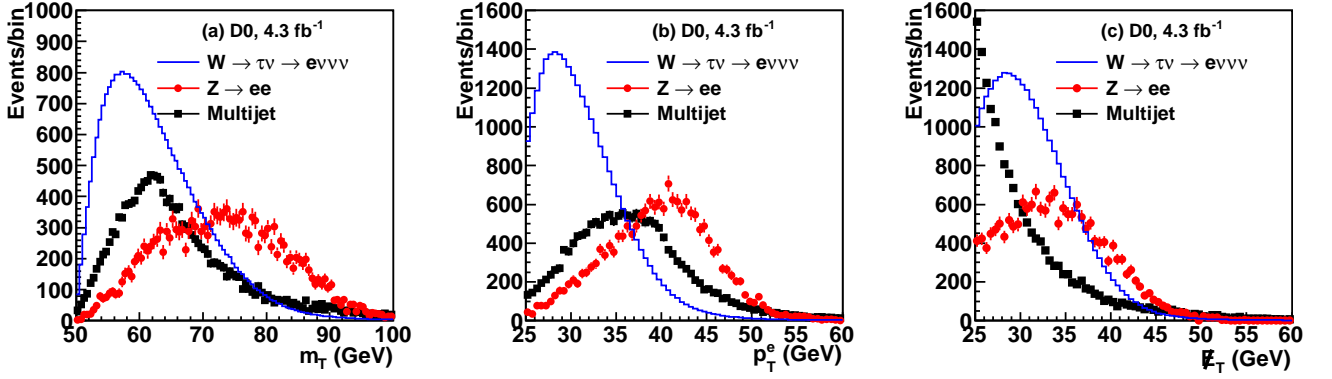}
  \caption{[color online] The (a) $\mt$, (b) $\pte$, and (c) $\met$ distributions for the three backgrounds $\zee$ (red), multijet (black) and $W\to\tau\nu$ (blue) with absolute
    normalization.\label{f:bkg}}
\end{figure*}

Propagated $M_W$ uncertainties are at most 1 MeV for both MJ and $W\to \tau \nu$ backgrounds for all three observables, and 1~MeV, 2~MeV, and 1~MeV for the $m_T$, $p^e_T$, and $\met$ observables for the $Z\to ee$ background. Distributions of the three background contributions are shown in Fig.~\ref{f:bkg}.

\section{Results}
\label{sec:results}

\begin{figure}[b]
  \includegraphics[width=\linewidth]{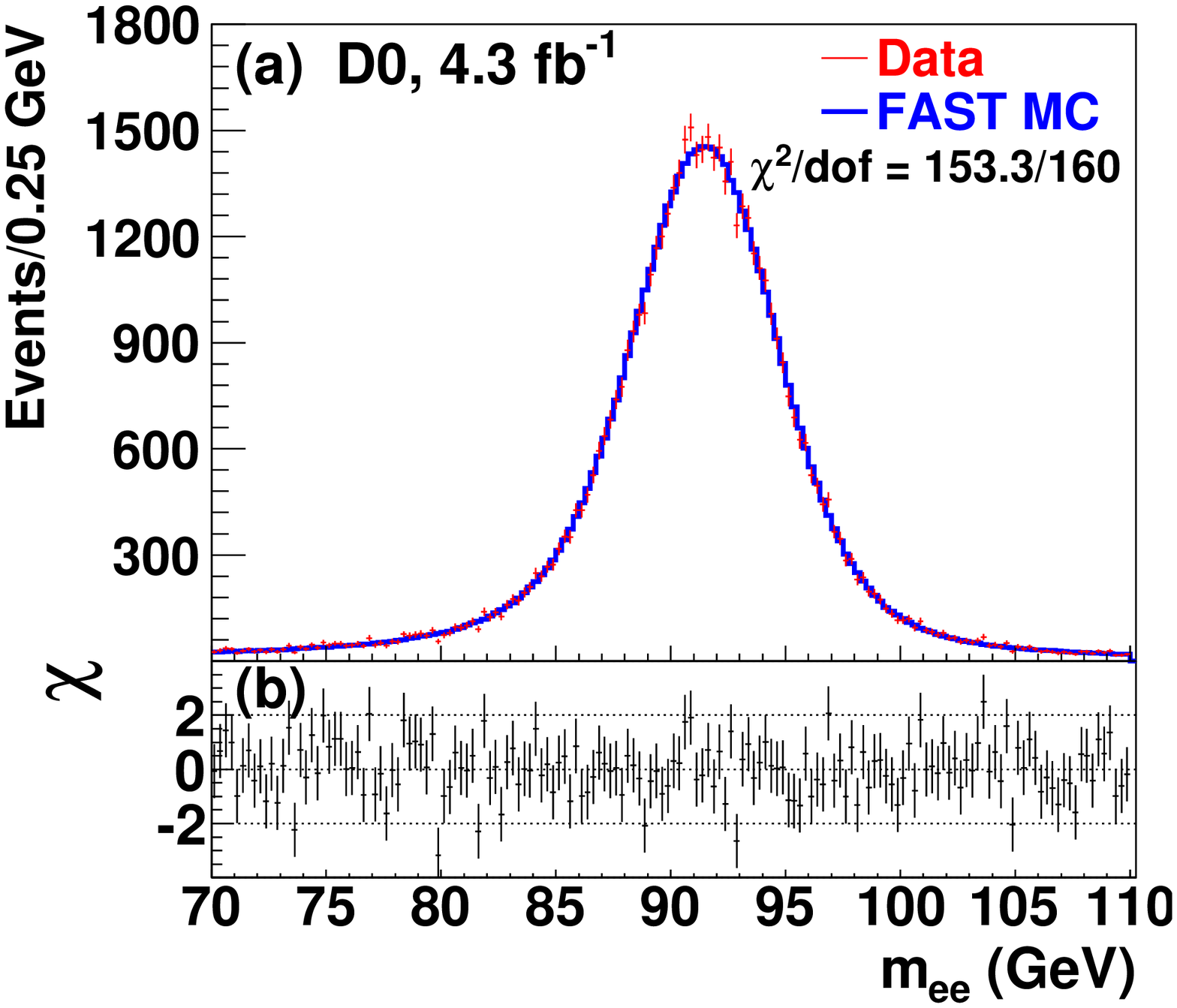}
  \caption{[color online] (a) The dielectron invariant mass distribution in $\zee$ data compare to the fast MC and (b) the $\chi$ values, where $\chi_i = \Delta N_i/\sigma_i$ for each bin in the distribution. $\Delta N_i$ is the difference between the number of events for data and fast MC and $\sigma_i$ is the statistical uncertainty in bin $i$.
    \label{fig:zfinal}}
\end{figure}

\begin{figure*}[hbpt]
  \includegraphics[width=0.32\linewidth]{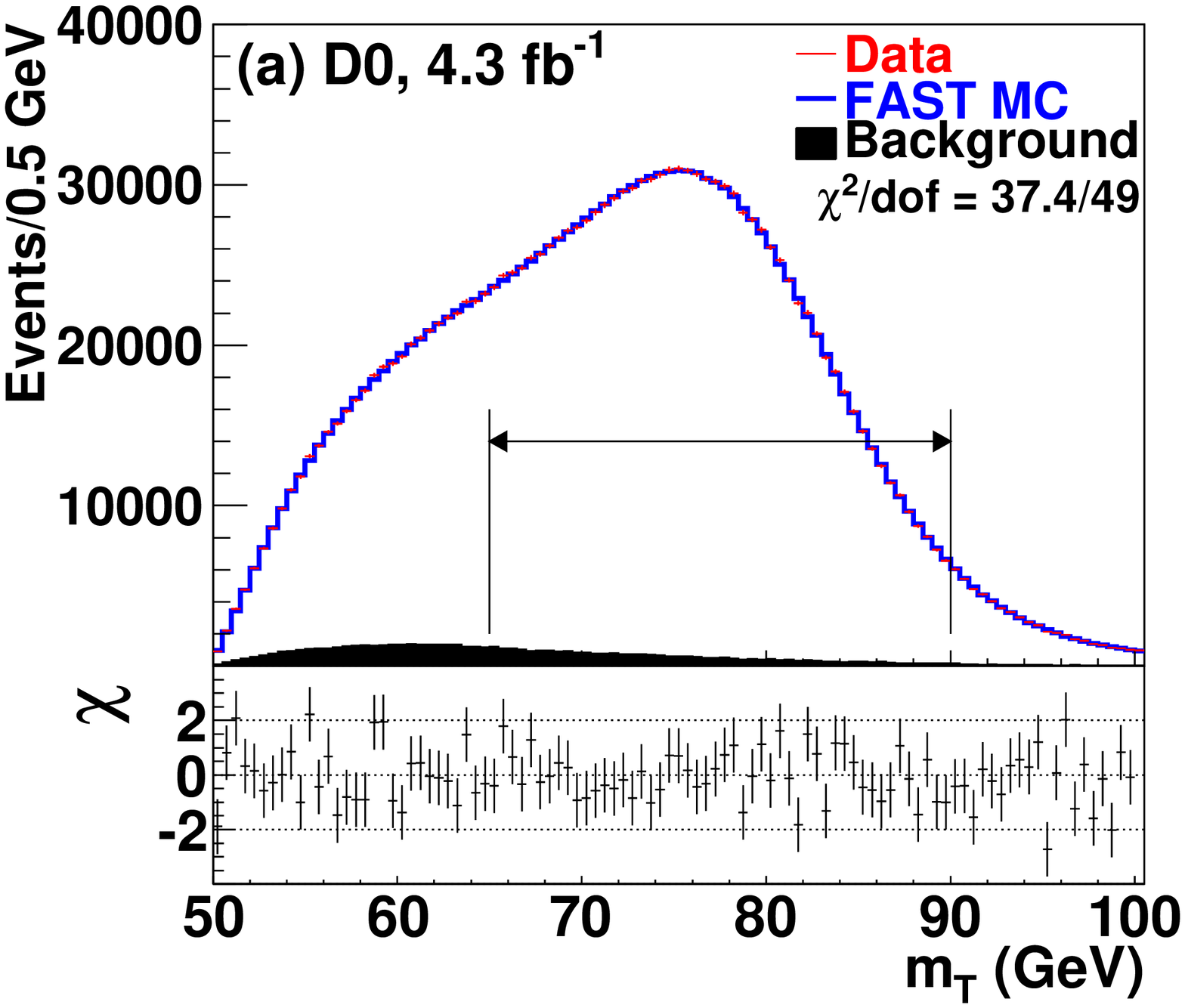}
  \includegraphics[width=0.32\linewidth]{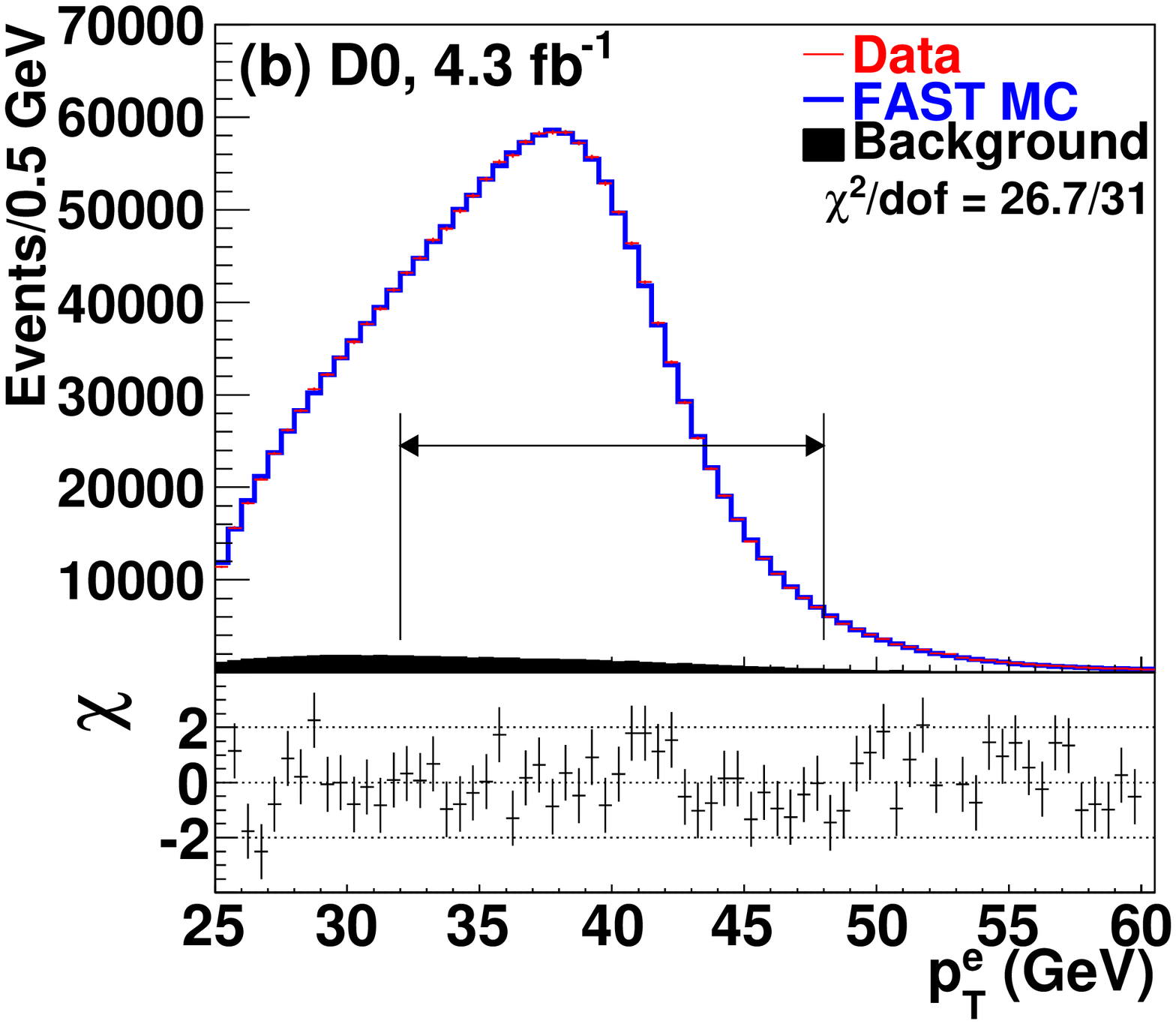}
  \includegraphics[width=0.32\linewidth]{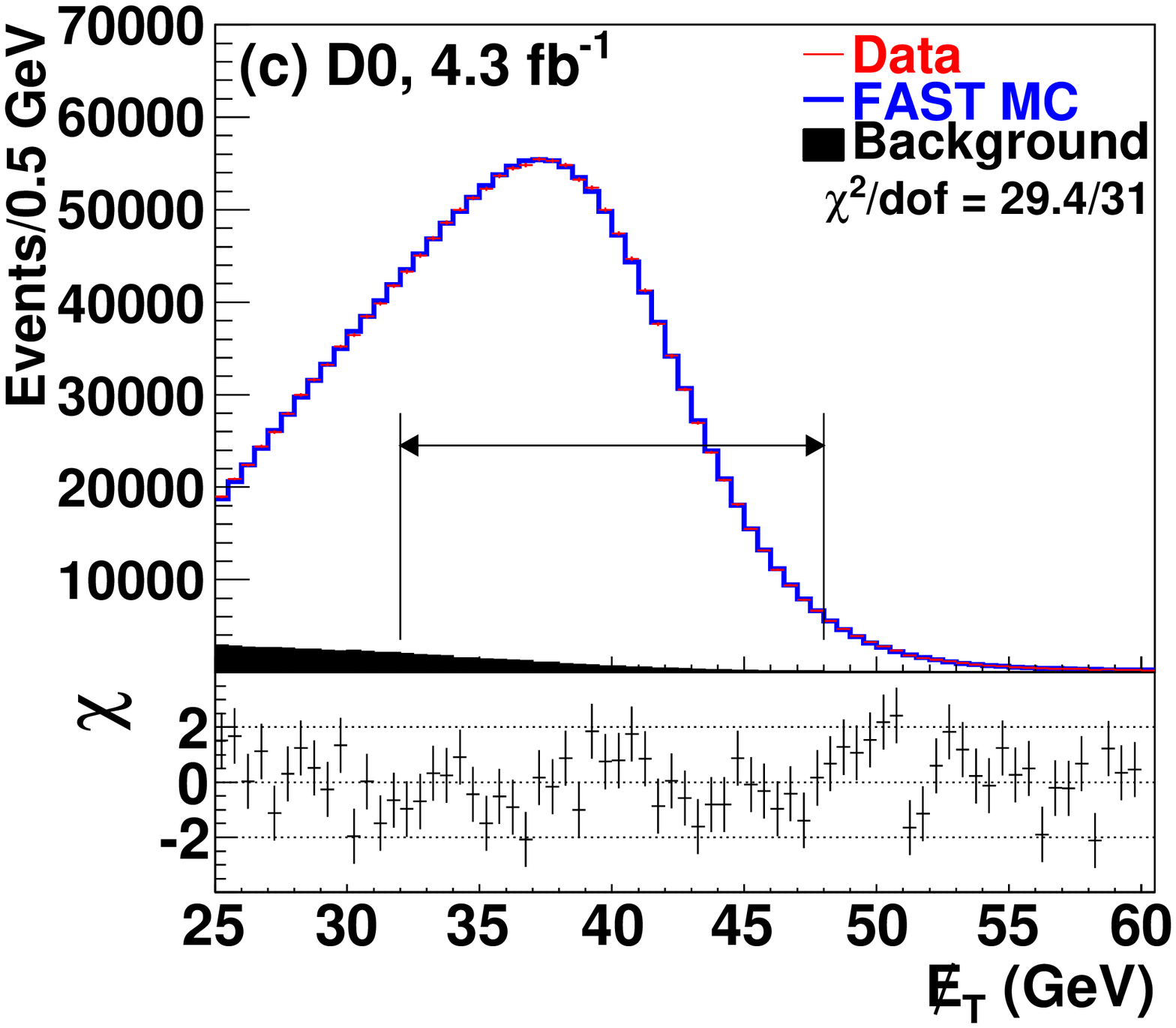}
  \caption{[color online] Distributions of (a) $\mt$, (b) $\pte$, and (c) $\met$ for data and fast MC with backgrounds. The $\chi$ values are shown below each distribution, where $\chi_i = \Delta N_i/\sigma_i$. $\Delta N_i$ is the difference between the number of events for data and fast MC and $\sigma_i$ is the statistical uncertainty in bin $i$. The fit ranges are indicated by the double-ended horizontal arrows.\label{fig:fits}}
\end{figure*}

Figure~\ref{fig:zfinal} shows the agreement between data and fast MC in fitting the invariant mass distribution of $\zee$ events. For an input value $M_Z = 91.188\,\text{GeV}$ used in the fast MC tuning, the value returned from the post-tuning was $91.193 \pm 0.017\,\text{(stat)}\,\text{GeV}$. Figure~\ref{fig:fits} shows comparisons of the data to the fast MC for the distributions we use to measure the $M_W$ including the fitting range used. The fitting ranges are determined by minimizing the sum in quadrature of the PDF and the expected statistical uncertainties from pseudo-experiments, which are the most sensitive uncertainties to the choice of fitting range.

The $W$ boson mass is determined by comparing the $\mt$, $\pte$, and $\met$ distributions from data to templates generated using the fast MC with different $M_W$ hypotheses in 10~MeV steps.  The backgrounds are added to the simulated distributions. The binned data distributions are compared to the binned template distributions for each $M_W$ hypothesis to obtain a log-likelihood value for each mass hypothesis. The log--likelihood distribution is interpolated between the individual mass hypotheses and the measured value of $M_W$ is determined as the mass value which maximizes the distribution. The measurements are performed separately for the $\mt$, $\pte$, and $\ \met$ distributions. The $W$ boson mass fit results from data are given in Table~\ref{t:answ}~\cite{OurNewPRL}.

\begin{table}[hbtp]
\begin{center}
  \caption{Results from the fits to data.  The uncertainty is the statistical uncertainty on the $W$ boson data sample.  The $\chi^2/$d.o.f. values are computed over the fit range.\label{t:answ}}
  \begin{tabular}{c|c|cc}\hline\hline
     Variable & Fit Range (GeV) &         Result (GeV)       & $\ \ \ \chi^2$/d.o.f. \\ \hline
      $\mt$   & $65<\mt<90$     & $\ \ \ 80.371\pm0.013\ \ \ $ &  37/49  \\
     $\pte$   & $32<\pte<48$    & $      80.343\pm0.014      $ &  27/31  \\ 
     $\met$   & $32<\met<48$    & $      80.355\pm0.015      $ &  29/31  \\\hline\hline
  \end{tabular}
\end{center}
\end{table}

The systematic uncertainties in the $W$ boson mass measurement arise from a variety of sources, but can be categorized as arising from experimental sources or from the $W$ and $Z$ production models.  The systematic uncertainties are determined from error propagation using MC pseudo-experiments, as described in Sec.~\ref{sec:syst}, or from ensemble test. They are summarized in Table~\ref{t:syst}.  The largest uncertainty of $(16, 17, 16)$~MeV for the $M_W$ measurements with the $(\mt, \pte, \met)$ distributions arises from the precision of knowledge of the absolute electron energy scale, which is limited by the size of the $\zee$ sample. It is expected to improve with more data.

\begin{table*}

    \caption{Systematic uncertainties on $M_W$ (in~MeV). The section of this paper where each uncertainty is discussed is given in the Table.\label{t:syst}}
  \begin{tabular}{ l | l | c | c| c} \hline\hline
   Source &Section\hphantom{000} &  \hphantom{000}$\mt$\hphantom{000} & \hphantom{000}$\pte$\hphantom{000} & \hphantom{000}$\met$\hphantom{000}\T\B\\
  \hline 
  \multicolumn{1}{c|}{Experimental} & & & &\T\\
  Electron Energy Scale                 & \ref{sec:elec_energy} & 16 &  17 & 16 \\
  Electron Energy Resolution        & \ref{sec:ConstantTerm} &  2 &   2 &  3 \\
  Electron Shower Model              & \ref{sec:SystElecNonLin} & 4 &   6 &  7 \\
  Electron Energy Loss                 & \ref{sec:QElossCorr} &  4 &   4 &  4 \\
  Recoil Model                      & \ref{sec:recoiluncertainty} &  5 &   6 & 14 \\
  Electron Efficiencies             & \ref{sec:sys_eff} &  1 &   3 &  5 \\
  Backgrounds                       & \ref{sec:background}    &  2 &   2 &  2 \\ \hline
  $\sum$(Experimental)       &    & 18 &  20 & 24 \T\B\\ \hline

  \multicolumn{1}{c|}{$W$ Production and Decay Model} & & & &\T\\
                                                                     
  PDF                          & \ref{sec:PDF} & 11 &  11 & 14 \\
  QED                          & \ref{sec:EWcorrections} &  7 &   7 &  9 \\
  Boson $p_T$                  & \ref{sec:bosonpT} &  2 &   5 &  2 \\ \hline
  $\sum$(Model) &    & 13 &  14 & 17 \T\B\\ \hline
 Systematic Uncertainty (Experimental and Model)\hphantom{000} &    & 22 &  24 & 29 \T\B\\ \hline
  $W$ Boson Statistics               & \ref{sec:results} & 13 &  14 & 15 \T\B\\ \hline
 Total Uncertainty        &    & 26 &  28 & 33 \T\B\\ \hline\hline
  \end{tabular}
\end{table*}    

\section{Combination}
\label{sec:combination}

The measurements from the three observables are correlated.  Correlation matrices for the $W$ boson data sample statistical uncertainties, the electron energy scale, the recoil scale and resolution, and the PDFs are determined using ensemble tests and standard uncertainty propagation.  The resulting correlation matrices are shown in Table~\ref{t-corr}.  The other model uncertainties besides PDF listed in Table~\ref{t:syst} are assumed to have a 100\% correlation among the $\mt$, $\pte$ and $\met$ fits. The electron energy scale uncertainty is shown to also be fully correlated among the three results. The different sources of uncertainty are assumed to be uncorrelated with each other. 

\def\hsp{\hphantom{00}}
\begin{table}[hbtp]
 \caption{Correlation matrices for the $W$ boson statistical, recoil scale and resolution, and the PDF uncertainties determined for the $4.3\,\text{fb}^{-1}$ data sample. The correlation matrices use the same ordering as in Eq.~\ref{e:cmdef}.\label{t-corr}}
\begin{tabular}{lc}\hline\hline
   Source & \hsp 4.3~fb$^{-1}$ Correlation Matrices \hsp \T\B\\ \hline

$W$ boson statistics  &
     $\left(
      \begin{array}{ccc}
  1 &  0.658 &  0.744 \\
  0.658 &  1 &  0.436 \\
  0.744 &  0.436 &  1 \\
       \end{array}
      \right)$
\rule{0pt}{7.0ex}  \\

 \hphantom{SPACE} & \\

Recoil scale and resolution  &
     $\left(
      \begin{array}{ccc}
  1 &  0.754 &  0.571 \\
  0.754 &  1 &  0.128 \\
  0.571 &  0.128 &  1 \\
      \end{array}
     \right)$
   \\

 \hphantom{SPACE} & \\

PDF  &

     $\left(
      \begin{array}{ccc}
   1    &  0.990 &  1 \\
  0.990 &  1     &  0.988 \\
   1    &  0.988 &  1\\
      \end{array}
      \right)$
\rule[-6.0ex]{0pt}{0pt}\\\hline\hline

 \end{tabular}
\end{table}

The total correlation matrix including all uncertainties is
\begin{equation}
\bordermatrix{
           & \mt   & \pte   & \met \cr
 \mt       & 1   &  0.89 & -0.86 \cr
 \pte      & 0.89  &  1  & -0.75 \cr
 \met      & -0.86 & -0.75 &  1}.
\label{e:cmdef}
\end{equation}


The three measurements can be combined using the BLUE method~\cite{blue_1, blue_2}. Using the correlation matrices from Table~\ref{t-corr} and the uncertainties from Tables~\ref{t:answ} and \ref{t:syst}, we find weights of 1.08, 0.11, and -0.19 for the $\mt$, $\pte$, and $\met$ measurements, respectively. The negative weight for the $\met$ measurement arises from the large correlation it has with the other measurements, as well as its relatively larger uncertainty. The values of the correlations between the $\met$ measurement and the other two receive large contribution from the assumed 100\% correlation in the $W$ production and decay model uncertainties. Because of the relatively larger uncertainty, the inclusion of the $\met$ measurement in the combination would not modify the final uncertainty. Thus, we choose to combine only the $\mt$ and the $\pte$ measurements, which despite being strongly correlated, have similar systematic uncertainties. With this choice, the weights for the combination are 0.87 and 0.13 for the $\mt$ and $\pte$ measurements, respectively. We obtain:


\begin{equation}
\begin{split}
M_W &= 80.367 \pm 0.013\thinspace \text{(stat)} \pm 0.022\thinspace \text{(syst)\ GeV}\\
         &= 80.367 \pm 0.026\, \text{GeV}.
\end{split}
\end{equation}

The $\chi^2$ probability of this combination is 2.8\%. The inclusion of the $\met$ measurement would give a negligible change in the average value of $M_W$. This result is combined with an earlier D0 measurement~\cite{OurPRL} to give the new D0 Run~II result of

\begin{equation}
M_W = 80.375 \pm 0.023\ \text{GeV}.
\end{equation}
For the combination of this new measurement and the measurement in
Ref.~\cite{OurPRL}, the production model uncertainties are treated as
fully correlated between the two measurements, and all other uncertainties,
dominated by statistics, are assumed to be uncorrelated.  


\section{Consistency Checks}
\label{sec:checks}

In this section we present consistency checks of the analysis. Two kinds of checks are performed.  For the first, we vary the fit ranges shown in Table~\ref{t:answ} used in the final $M_W$ fits.  For the second, we determine the $W$ and $Z$ boson masses for many different subsets of the data. We then determine whether the ratio of the $W$ boson mass to the $Z$ boson mass is stable. The subsets are defined using variables that are {\em a priori} considered to be difficult to describe or which have critical impact on the result. After consideration of the systematic uncertainties and their correlations, we find that each of these consistency checks shows good agreement among the subsets of data.

\subsection{Fitting Range}
\begin{figure}[ht]
  \includegraphics[width=0.94\linewidth]{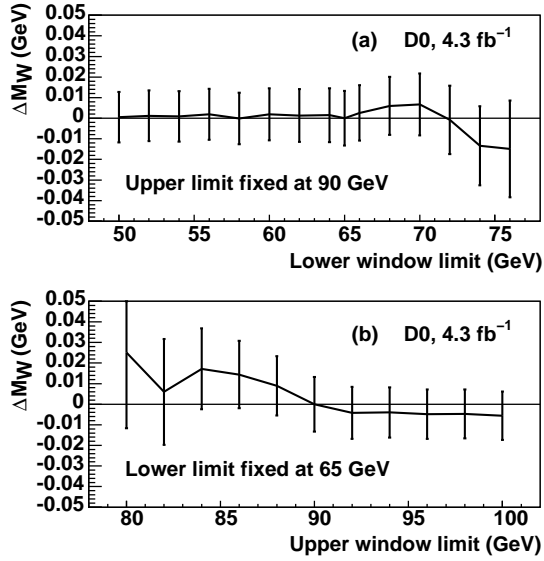}
  \caption{Variations in $M_W$ determined from fits to the $\mt$ spectrum as the fit range is changed. (a) Impact of varying the lower edge of the $m_T$ fit range, and (b) the impact of varying the upper edge. For each of the variations the differences between the result from the varied range and the result from the nominal range are shown.  The uncertainties represent the statistical uncertainties of the varied range fits.\label{f:range}}
\end{figure}

To study the impact of the fit ranges shown in Fig.~\ref{fig:fits} and used to
determine $M_W$, the $M_W$ measurements are repeated by changing the range.
Figure~\ref{f:range} shows the variation resulting from these tests applied to
the $\mt$ distribution.  The result is stable within the uncertainty as the
fit range is varied. Similar studies of the fit ranges for $\pte$ and $\met$ also show stable results.

\subsection{Instantaneous Luminosity}

\begin{figure}[ht]
\centering
\includegraphics [width=0.94\linewidth] {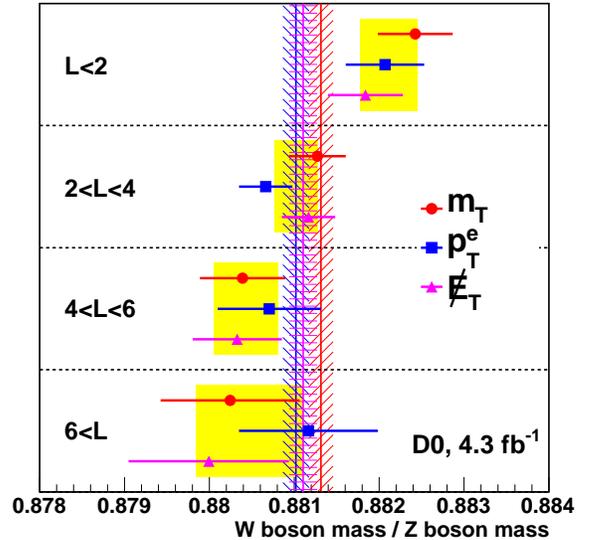}
\caption{
  [color online] The measured $M_W/M_Z$, separately for the $\mt$, $\pte$, and $\met$ observables and in four bins of instantaneous luminosity, in units of $36\times 10^{30}\,\text{cm}^{-2}\text{s}^{-1}$. The error bars for each observable represent the statistical uncertainty due to limited size of the $W$~boson sample. The yellow bands indicate the contribution from the $Z$~boson statistics, which is fully correlated for the three observables. The three vertical lines with hashed bands indicate the results from the three observables for the full data sample. When systematic uncertainties are considered, the measured $M_W/M_Z$ values are consistent.}
\label{fig:CheckLumi}
\end{figure}

We divide the $W$ and $Z$ boson samples into four subsets of different instantaneous luminosity per bunch using the the same criteria as for the parametrization of the electron identification efficiencies (Sec.~\ref{sec:DataHack}) and for the tuning of the absolute EM energy scale (Sec.~\ref{sec:elec_energy}). The ratio of the $W$~boson mass and $Z$~boson mass measurements are shown in Fig.~\ref{fig:CheckLumi}.

\subsection{Data-Taking Period}
\begin{figure}[ht]
\centering
\includegraphics [width=0.94\linewidth] {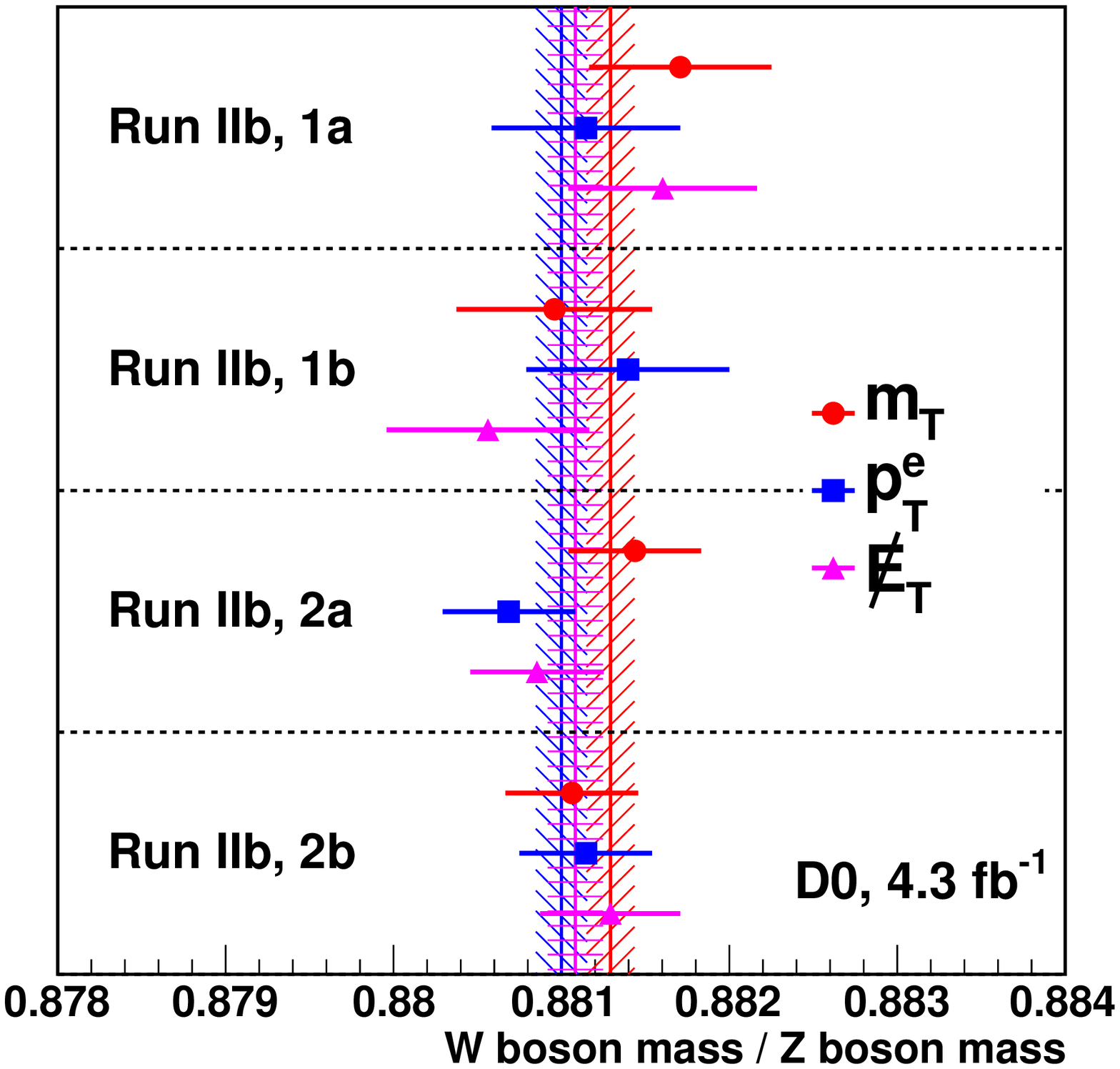}
\caption{
  [color online] The measured ratio $M_W/M_Z$, separately for the $\mt$, $\pte$, and $\met$ observables and for four data-taking periods. The uncertainties for each observable represent the combined statistical uncertainty due to limited $W$~statistics and $Z$~statistics. The three vertical lines with hashed uncertainties indicate the results from the three observables for the full data sample.}
\label{fig:CheckTime}
\end{figure}

We divide the data into four data-taking periods.  The first two and the last two are separated by a one-month accelerator shutdown. Each half is divided into two periods with equal integrated luminosities. The results are given in Fig.~\ref{fig:CheckTime}.

\subsection{Electron {\boldmath $\eta_{\rm det}$}}
\begin{figure}[ht]
\centering
\includegraphics [width=0.94\linewidth] {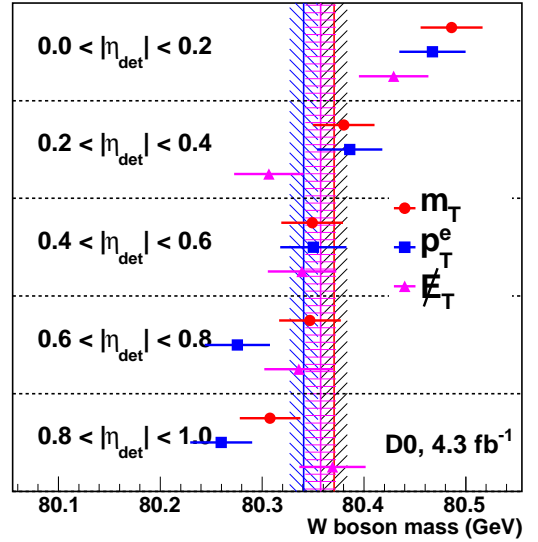}

\caption{
 [color online] Measured $M_W$ from the $\mt$, $\pte$, and $\met$ observables, separately for five different regions in electron~$|\eta_{\rm det}|$. The error bars for each observable represent the statistical uncertainty of the $W$~boson sample. The three vertical lines with hashed bands indicate the results from the three observables for the full data sample. When systematic uncertainties are considered, the measured $M_W$ values are consistent.}
\label{fig:CheckEta}
\end{figure}

We divide the data into five samples as defined in Table~\ref{table:StandardEtaBins}. This is the same categorization that is used in the determination of the $\eta_{\rm det}$~dependence of the EM energy scale (Sec.~\ref{sec:EMscaleEtaAdj}). The measured $W$~boson mass for each of the five sub-samples is shown in Fig.~\ref{fig:CheckEta}.  We do not show the mass ratio because we apply an explicit $\eta_{\rm det}$ dependent calibration and there are two electrons in each $\zee$ event. 

The electron energy scale in a single $\eta_{\rm det}$ region is determined from $\zee$ events in which one decay electron in the given region but the other can be anywhere else in the CC. Therefore, there are systematic anti-correlations between the $\eta_{\rm det}$ bins whose precise values are difficult to calculate since it would involve a simultaneous 10-dimensional fit for the parameters in the electron energy response model (Sec.~\ref{sec:elec_energy}) in each of the five sub-samples. If this simultaneous determination could be performed, we could calculate the probability that a disagreement at least as extreme as the one observed in the data would happen assuming a common value for $M_W$ across the five $\eta_{\rm det}$ bins. In the absence of the exact value, a lower bound on this probability can be given assuming no correlation between the five bins and that the systematic uncertainty in each bin scales as $\sqrt{5}\times 16\,\text{MeV}$. With these assumptions, considering the electron energy scale and PDF systematic uncertainties together with the statistical uncertainties, we find lower bounds on the probability of 35\%, 26\%, and 81\% for the $\mt$, $\pte$, and $\met$ fits, respectively, which shows consistency among the $\eta_{\rm det}$ regions.

\subsection{Hadronic Recoil {\boldmath \upara}}
\begin{figure}[ht]
\centering
\includegraphics [width=0.94\linewidth] {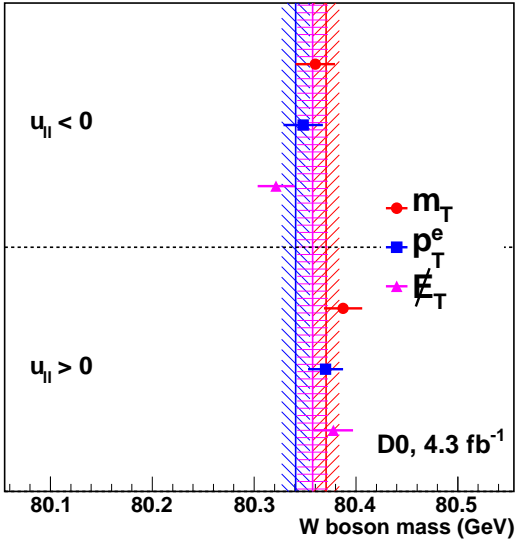}
\caption{[color online] The measured $M_W$ from the $\mt$, $\pte$ and $\met$ observables, separately for positive and negative~$\upara$. The three vertical lines with hashed bands indicate the results from the three observables for the full data sample.
  }
\label{fig:CheckUpar}
\end{figure}

We split the $W$~boson sample into a sample with $\upara < 0$ and a sample with $\upara > 0$. There are no equivalent splitting for the $Z$~boson sample because the two electrons from each $Z$~boson decay are reconstructed in approximately opposite directions in the transverse plane. We therefore show only the $M_W$ fits in Fig.~\ref{fig:CheckUpar}.

\subsection{Electron {\boldmath $\phi_{\text{mod}}$} Fiducial Requirement}
\begin{figure}[ht]
\centering
\includegraphics [width=0.94\linewidth] {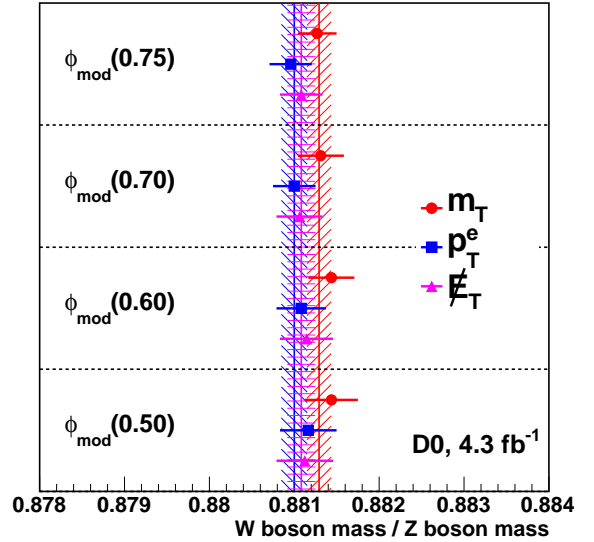}
\caption{
  [color online] The measured ratio $M_W/M_Z$, separately for the $\mt$,
  $\pte$, and $\met$ observables and for four $\phi_{\mathrm{mod}}$ selection variations. The numbers in parenthesis indicate which fraction of the CC~EM~module around its center is included in the electron fiducial region. The three vertical lines with hashed bands indicate the results from the three observables for the full data sample.
  }
\label{fig:CheckPhiMod}
\end{figure}

The nominal requirement, $0.1 \leq \phi_{\mathrm{mod}} \leq 0.9$, removes 10\% of the phase space at each edge of each CC EM~module (Sec.~\ref{sec:eff_phimod}). We also study four tighter versions of the requirement, namely $0.125 \leq \phi_{\mathrm{mod}} \leq 0.875$, $0.15 \leq \phi_{\mathrm{mod}} \leq 0.85$, $0.2 \leq \phi_{\mathrm{mod}} \leq 0.8$, and $0.25 \leq \phi_{\mathrm{mod}} \leq 0.75$, which remove 12.5\%, 15\%, 20\% and 25\%, respectively, of the acceptance at each edge of each CC EM~module. The effects of these variations are summarized in Fig.~\ref{fig:CheckPhiMod}. The measured $M_W/M_Z$ values are consistent for all variations.

\subsection{Hadronic Recoil {\boldmath $u_T$} Requirement}
\begin{figure}[ht]
\centering
\includegraphics [width=0.94\linewidth] {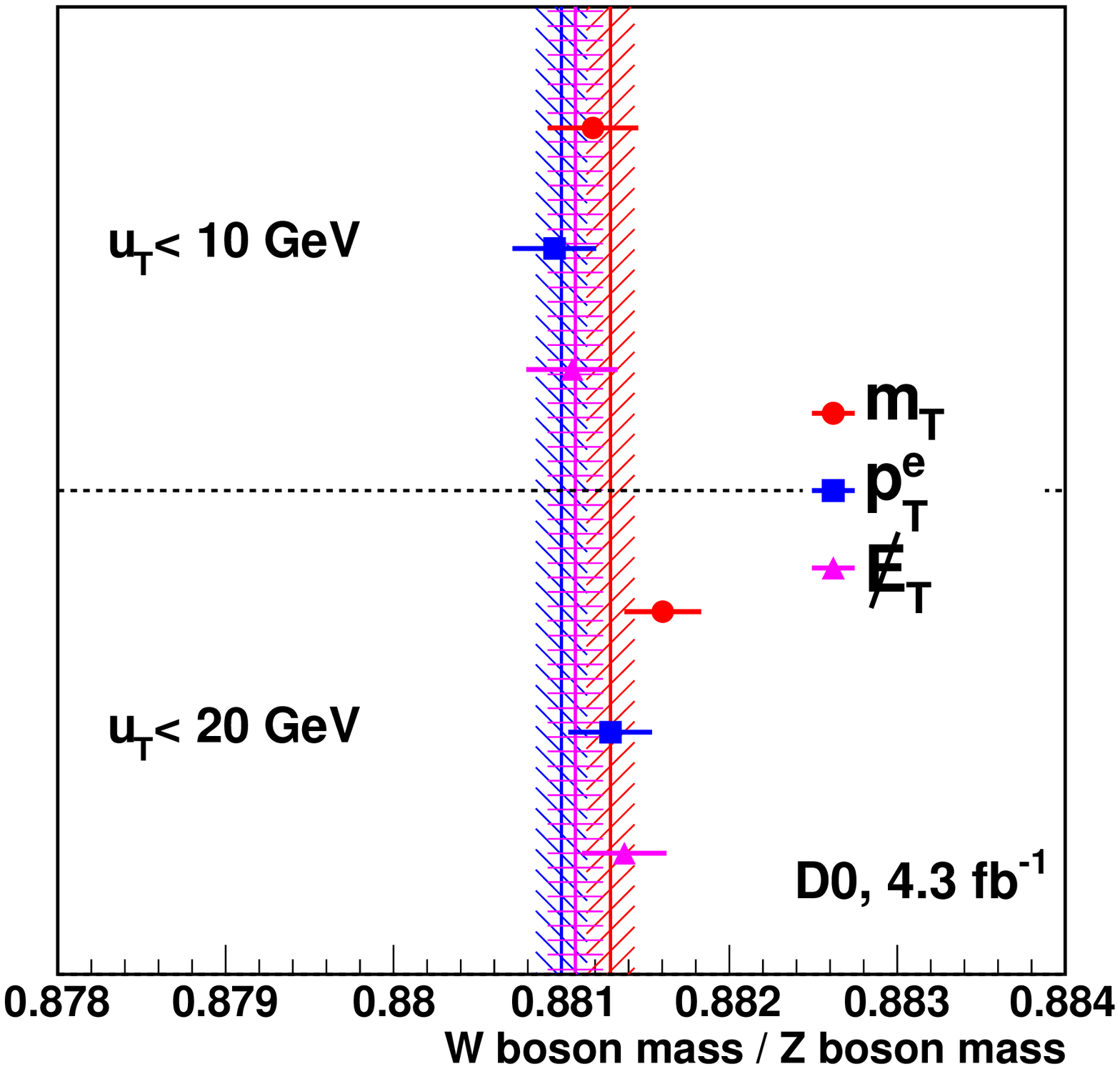}
\caption{
  [color online] The measured ratio $M_W/M_Z$, separately for the $\mt$, $\pte$, and $\met$ observables and for two $u_T$ variations. The three vertical lines with hashed bands indicate the results from the three observables with the nominal $u_T$ requirement.
  }
\label{fig:CheckUt}
\end{figure}

The nominal requirement of $u_T < 15$~GeV is changed to $u_T < 10$~GeV, and $u_T < 20$~GeV. The effects of these variations are summarized in Fig.~\ref{fig:CheckUt}. We find that, for both variations of the maximum $u_T$ requirement, the measured values of $M_W$ are consistent of the nominal one.

\subsection{Hadronic Recoil {\boldmath $\phi$}}

\begin{figure}[ht]
\centering
\includegraphics [width=0.94\linewidth] {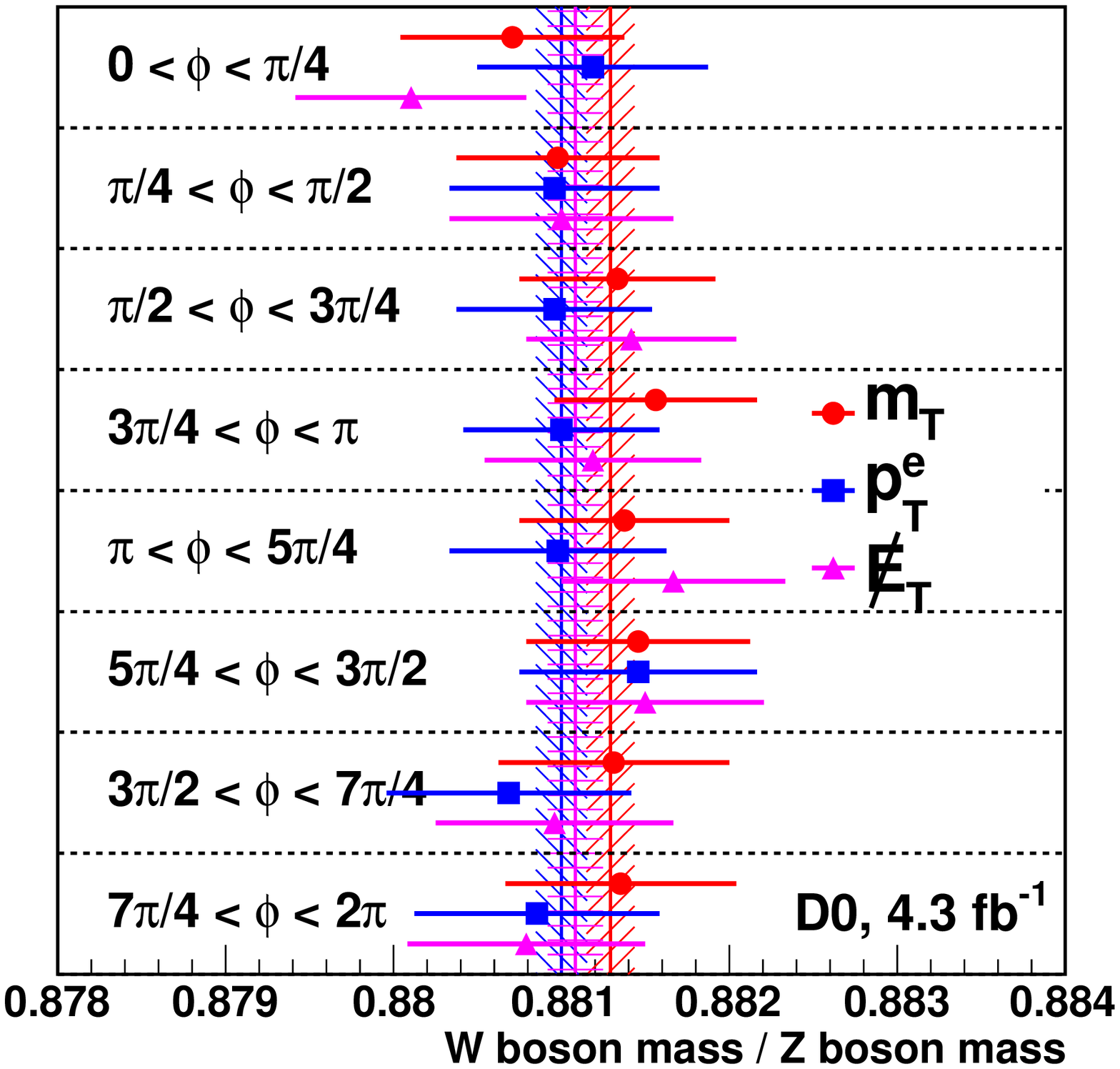}
\caption{
  [color online] The measured ratio $M_W/M_Z$, separately for the $\mt$,  $\pte$, and $\met$ observables and for eight bins in recoil~$\phi$. 
  }
\label{fig:CheckPhiRecoil}
\end{figure}

The last division is based on recoil~$\phi$. We divide the data sample into eight subsets, as defined in Fig.~\ref{fig:CheckPhiRecoil}.  The results of the ratio of the $W$ mass to the $Z$ mass are shown in the same figure. The measured $M_W/M_Z$ values are consistent for all regions.

\section{Conclusions}
\label{sec:conclusions}

We have presented a detailed description of the $W$ boson mass measurement using the $\wen$ mode and $4.3$~fb$^{-1}$ of D0 integrated luminosity recorded between 2006 and 2009. Three measurements are performed, using three kinematic variables $\mt$, $\pte$, and $\met$. The $\mt$ and $\pte$ measurements are combined to give the result
\begin{equation}
\begin{split}\nonumber
  M_W & = 80.367 \pm 0.013\thinspace \text{(stat)} \pm 0.022\thinspace \text{(syst)}\, \text{GeV} \\
      & = 80.367 \pm 0.026\,\text{GeV}.\nonumber
\end{split}
\end{equation}
This result is combined with an earlier D0 measurement based on $1\,\text{fb}^{-1}$ of data and similar analysis techniques to give
\begin{equation}\nonumber
  M_W = 80.375 \pm 0.023\, \text{GeV}.
\end{equation}
This measurement is in agreement with other measurements and has a precision equal to the world average prior to this paper and the most recent CDF measurements~\cite{combD0CDF}.

\begin{figure}
  \includegraphics[width=\linewidth]{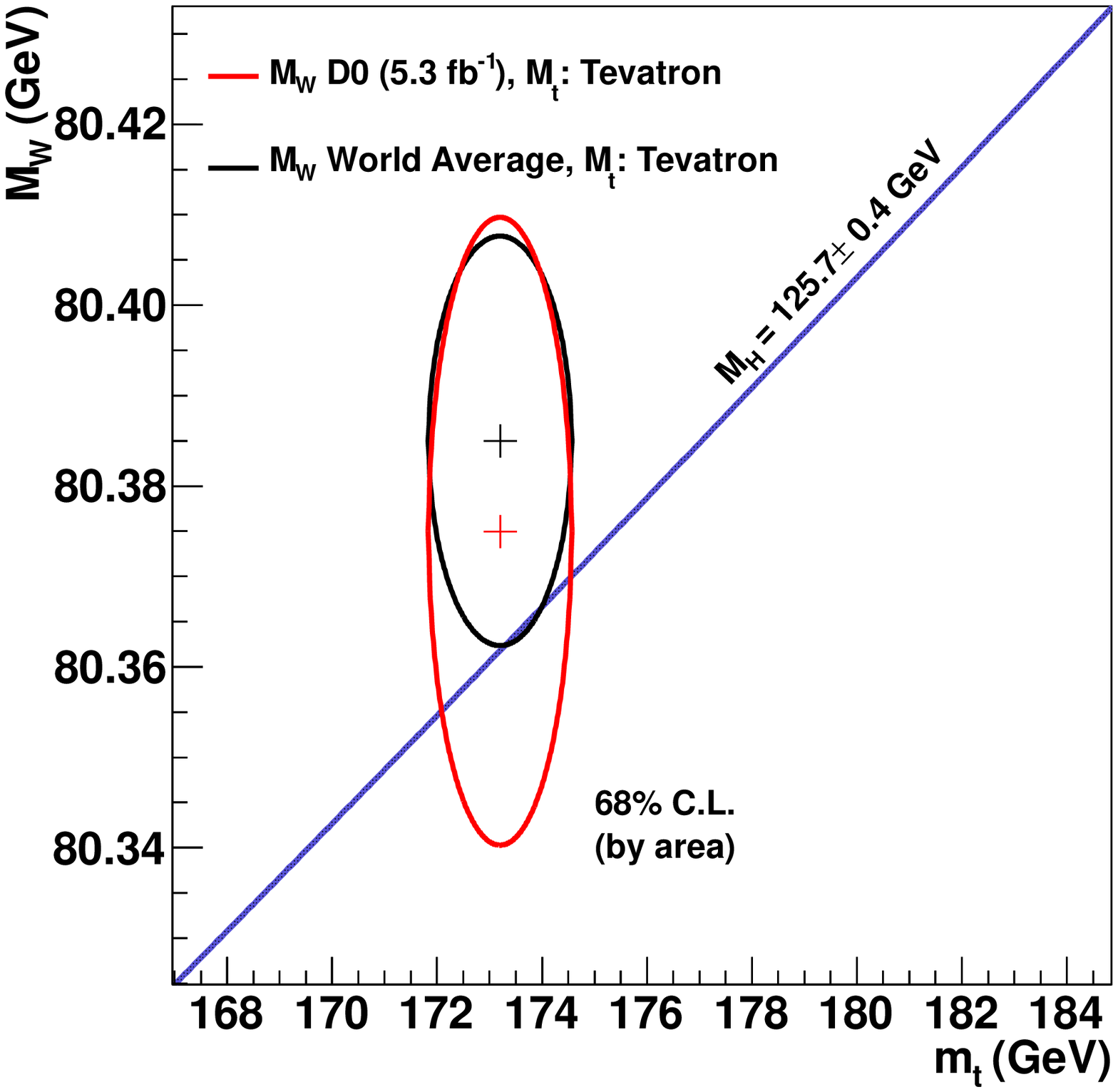}
  \caption{[color online] The D0 Run II measurement of $M_W$ shown with the world-average mass of the top quark $m_t$~\cite{Aaltonen:2012ra} at 68\% C.L. by area. The new world-average for $M_W$~\cite{combD0CDF} is also shown. The thin blue band is the prediction of $M_W$ in the Standard Model given by Eq.~\ref{eq:SMWmassPred}, assuming $M_H=125.7\pm 0.4\,\text{GeV}$\label{fig:higgs12}}
\end{figure}

Figure~\ref{fig:higgs12} shows this combined measurement, the world average top quark mass measurement~\cite{Aaltonen:2012ra}, and the consistency among these and a Higgs boson mass of $M_H = 125.7\,\text{GeV}$.

We thank the staffs at Fermilab and collaborating institutions,
and acknowledge support from the
DOE and NSF (USA);
CEA, CNRS/IN2P3 and CIMENT (France);
MON, NRC KI and RFBR (Russia);
CNPq, FAPERJ, FAPESP and FUNDUNESP (Brazil);
DAE and DST (India);
Colciencias (Colombia);
CONACyT (Mexico);
NRF (Korea);
FOM (The Netherlands);
STFC and the Royal Society (United Kingdom);
MSMT and GACR (Czech Republic);
BMBF and DFG (Germany);
SFI (Ireland);
The Swedish Research Council (Sweden);
and
CAS and CNSF (China).

\setcounter{secnumdepth}{0}
\section*{Appendix: Plots for {\boldmath $Z$} and {\boldmath $W$} Events}
\label{sec:Zcomparison}

\begin{figure}[pt]
\includegraphics [width=0.9\linewidth] {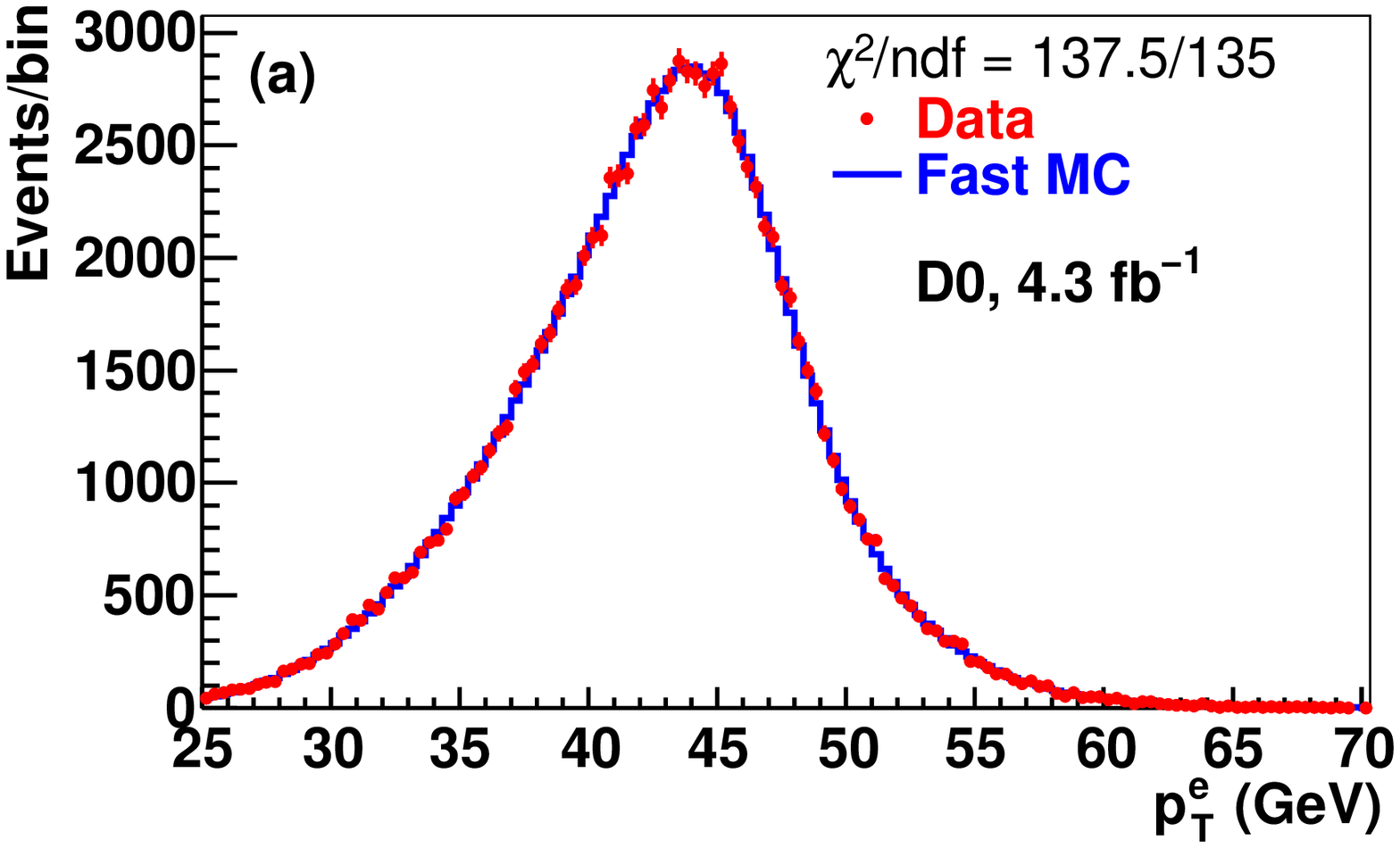}\\
\includegraphics [width=0.9\linewidth] {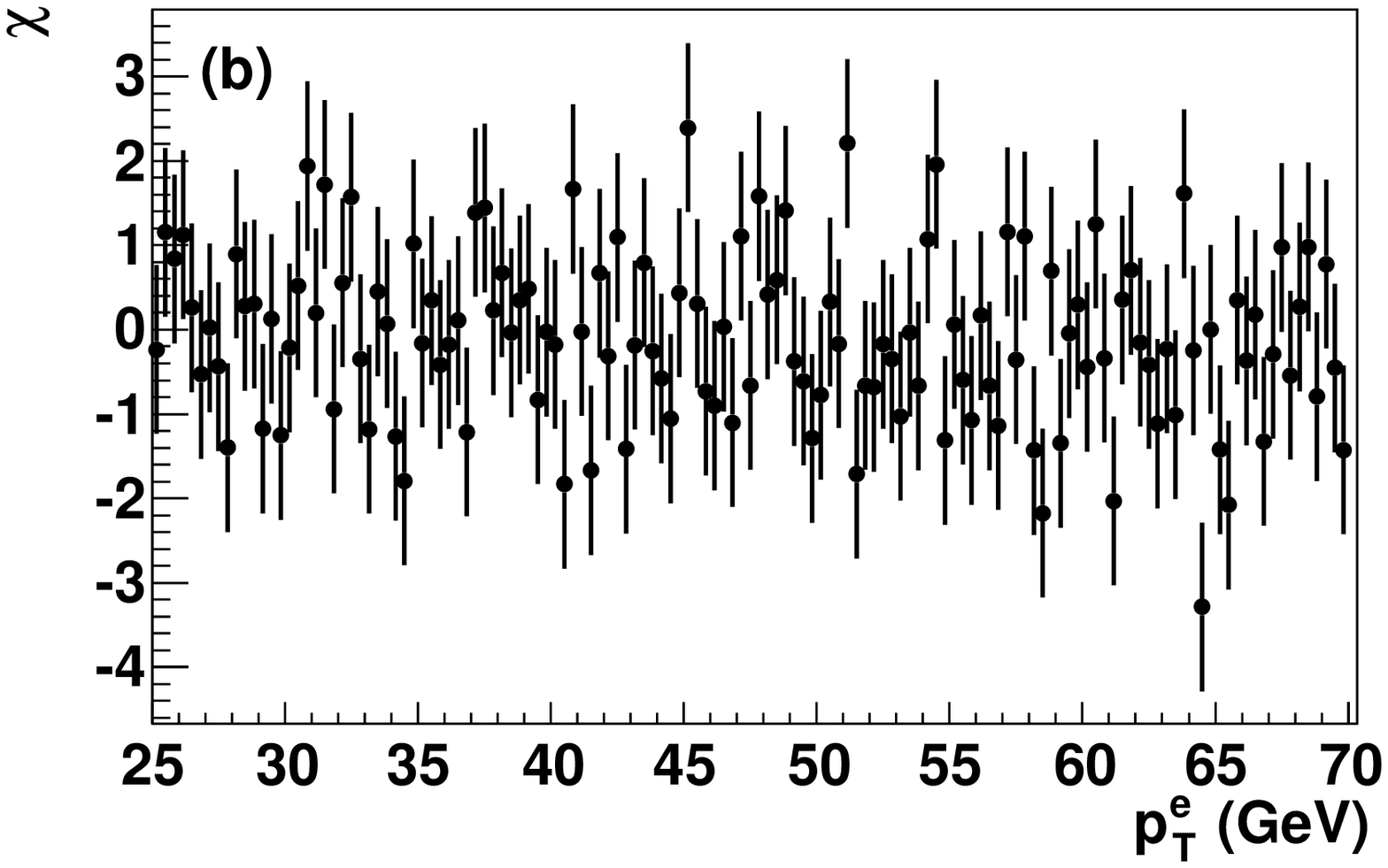}
\caption{[color online] (a) Comparison between data and fast MC for the $p^e_T$~distribution of the electrons from $\zee$, and (b) $\chi$ value per bin, where $\chi_i = \Delta N_i/\sigma_i$. $\Delta N_i$ is the difference between the number of events for data and fast MC and $\sigma_i$ is the statistical uncertainty in bin $i$.}
\label{fig:Zpte}
\end{figure}

\begin{figure}[htbp]
\includegraphics [width=0.9\linewidth] {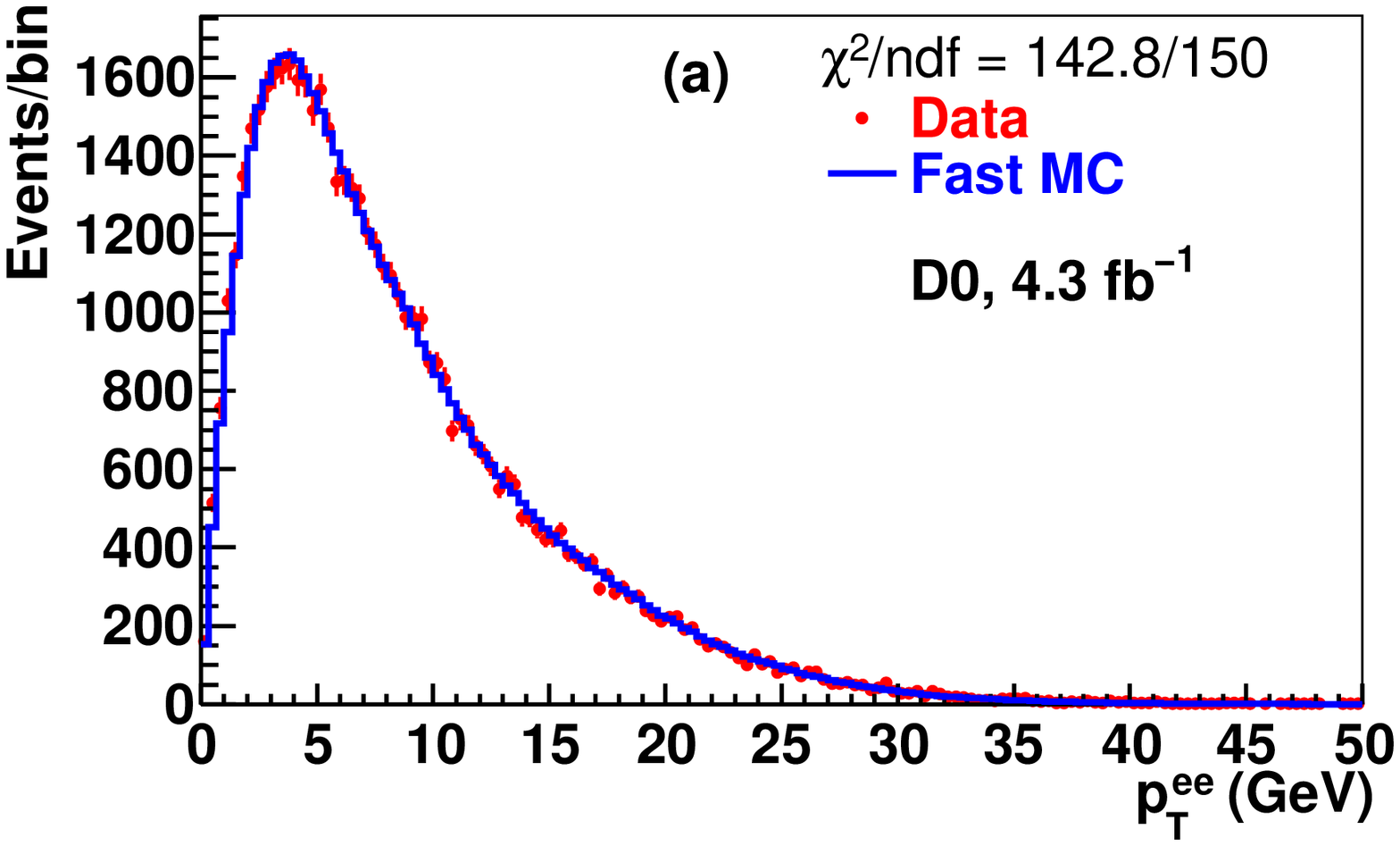}\\
\includegraphics [width=0.9\linewidth] {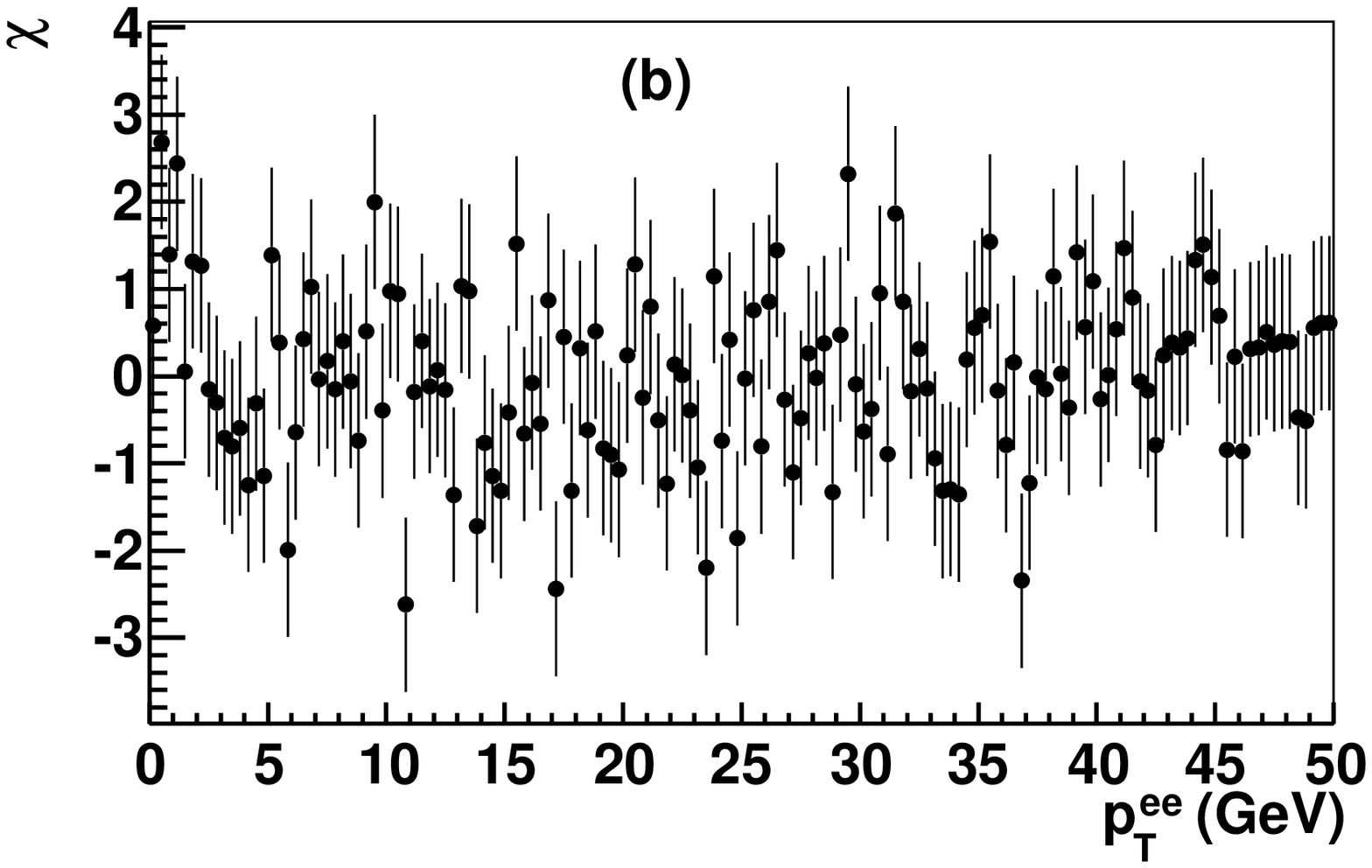}
\caption{[color online] (a) Comparison between data and fast MC for the $p^Z_T$~distribution of $\zee$ events, and (b) $\chi$ value per bin, where $\chi_i = \Delta N_i/\sigma_i$. $\Delta N_i$ is the difference between the number of events for data and fast MC and $\sigma_i$ is the statistical uncertainty in bin $i$.}
\label{fig:Zpt}
\end{figure}

\begin{figure}[htbp]
\includegraphics [width=0.9\linewidth] {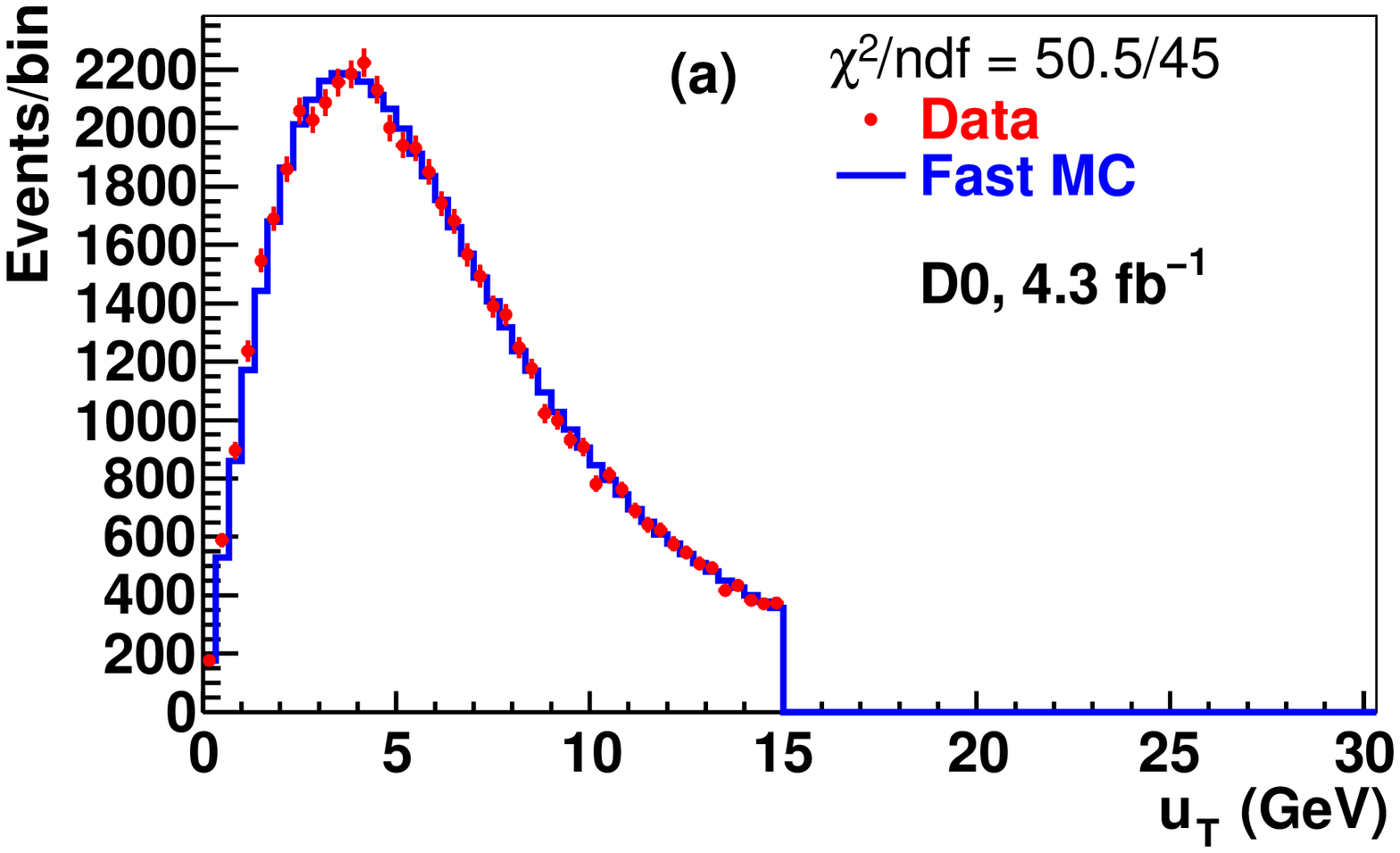}\\
\includegraphics [width=0.9\linewidth] {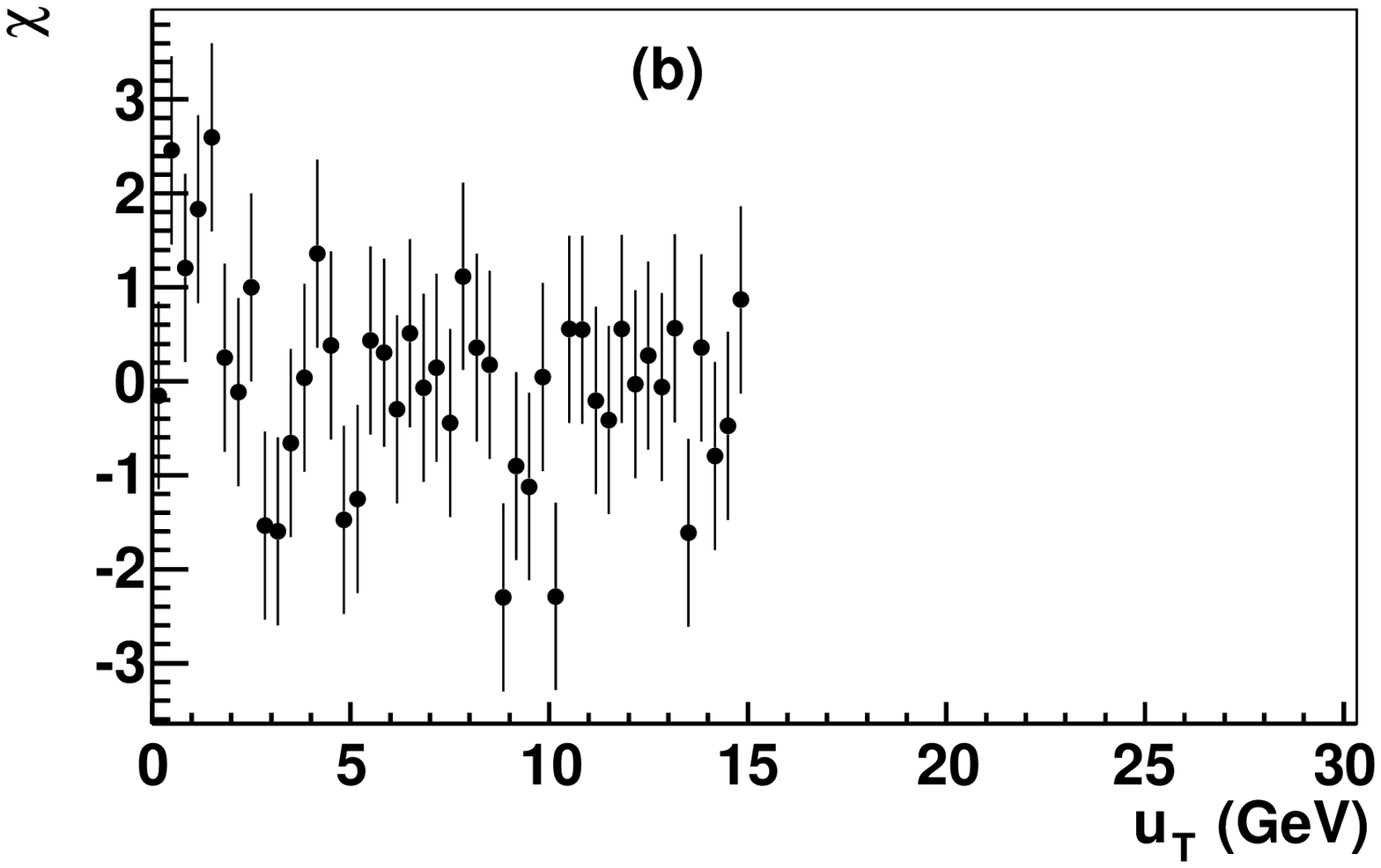}
\caption{[color online] (a) Comparison between data and fast MC for the $u_T$~distribution of $\zee$ events, and (b) $\chi$ value per bin, where $\chi_i = \Delta N_i/\sigma_i$. $\Delta N_i$ is the difference between the number of events for data and fast MC and $\sigma_i$ is the statistical uncertainty in bin $i$.}
\label{fig:Zut}
\end{figure}

\begin{figure}[htbp]
\includegraphics [width=0.9\linewidth] {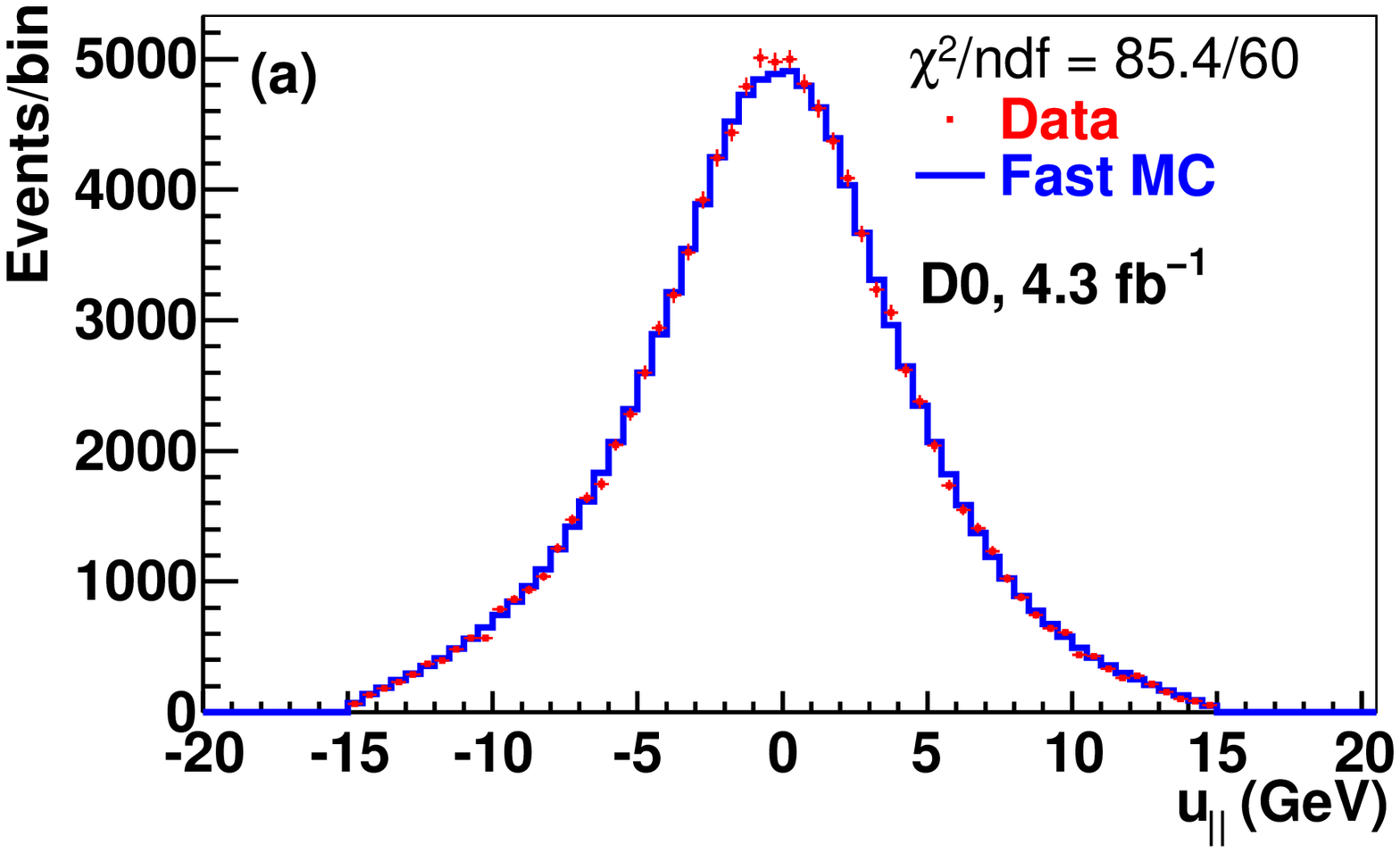}\\
\includegraphics [width=0.9\linewidth] {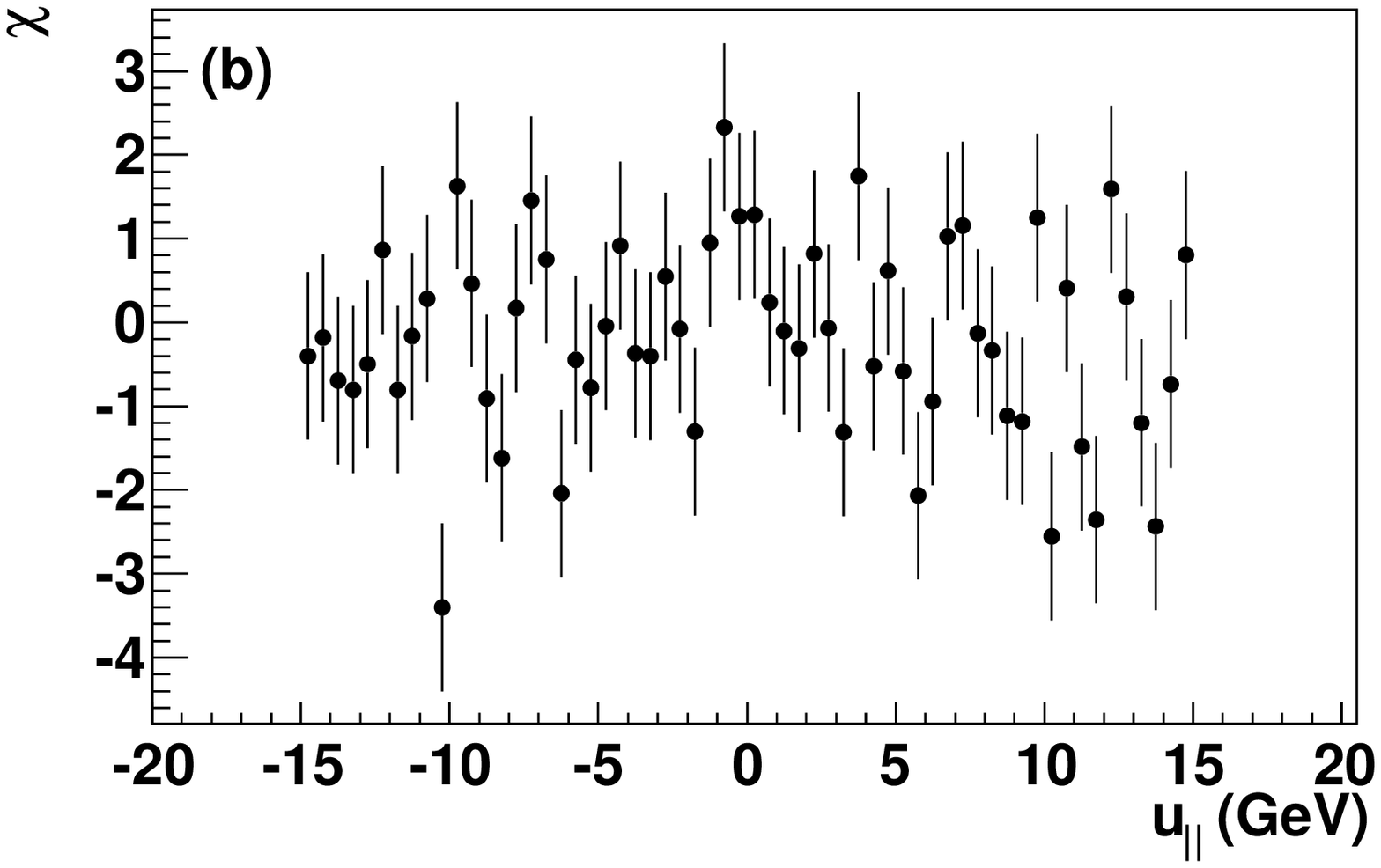}
\caption{[color online] (a) Comparison between data and fast MC for the $u_\parallel$~distribution of electrons in $\zee$ events, and (b) $\chi$ value per bin, where $\chi_i = \Delta N_i/\sigma_i$. $\Delta N_i$ is the difference between the number of events for data and fast MC and $\sigma_i$ is the statistical uncertainty in bin $i$.}
\label{fig:Zupara}
\end{figure}

\begin{figure}[htbp]
\includegraphics [width=0.9\linewidth] {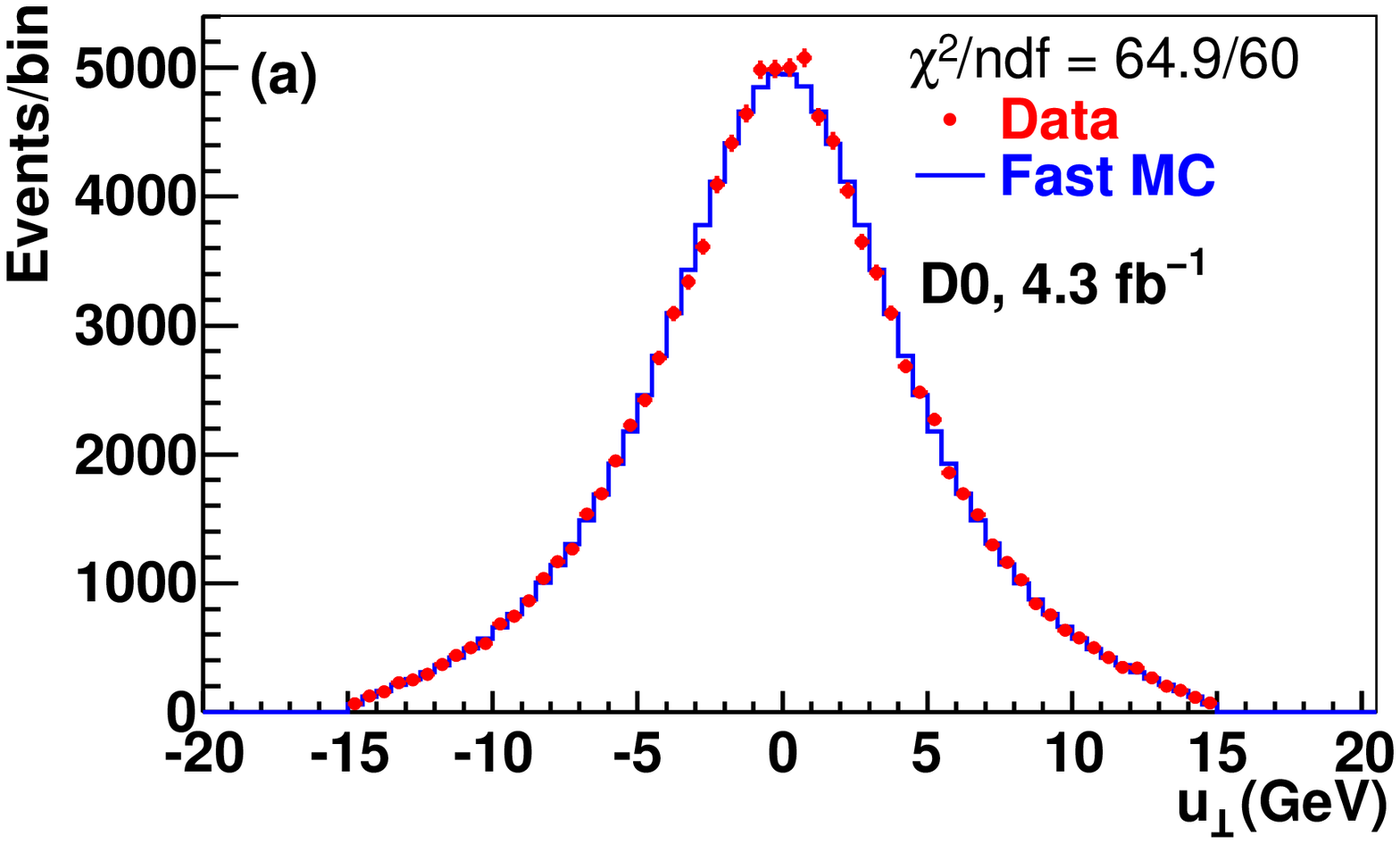}\\
\includegraphics [width=0.9\linewidth] {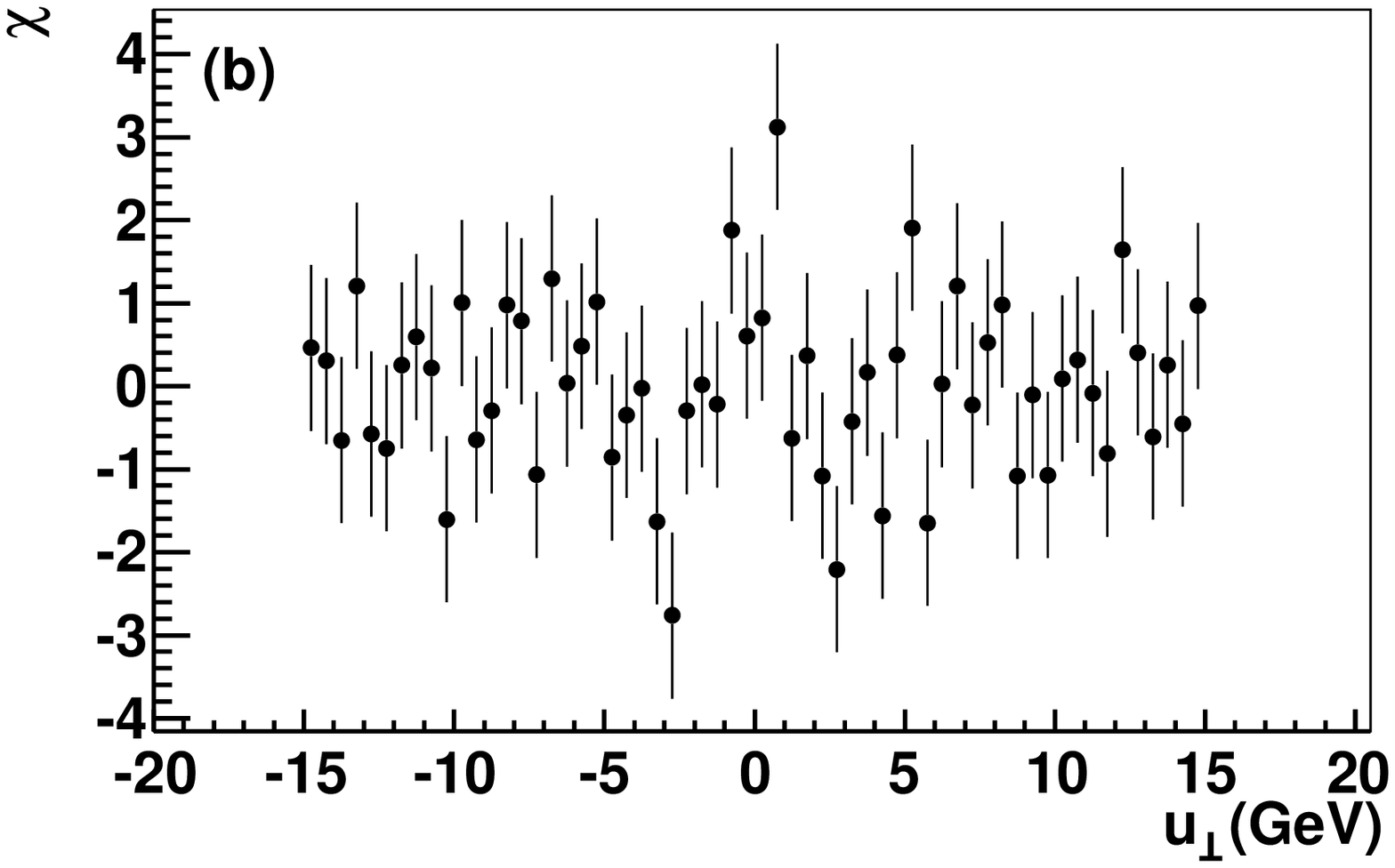}
\caption{[color online] (a) Comparison between data and fast MC for the $u_\perp$~distribution of electrons in $\zee$ events, and (b) $\chi$ value per bin, where $\chi_i = \Delta N_i/\sigma_i$. $\Delta N_i$ is the difference between the number of events for data and fast MC and $\sigma_i$ is the statistical uncertainty in bin $i$.}
\label{fig:Zuperp}
\end{figure}

\begin{figure}[htbp]
\includegraphics [width=0.9\linewidth] {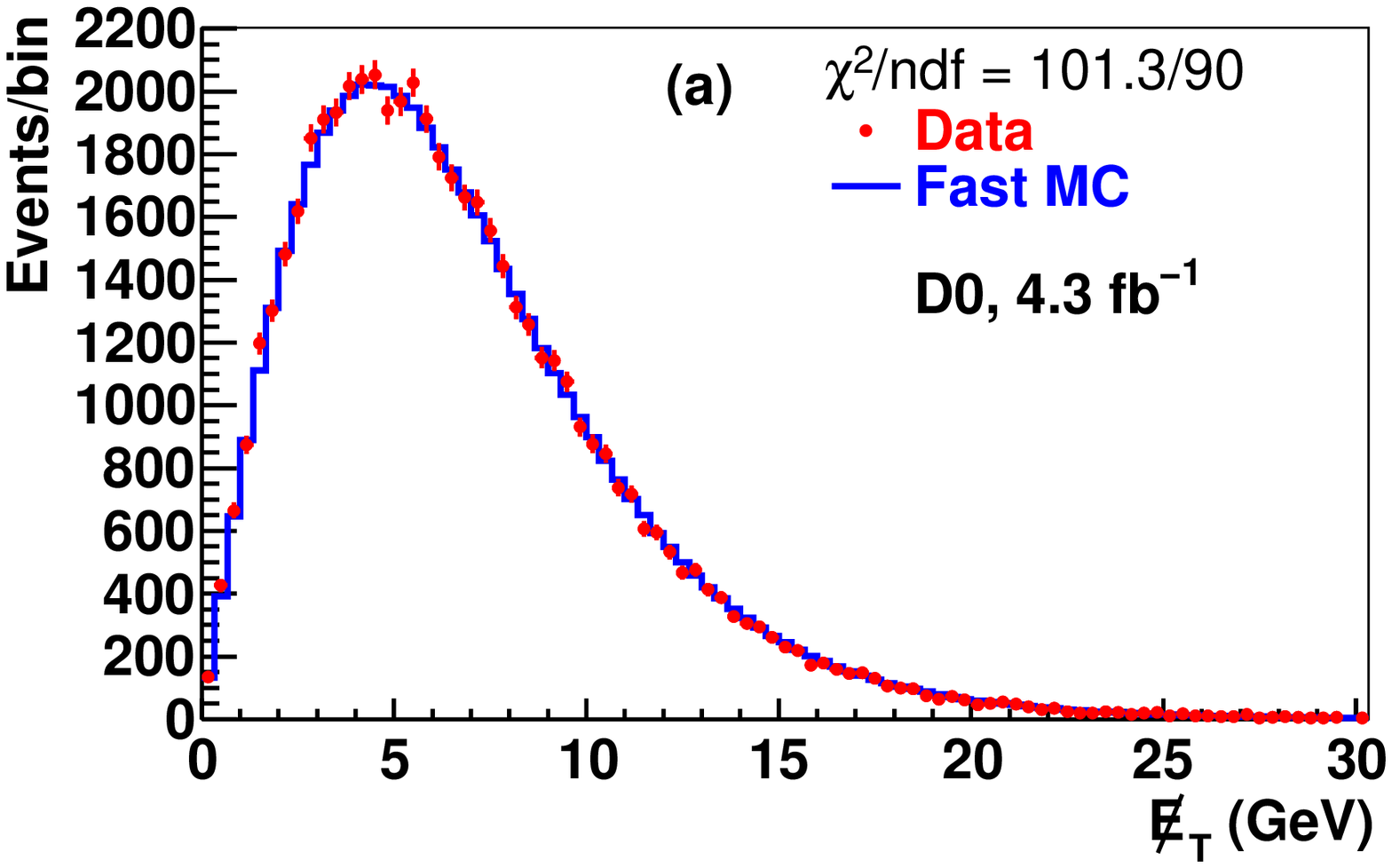}\\
\includegraphics [width=0.9\linewidth] {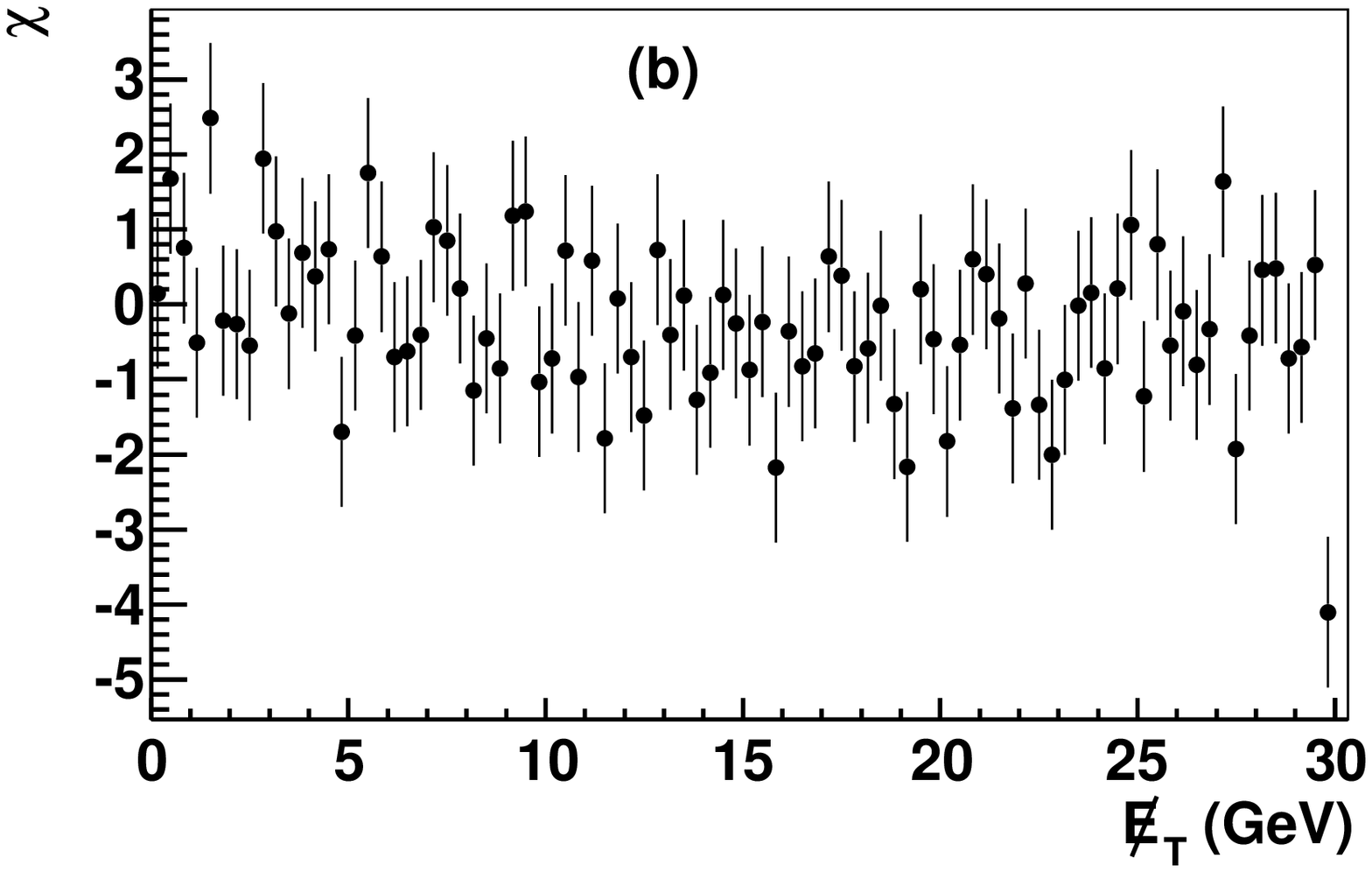}
\caption{[color online] (a) Comparison between data and fast MC for the $\met$~distribution of in $\zee$ events, and (b) $\chi$ value per bin, where $\chi_i = \Delta N_i/\sigma_i$. $\Delta N_i$ is the difference between the number of events for data and fast MC and $\sigma_i$ is the statistical uncertainty in bin $i$.}
\label{fig:Zmet}
\end{figure}

Since $Z$~boson decays are used as the main control sample to parameterize the fast
MC, it is important to check the agreement between the fast MC and the data
sample used for tuning. Figures \ref{fig:Zpte},
\ref{fig:Zpt}, and \ref{fig:Zut} show the $p^e_{T}$, $p^Z_T$, and $u_{T}$ distributions compared
between data and fast MC.  Figures \ref{fig:Zupara}, \ref{fig:Zuperp}, and
\ref{fig:Zmet} show the $u_\parallel$ and $u_\perp$ distributions of electrons from $\zee$ events,
and the $\met$ in $Z$ events, respectively.  The overall agreement is good.
The $\met$ distribution shows that the D0 calorimeter system worked well.


\begin{figure}[htpb]
\includegraphics [width=0.9\linewidth] {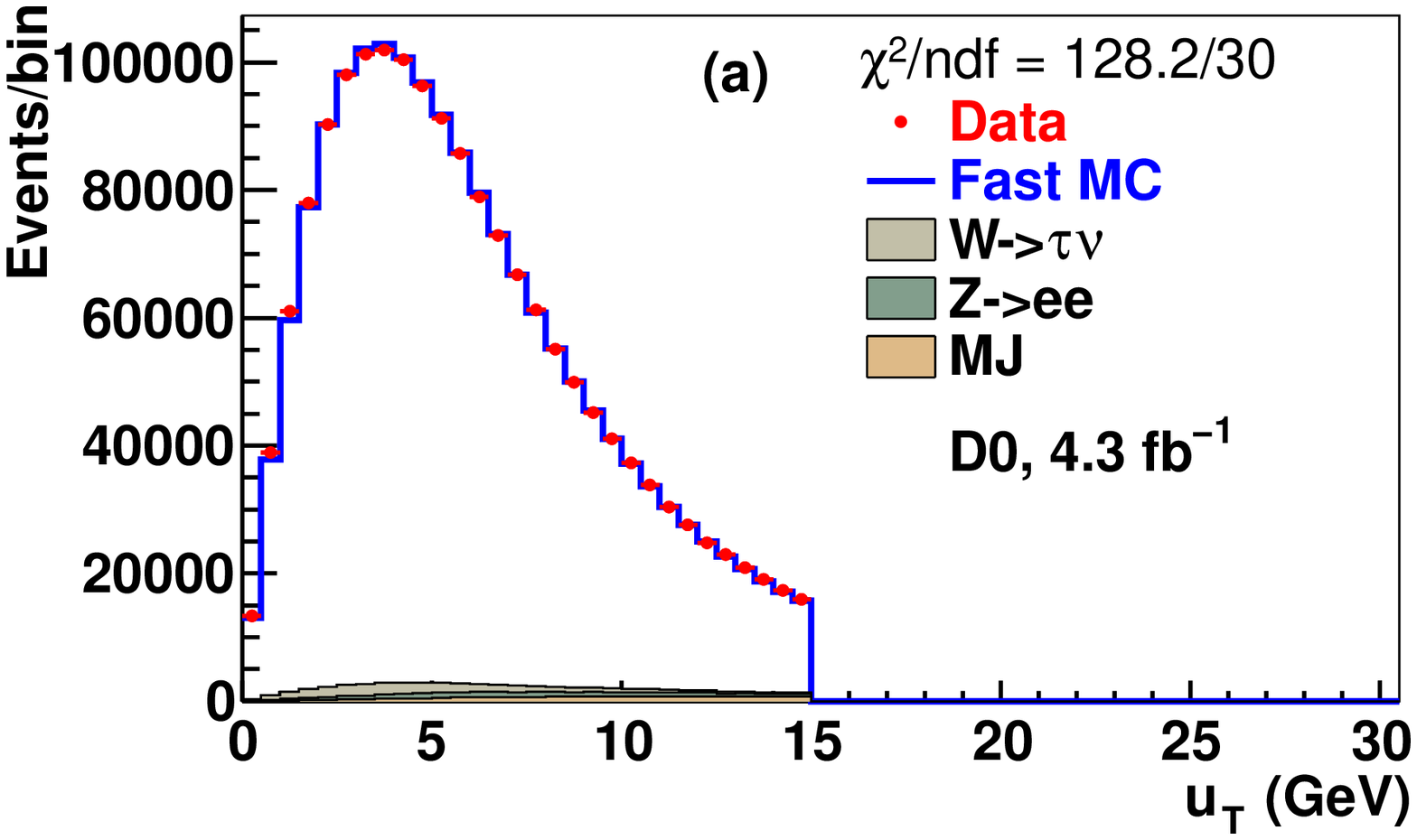}\\
\includegraphics [width=0.9\linewidth] {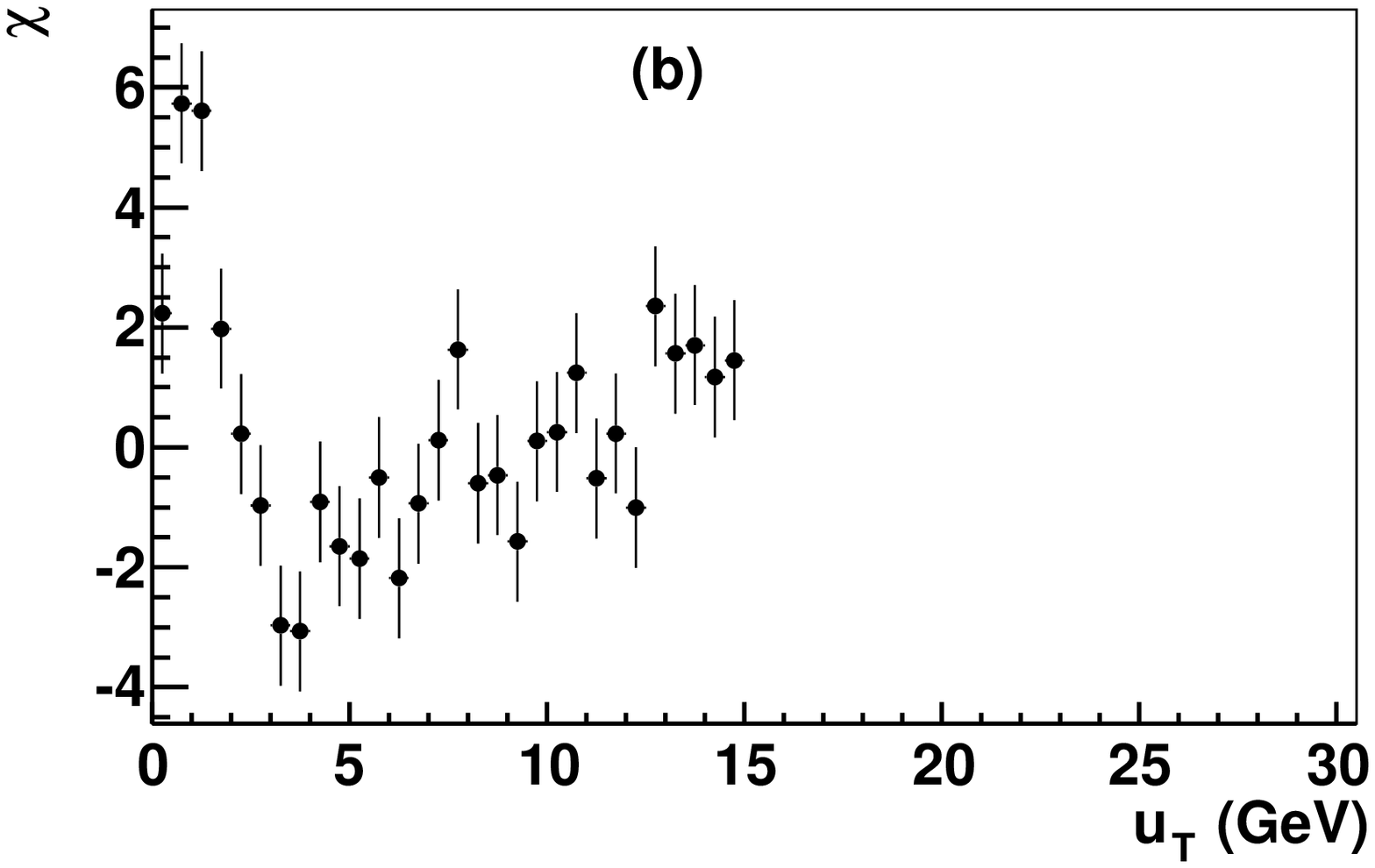}
\caption{[color online] (a) Comparison between data and fast MC for the $u_T$~distribution in $\wen$ data, and (b) $\chi$ value per bin, where $\chi_i = \Delta N_i/\sigma_i$. $\Delta N_i$ is the difference between the number of events for data and fast MC and $\sigma_i$ is the statistical uncertainty in bin $i$.}
\label{fig:Wut}
\end{figure}

\begin{figure}[htbp]
\includegraphics [width=0.9\linewidth] {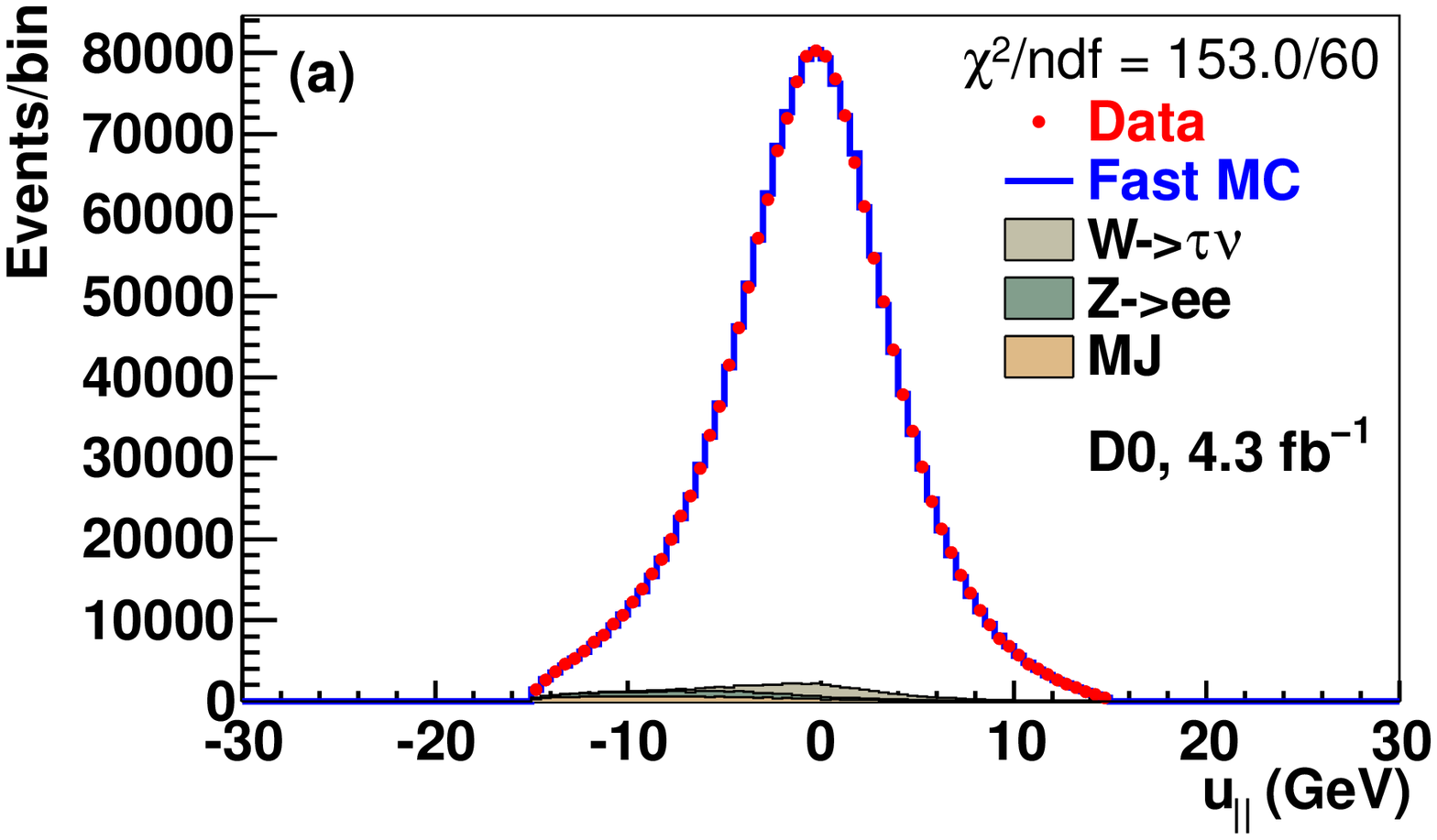}\\
\includegraphics [width=0.9\linewidth] {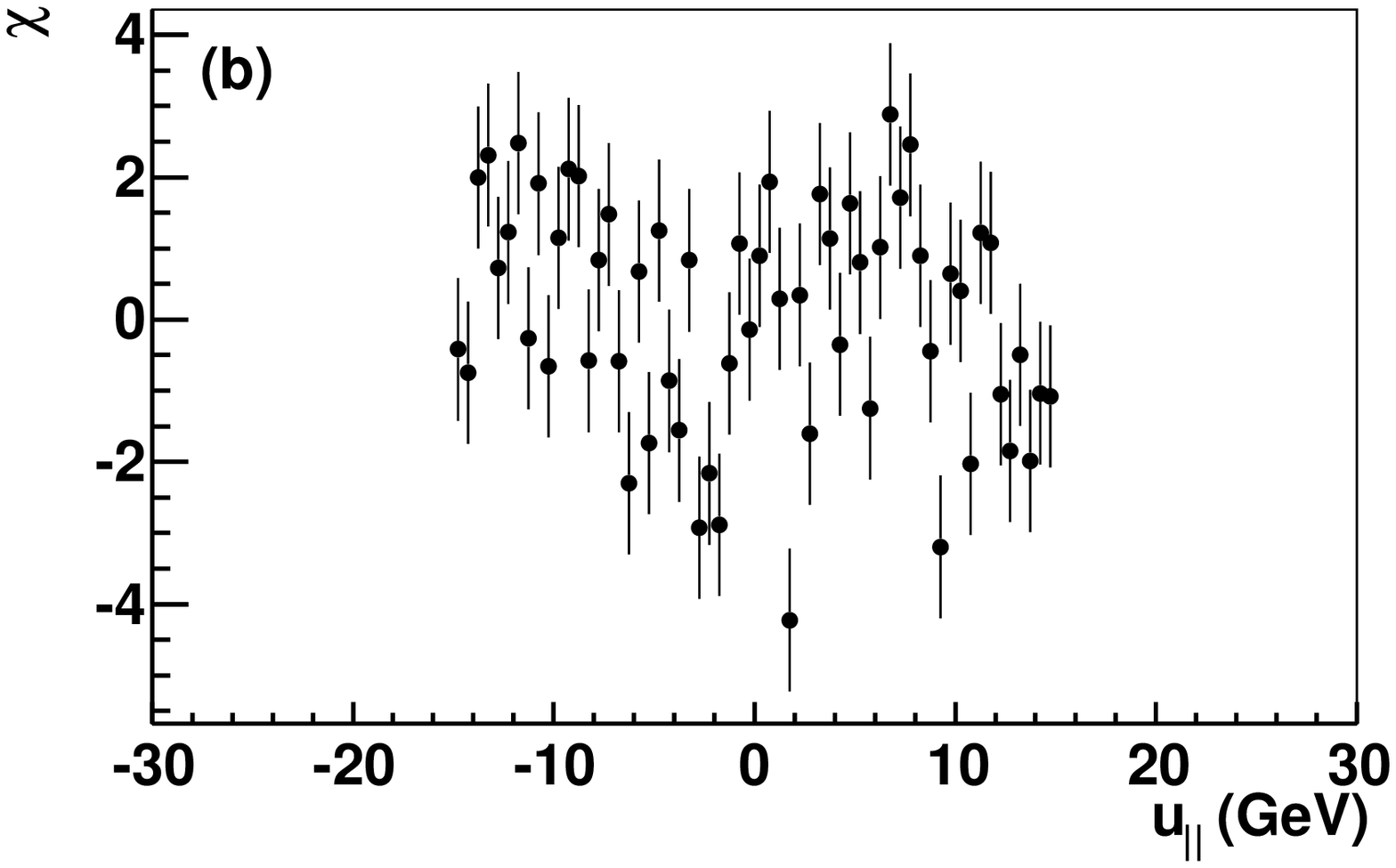}
\caption{[color online] (a) Comparison between data and fast MC for the $u_\parallel$~distribution in $\wen$ data, and (b) $\chi$ value per bin, where $\chi_i = \Delta N_i/\sigma_i$. $\Delta N_i$ is the difference between the number of events for data and fast MC and $\sigma_i$ is the statistical uncertainty in bin $i$.}
\label{fig:Wupara}
\end{figure}

\begin{figure}[htbp]
\includegraphics [width=0.9\linewidth] {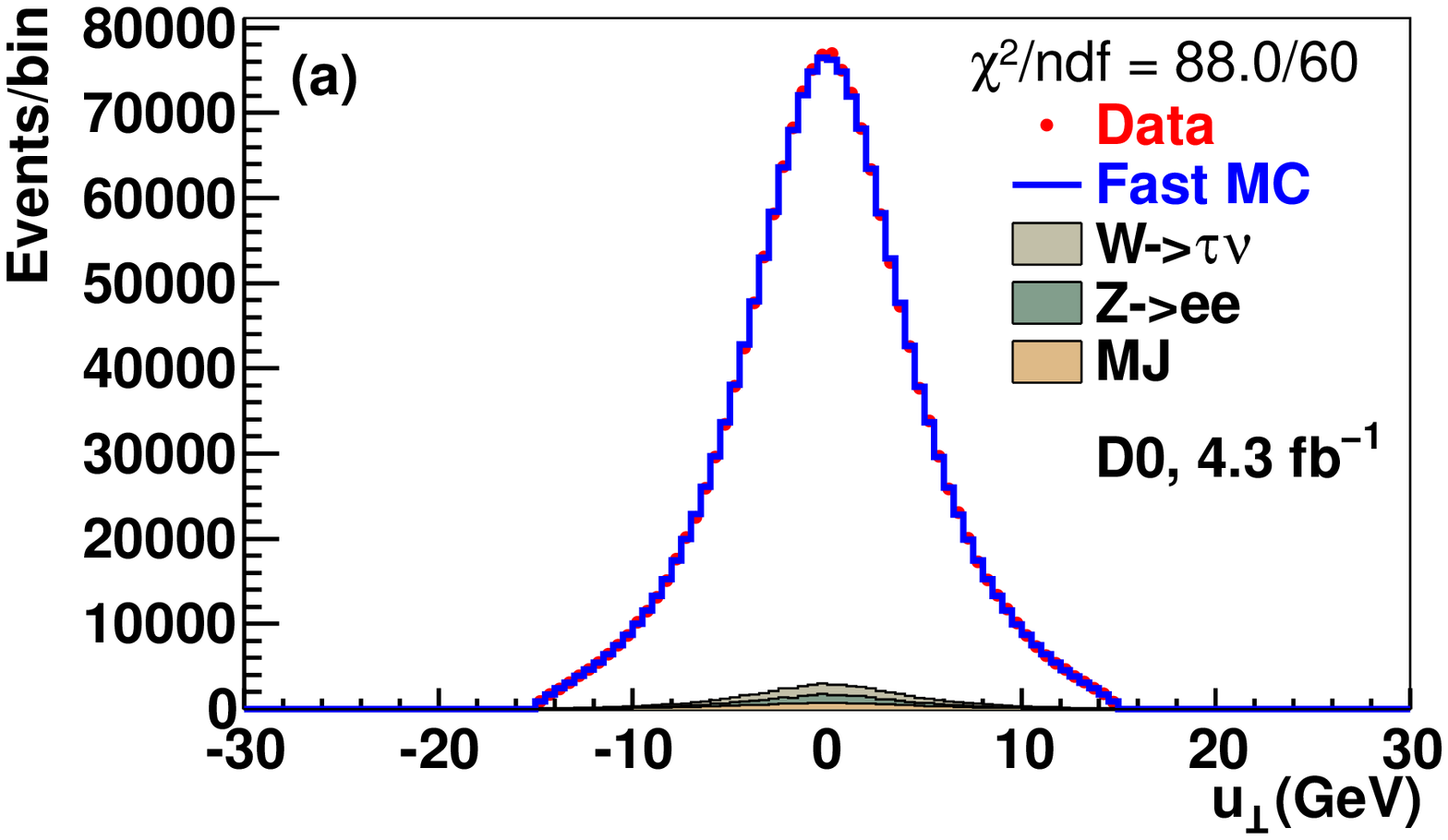}\\
\includegraphics [width=0.9\linewidth] {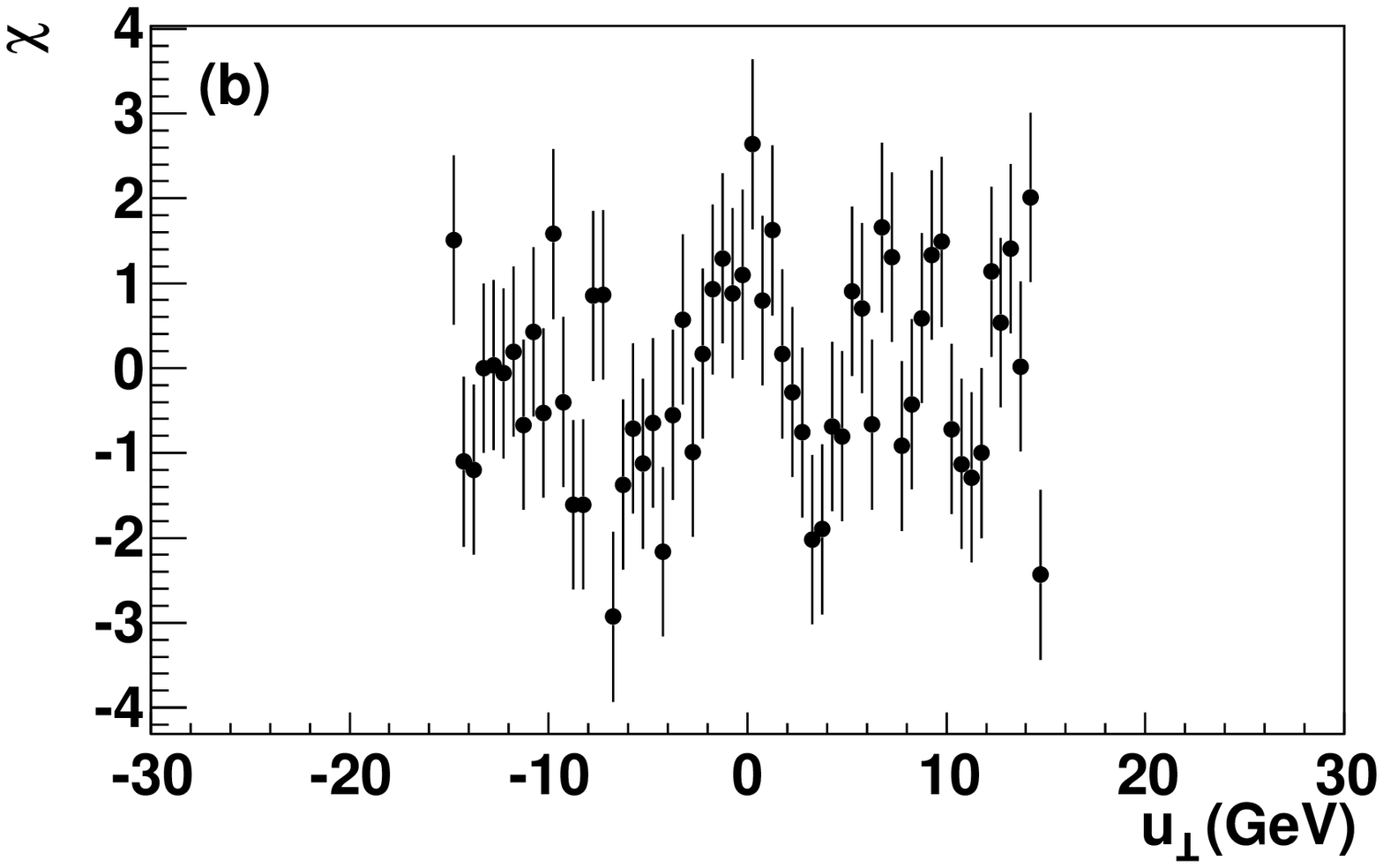}
\caption{[color online] (a) Comparison between data and fast MC for the $u_\perp$~distribution in $\wen$ data, and (b) $\chi$ value per bin, where $\chi_i = \Delta N_i/\sigma_i$. $\Delta N_i$ is the difference between the number of events for data and fast MC and $\sigma_i$ is the statistical uncertainty in bin $i$.}
\label{fig:Wuperp}
\end{figure}

\begin{figure}[htbp]
\includegraphics [width=0.9\linewidth] {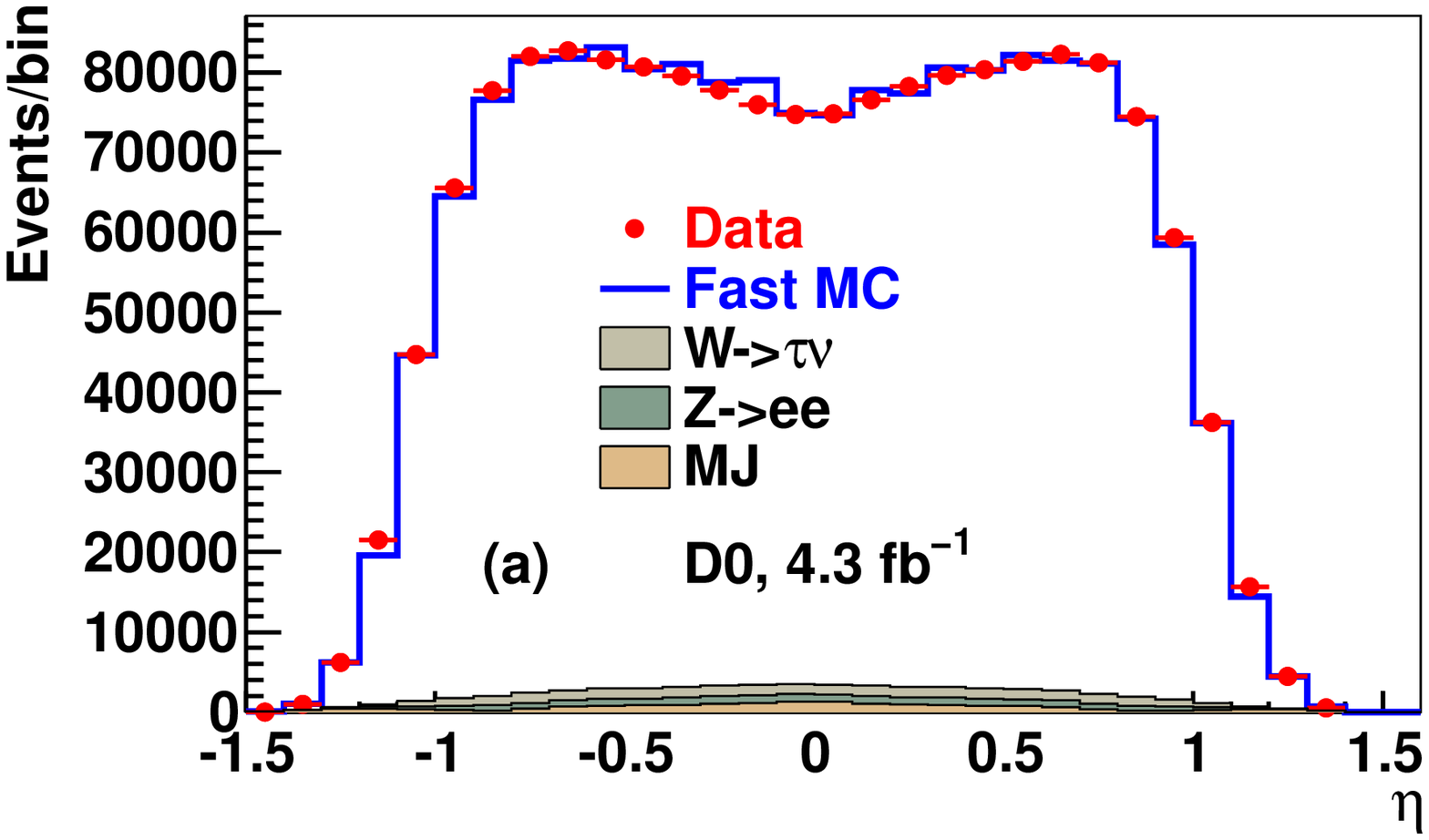}\\
\includegraphics [width=0.9\linewidth] {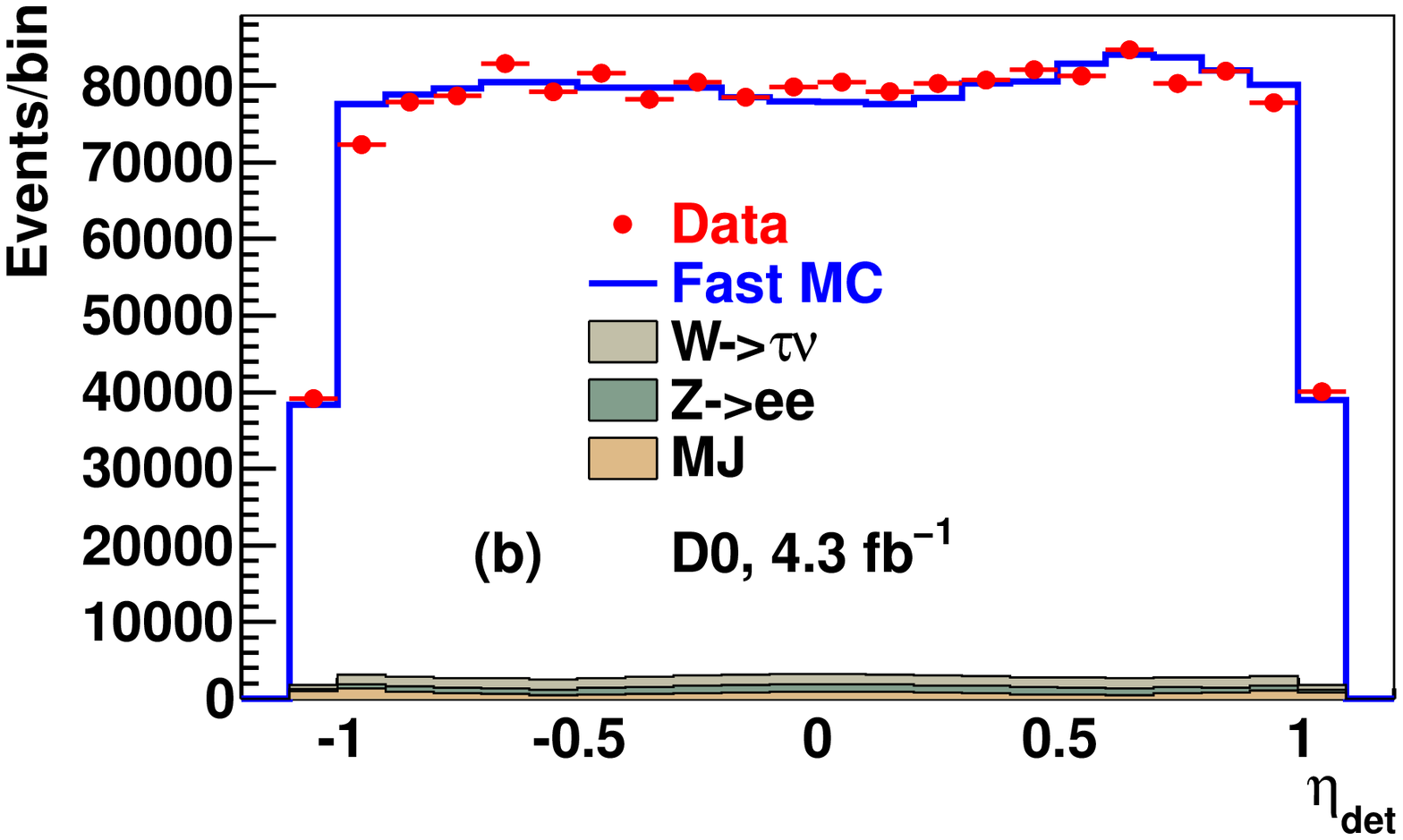}
\caption{[color online] Comparison between data and fast MC for the (a) $\eta$ distribution of electrons from $\wen$, and (b) $\eta_{\text{det}}$ distribution of electrons from $\wen$.}
\label{fig:Weta}
\end{figure}

\begin{figure}[htbp]
\includegraphics [width=0.9\linewidth] {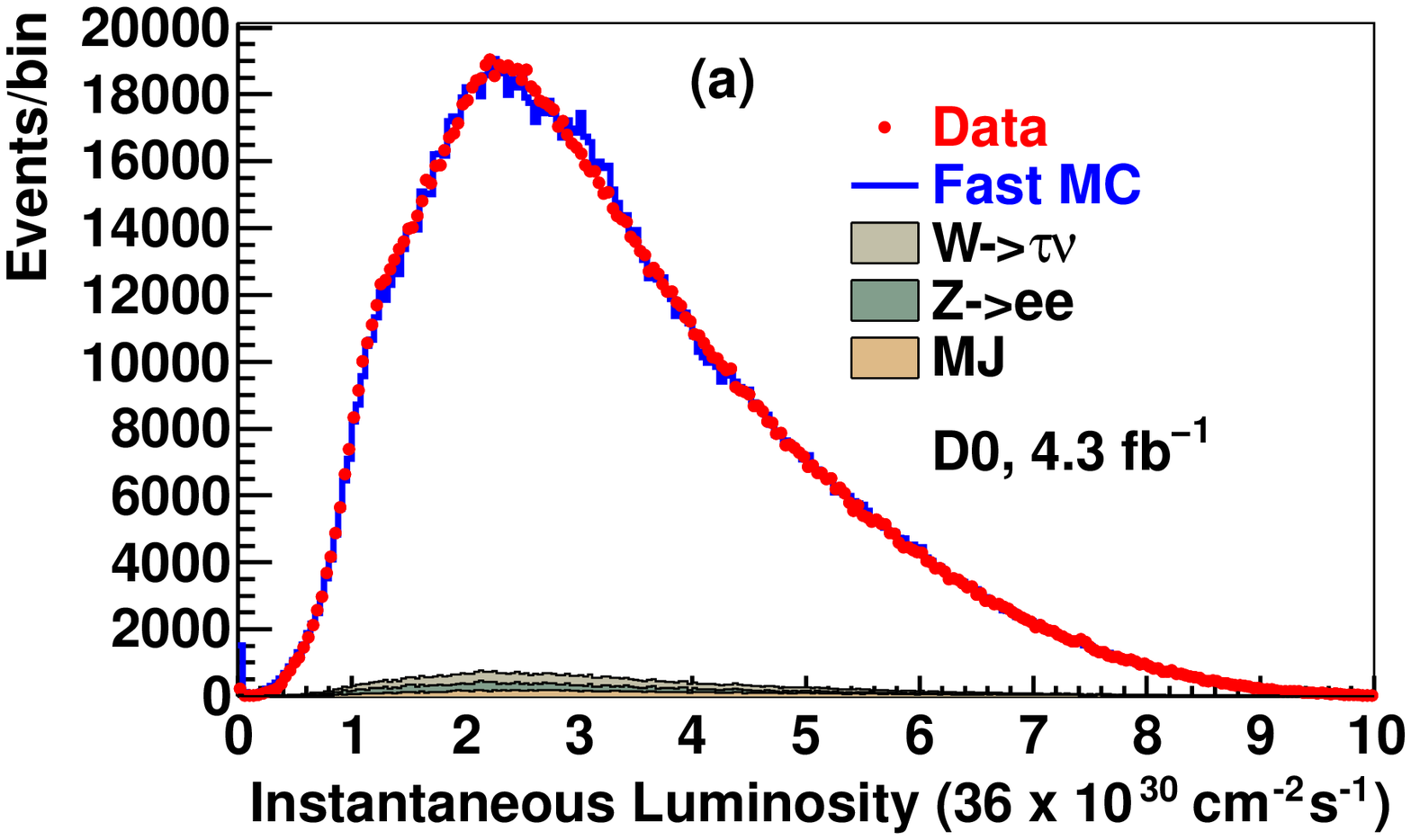}\\
\includegraphics [width=0.9\linewidth] {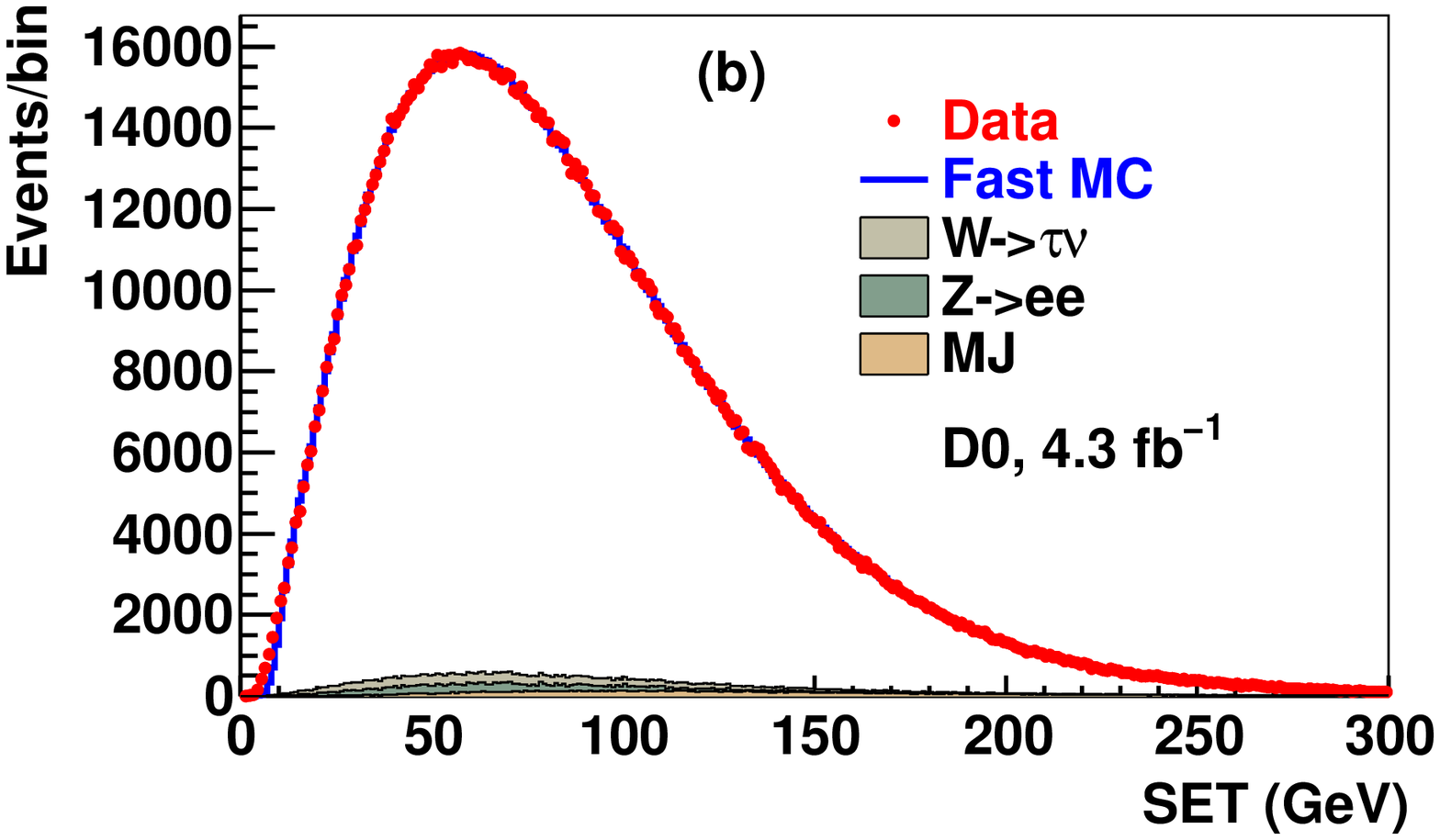}
\caption{[color online] Comparison between data and fast MC for the (a) instantaneous luminosity distribution of $\wen$ events, and (b) SET distribution of $\wen$ events.}
\label{fig:Wlumiset}
\end{figure}

The agreement between data and fast MC for the ($m_T$,$p^e_T$,\ $\met$) distributions, which are used to measure $M_W$, are shown in Fig.~\ref{fig:fits}.  In these distributions there are typically 50,000 events per 0.5 GeV bin yielding statistical uncertainty of $0.5\%$ in a bin.  We present further comparison plots of the $u_T$, $u_\parallel$, $u_\perp$, $\eta$, $\eta_{\text{det}}$, $L$, and SET distributions in $W$ events in Figs.~\ref{fig:Wut}, \ref{fig:Wupara}, \ref{fig:Wuperp}, \ref{fig:Weta}, and \ref{fig:Wlumiset}.  The agreement is not as good for some of these variables (particularly $u_T$, $L$, and SET) but satisfactory considering the statistical precision of the data and the fact that the fast MC is tuned using $Z$~boson and not $W$~boson events.  The agreement between data and fast MC is sufficient for these distributions, since we do not use them to directly measure $M_W$ and the residual disagreements have negligible impact on the measured value of $M_W$.

\newpage
\label{sec:thebib}


\begin{thebibliography}{99}
\bibitem{molierefootnote} In the standard algorithm, {\sc geant}3 automatically sets the lower bound of step sizes when considering Moli\`ere scattering. For low energy electrons, the value chosen results in steps which are sufficiently large to invalidate the Moli\`ere multiple scattering theory used by {\sc geant}3. After turning off the \texttt{AUTO} option in \textsc{geant}3, we set the following parameter values: \texttt{DCUTE} = 10~keV, \texttt{DEEMAX} = 0.005, \texttt{STMAX} = 0.1~cm, \texttt{STMIN} = $10^{-7}$~cm, and \texttt{EPSIL} = $10^{-3}$~cm.
\bibitem{slowgeantfootnote} A consequence of this set of changes is a factor of eighty increase in the processing time for {\sc geant}3 event simulation.
\end{thebibliography}
\end{document}